\RequirePackage{pdf14}
\pdfoutput=1
\documentclass[11pt]{article}
\usepackage[usenames,dvipsnames,table]{xcolor}
\usepackage[utf8]{inputenc}
\usepackage{graphicx}
\usepackage{caption}
\usepackage{subcaption}
\usepackage{bbold,array,setspace,mathrsfs,amsthm,amsfonts,amsmath,yfonts,dsfont,bbm,colonequals,tikz}
\usepackage{./aux/jheppub}
\usepackage{afterpage}
\usepackage{xspace}
\usepackage{braket}
\usepackage[normalem]{ulem}
\usepackage{multirow}
\usepackage{multicol}
\usepackage{bigstrut}

\newtheorem*{result*}{Result}

\newtheorem*{conj*}{Conjecture}
\newtheorem{claim}{Claim}
\newtheorem*{claim*}{Claim}

\def\ii{\mathrm{i}}
\def\asecond{{a^{\circlearrowleft/\circlearrowright}}}
\def\Csecond{{\CC^{\circlearrowleft/\circlearrowright}}}
\def\dDisc{\mathrm{dDisc}}
\def\zb{{\bar{z}}}
\def\so{{\mathfrak{so}}}
\def\hb{{\bar{h}}}

\def\AA{{\mathcal{A}}}
\def\BB{{\mathcal{B}}}
\def\DD{{\mathcal{D}}}
\def\CC{{\mathcal{C}}}

\def\FF{{\mathcal{F}}}

\def\NN{{\mathcal{N}}}
\def\GG{{\mathcal{G}}}
\def\LL{{\mathcal{L}}}
\def\XX{{\mathcal{X}}}
\let\a=\alpha \let\b=\beta  
\let\d=\delta 
    \let\k=\kappa
\let\l=\lambda    
     
  \let\D=\Delta  
    
    \let\G=\Gamma
\let\del=\partial

\def\eg{\emph{e.g.}}
\newcommand{\be}{\begin{equation}}
\newcommand{\ee}{\end{equation}}
\newcommand{\bea}{\begin{eqnarray}}
\newcommand{\eea}{\end{eqnarray}}
\newcommand{\nn}{\nonumber}

\newcommand{\scp}{\text{s.c.p.}}

\title{Regge trajectories for the $(2,0)$ theories}

\author[a]{Madalena Lemos,}
\author[b]{Balt C. van Rees,}
\author[b]{Xiang Zhao}

\affiliation[a]{Department of Mathematical Sciences, Durham University, Lower Mountjoy, DH1 3LE Durham, United Kingdom.}
\affiliation[b]{CPHT, CNRS, Ecole Polytechnique, Institut Polytechnique de Paris, Route de Saclay, 91128 Palaiseau, France.}

\emailAdd{madalena.lemos@durham.ac.uk}
\emailAdd{balt.van-rees@polytechnique.edu}
\emailAdd{xiang.zhao@polytechnique.edu}

\preprint{}

\bigskip
\abstract{
We investigate the structure of conformal Regge trajectories for the maximally supersymmetric $(2,0)$ theories in six dimensions. The different conformal multiplets in a single superconformal multiplet must all have similarly-shaped Regge trajectories. We show that these super-descendant trajectories interact in interesting ways, leading to new constraints on their shape. For the four-point function of the stress tensor multiplet supersymmetry also softens the Regge behavior in some channels, and consequently we observe that `analyticity in spin' holds for all spins greater than $-3$. All the physical operators in this correlator therefore lie on Regge trajectories and we describe an iterative scheme where the Lorentzian inversion formula can be used to bootstrap the four-point function. Some numerical experiments yield promising results, with OPE data approaching the numerical bootstrap results for all theories with rank greater than one.
}


\begin{document}
\setcounter{tocdepth}{2}
\maketitle
\setcounter{page}{1}

\section{Introduction}
\label{sec:intro}
The advent of Regge theory in the 1960s led to a profound improvement in our understanding of relativistic scattering amplitudes, relating in particular their high-energy behavior to the spectrum of resonances and bound states. Holography led to expectations that a similar structure should exist in conformal field theories (CFTs) \cite{Brower:2006ea,Cornalba:2007fs,Cornalba:2008qf,Costa:2012cb}, but it was only recently that these ideas to become formalized non-perturbatively in a seminal paper by Caron-Huot \cite{Caron-Huot:2017vep}. 

The results of \cite{Caron-Huot:2017vep} indicate that a CFT spectrum organizes itself in \emph{Regge trajectories} with spectra and OPE coefficients that are smooth functions of the spin $\ell$. This picture elucidates the remarkable smoothness of numerically obtained OPE data, for example that of the three-dimensional Ising model analyzed in \cite{Alday:2015ewa,Simmons-Duffin:2016wlq}, and goes some way towards explaining the success of large spin perturbation theory \cite{Fitzpatrick:2012yx,Komargodski:2012ek}, see for example \cite{Albayrak:2019gnz,Liu:2020tpf}. As has become customary in the literature, we will use the expression ``analyticity in spin'' (of the OPE data) to refer to this circle of ideas.

One important caveat concerns the behavior for the lowest possible spins. For spins below some critical value $\ell^\star$ it becomes much harder to use analyticity in spin to extract concrete information about the spectrum of the theory, see for example \cite{Caron-Huot:2020ouj,Liu:2020tpf} for some attempts for the three-dimensional Ising and $O(2)$ models. According to the analysis in \cite{Caron-Huot:2017vep} the exact value of $\ell^*$  is related to the behavior of the correlation function in the Regge limit. For a generic unitary CFT, it was deduced in \cite{Caron-Huot:2017vep} that $\ell^* \leqslant 1$ because its (suitably normalized) correlation functions are bounded by a constant in the Regge limit. A priori the spin 1 and spin 0 OPE data therefore do not need to smoothly connect to the higher spin OPE data.

In supersymmetric theories everything is better, and it is therefore worthwhile to investigate how analyticity in spin for conformal theories combines with supersymmetry. A first positive sign is that the analyticity in spin can extend also to lower spins in the theory, simply because a superconformal primary of spin $\ell \leqslant \ell^*$ can have conformal primary descendants of spin $\ell > \ell^*$ and OPE coefficients related by supersymmetry. Since the scaling dimensions of these descendants are simply integer-shifted compared to that of the primary, and the coefficients of each descendant conformal block are often equal to that of the primary times a simple rational function of $\Delta$ and $\ell$, analyticity in spin of such a non-vanishing descendant trajectory would imply analyticity in spin of the primary trajectory\footnote{We caution the reader that this does not automatically ensure analyticity in spin of the full OPE data, \emph{i.e.} including that of descendants, below $\ell^*$. We will provide concrete counterexamples below.} also below $\ell^*$! %

In this paper we undertake a study of the non-perturbative implications of analyticity in spin for supersymmetric conformal field theories (SCFTs). We have chosen to focus on the six-dimensional $(2,0)$ theories, but at a qualitative level our results certainly extend to theories in lower dimensions and likely also to theories with less supersymmetry -- we comment on this further in section~\ref{sec:conc}.\footnote{At a perturbative level the Lorentzian inversion formula of \cite{Caron-Huot:2017vep} has already been widely used for superconformal theories. For the $(2,0)$ theories there are for large $c$ computations in \cite{Alday:2020tgi}, while $\NN=4$ SYM was studied in an expansion in $1/c$ in \cite{Alday:2017vkk,Caron-Huot:2018kta}. An approximate spectrum for large, but finite $\ell$ was also studied through the inversion formula for certain $\mathcal N= 2$ theories in \cite{Cornagliotto:2017snu}.} The particular four-point function we analyze is that of the superconformal primary of the stress tensor multiplet. This is the same four-point function as was analyzed holographically in \cite{Arutyunov:2002ff,Heslop:2004du}, and with numerical bootstrap methods in \cite{Beem:2015aoa}. Perturbatively, in a large $c$ expansion, this correlator was also studied in \cite{Alday:2020tgi}, taking as input the tree-level results of \cite{Rastelli:2017ymc,Zhou:2017zaw}; and in \cite{Heslop:2017sco,Chester:2018dga,Abl:2019jhh}.
The form of the superconformal Ward identities and superconformal blocks can be extracted from the more general analysis of \cite{Dolan:2004mu}. 

\subsection{Summary of results}

\begin{figure}
\begin{center} 
\includegraphics[width=15cm]{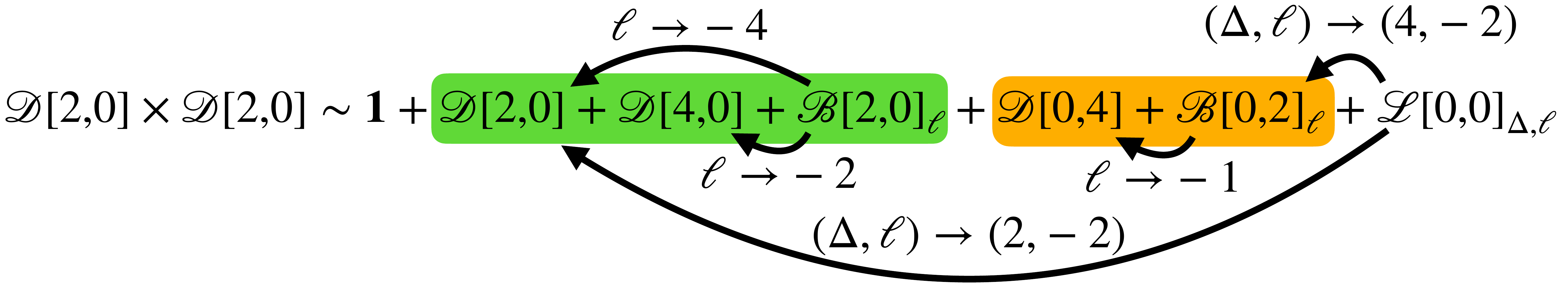}
\caption{\label{fig:opedec}The self-OPE of the stress tensor multiplet. See the main text for further explanations.}
\end{center}
\end{figure}

Sections \ref{sec:reggetrajectoriesAR} and \ref{sec:numerics} contain our key results. The results of section \ref{sec:reggetrajectoriesAR} are summarized in figure \ref{fig:opedec}. Unfortunately its explanation requires a minimal understanding of the rather technical superconformal block decomposition. We review this in section \ref{sec:4ptfn} and provide the essentials in the next paragraph.

We follow the notation of \cite{Beem:2015aoa} and denote superconformal multiplets as $\mathcal{X}[p,q]_{\Delta,\ell}$ with $(\Delta,\ell)$ and $[p,q]$ respectively corresponding to the conformal representation and the $\mathfrak{so}(5)$ R-symmetry Dynkin labels of the superconformal primary, and with $\mathcal{X} \in \{\LL,\AA,\BB,\CC,\DD\}$ denoting the type of shortening condition. Long multiplets are denoted $\LL$ and maximally short (half-BPS) multiplets are denoted $\DD$. For short multiplets we do not write $\Delta$ because it is fixed by the other quantum numbers, and similarly we omit $\ell$ for the $\DD$-type multiplets because it is always zero. Using this language the stress tensor multiplet is known to be a $\DD[2,0]$ multiplet and in its self-OPE we find the six non-trivial multiplets shown in figure \ref{fig:opedec}. (The Ward identities also allow for $\BB[0,0]_\ell$ multiplets but these contain higher spin currents and appear only in the free theory, as well as for $\DD[2,2]$ and $\DD[0,2]$ multiplets which are excluded by Bose symmetry.) Out of the five types of short multiplets there are three, shaded in green, whose OPE coefficients are completely fixed by virtue of the chiral algebra construction of \cite{Beem:2013sza,Beem:2014kka}. For the other short multiplets, which are shaded in orange, the coefficients are generally not known. For the long multiplets we know neither the scaling dimensions nor the coefficients.

Given this rather involved structure of the superconformal OPE, it is a natural question to ask how the superconformal blocks organize themselves into Regge trajectories. The answer to this question turns out to be surprisingly involved. First of all, it is important to realize that a single Regge trajectory for superconformal primaries will induce many Regge trajectories corresponding to the conformal primary superdescendants. And because some of those descendant trajectories have higher spin, it becomes necessary to extend the superconformal trajectories to \emph{negative spin} to get a complete picture. In doing so some obvious connections appear: the results that we review in section \ref{sec:4ptfn} (in particular table \ref{tab:superblocks}) indicate that the $\DD[4,0]$ and $\DD[0,4]$ multiplets find a natural place in the (straight) Regge trajectories of the $\BB$-type multiplets: the latter as the continuation of the $\BB[0,2]_{\ell}$ multiplet to spin $-1$, and the former as the continuation of the $\BB[2,0]_\ell$ multiplet to spin $-2$. What is less obvious, however, is that the $\DD[2,0]$ multiplet itself does not fit into these trajectories. Instead, one finds that it is a combination of (i) the continuation to spin $-2$ of a long, unprotected Regge trajectory, and (ii) the continuation to spin $-4$ of the straight $\BB[2,0]_\ell$ trajectory. Finally, the continuation of the $\BB[0,2]_{\ell}$ trajectory to spin $-3$ induces unwanted (descendant) blocks which can be canceled by the continuation to spin $-2$ of another unprotected long trajectory.

Altogether, then, analyticity in spin intertwines the straight and curved Regge trajectories in intricate ways, and multiplets with protected OPE coefficients or dimensions can appear on unprotected trajectories. This results in a non-trivial interplay between supersymmetry and analyticity in spin.\footnote{A somewhat orthogonal result is presented in section \ref{subsec:shadowsy}: we find new constraints on the coefficients for the conformal block decomposition of the superconformal blocks themselves. These coefficient are of course already determined by the superconformal algebra, but our constraints arise from analyticity in spin and shadow symmetry, and we checked they are satisfied by the superblocks. Similar constraints should hold for the blocks in any SCFT as well as for the decomposition of an ordinary conformal block in lower-dimensional conformal blocks as in \cite{Hogervorst:2016hal}.}

As we also review in section \ref{sec:4ptfn}, the unfixed OPE data for the four-point function of the $\DD[2,0]$ multiplet is captured in a single function $a(z,\zb)$. This function shares many similarities with an ordinary CFT four-point function and in particular has a standard conformal block decomposition in the $s$-channel. It is therefore natural to try to apply the Lorentzian inversion formula of \cite{Caron-Huot:2017vep} directly to this function. In section \ref{sec:inversion} we will set up the inversion procedure. We analyze numerous subtleties, leading ultimately to a picture of the analytic structure of the corresponding spectral density as shown in figure \ref{fig:analyticitycdeltaj} on page \pageref{fig:analyticitycdeltaj}. In line with our previous discussion, the $\ell$ axis extends to \emph{negative spins}: in fact, the rather soft Regge behavior of $a(z,\zb)$ leads one to conclude that $\ell_* \leqslant -3$! In the figure we also observe two straight Regge trajectories corresponding to the two types of protected operators of figure \ref{fig:opedec}, and the intersection of the leading long trajectory with a short trajectory at spin $-2$ as dictated by the analysis of section \ref{sec:reggetrajectoriesAR}. 

The observation that $\ell_* < 0$ means that, unlike in non-supersymmetric theories, \emph{all} the physical supermultiplets appearing in the $\DD[2,0]$ self-OPE are expected to be reachable via Regge trajectories. This leads one to the appealing prospect that this four-point function can (at least approximately) be bootstrapped: we iteratively apply the Lorentzian inversion formula to some initial trial spectrum until we hit a fixed point. Of course, with an OPE as in figure \ref{fig:opedec} the natural trial spectrum consists of the operators fixed by the chiral algebra.

In section \ref{sec:numerics} we present the initial results of such an approach. The first `inversion' of the protected data in the $t$-channel yields an approximate $s$-channel spectrum whose long multiplets are of double-twist type, with anomalous dimensions that we estimate. We then refine our estimate by inverting the \emph{leading} Regge trajectory several times until we hit a fixed point. Although this procedure ignores the subleading Regge trajectories, for sufficiently large $c$ the resulting scaling dimensions and OPE coefficients nicely track the numerical bootstrap bounds of \cite{Beem:2015aoa} --- see the numerous figures in section \ref{sec:numerics} starting on page \pageref{fig: long dimensions}. We therefore believe that a more complete iterative scheme would converge to the same `extremal' solutions as those found with numerical bootstrap methods.
\section{The four-point function}
\label{sec:4ptfn}

We consider the four-point function of the dimension four scalar which forms the bottom component of the $\DD[2,0]$ stress tensor multiplet in the six-dimensional $(2,0)$ theories. Knowledge of this correlator allows for the computation of any four-point function involving only operators in the stress tensor superconformal multiplet \cite{Dolan:2004mu}, thus making it a natural object to study. Our conventions are exactly those of \cite{Beem:2015aoa} from which we have lifted some of the equations and to which we refer the reader for more details. The essential summary is as follows. Our scalar transforms in the 14-dimensional $[2,0]$ representation of the $\mathfrak{so}(5)$ R-symmetry algebra. Since
\be
[2,0] \otimes [2,0] = [0,0] \oplus [2,0] \oplus [0,2] \oplus [4,0] \oplus [0,4] \oplus [2,2]\,,
\label{eq:tensorprod}
\ee
the four-point function features six different R-symmetry channels $A_{R}(z,\bar z)$ with $R$ labeling the representations on the right-hand side. Our conventions for the four-point function and the R-symmetry projectors are given in appendix~\ref{app:Rsym}. As usual, conformal symmetry and the operator product expansion dictate that each of these admits a decomposition into ordinary conformal blocks:
\be
A_{R}(z,\bar z) = \sum \lambda_{R\,\D,\ell}^2 \GG_{\D}^{(\ell)}(z,\zb)\,,
\label{eq:ARblockdecompositon}
\ee
but supersymmetry imposes far stricter constraints. First of all, for this correlator, the information in all six channels is completely encoded by two functions:
\be
a(z,\zb) \qquad \text{and} \qquad h(z)\,.
\ee
As indicated, the second function is independent of $\zb$. The relation is through simple second-order derivative operators. To fix ideas let us quote them here in full \cite{Dolan:2004mu}:
\begin{eqnarray}
A_{[4,0]}(z,\bar z)&=& \frac{1}{6} u^{4} \Delta_{2}\left[ u^2 a(z,\bar z) \right]~,\nn\\
A_{[2,2]}(z,\bar z)&=& \frac{1}{2} u^{4} \Delta_{2}\left[ u(v-1) a(z,\bar z)\right]~,\nn\\
A_{[0,4]}(z,\bar z)&=& \frac{1}{6} u^{4} \Delta_{2}\left[ u(3(v+1)-u) a(z,\bar z)\right]~,\nn\\
A_{[0,2]}(z,\bar z)&=& \frac{1}{2} u^{4} \Delta_{2}\left[ (v-1)\left((v+1)-\frac{3}{7}u\right) a(z,\bar z) \right] \nn\\
&& - u^2 \left( \frac{ (z-2) z h'(z)+(\bar z-2) \bar z h'(\bar z)}{2 (z-\bar z)^2} +(z+ \bar z- z \bar z) \frac{ h(z) -  h(\bar z) }{ (z-\bar z)^3} \right)~,\nn \\
A_{[2,0]}(z,\bar z)&=& \frac{1}{2} u^{4} \Delta_{2}\left[ \left((v-1)^2 - \frac{1}{3}u(v+1)+\frac{2}{27}u^2\right) a(z,\bar z) \right] \nn\\
&& + u^2 \left(z \bar z  \frac{  h(z)-  h(\bar z)}{(z-\bar z)^3} -\frac{ z^2 h'(z)+\bar z^2 h'(\bar z)}{2 (z-\bar z)^2} \right)~,\nn\\
A_{[0,0]}(z,\bar z)&=& \frac{1}{4} u^{4} \Delta_{2}\left[ \left((v+1)^2- \frac{1}{5}(v-1)^2-\frac{3}{5}u (v+1)+\frac{3}{35}u^2\right) a(z,\bar z)\right] \nn\\
&& - u^2 \frac{ (5 (1-z) + z^2) h'(z)+(5(1-\bar z) + \bar z^2) h'(\bar z)}{5 (z-\bar z)^2}\nn\\
&& + u^2  \left(2 z \bar z +5(1- z)+ 5(1- \bar z)\right) \frac{ h(z) - h(\bar z) }{5 (z-\bar z)^3}~.
\label{eq:Ainah}
\end{eqnarray}
with the operator
\begin{equation}
\D_2 f(z,\bar z) \colonequals\left( \frac{\del^2}{\del z \del \bar z} - \frac{2}{z - \bar z} \left( \frac{\del}{\del z} - \frac{\del}{\del \bar z} \right) \right) z \bar z f(z,\bar z)~.
\label{eq:diffops}
\end{equation}
and with $u =  z \bar z$ and $v = (1-z)(1-\bar z)$ as usual.

Equation \eqref{eq:Ainah} automatically resolves all the constraints of the superconformal Ward identities and was first published in \cite{Dolan:2004mu}. The deeper reason for the appearance of a meromorphic function $h(z)$ is the existence of a chiral algebra for the six-dimensional $(2,0)$ theories \cite{Beem:2013sza,Beem:2014kka}.

\subsection{Superconformal block decomposition}
A related consequence of supersymmetry is the grouping of ordinary conformal blocks into superconformal blocks. In the OPE under consideration there can appear eight types of supermultiplets \cite{Eden:2001wg,Ferrara:2001uj,Arutyunov:2002ff,Heslop:2004du}:
\be \label{D20tensorproduct}
\DD[2,0] \times \DD[2,0] \sim {\mathbf 1} + \DD[2,0] + \DD[4,0] + \DD[0,4] + \BB[2,0]_{\ell} + \BB[0,2]_{\ell} + \BB[0,0]_{\ell} + \LL[0,0]_{\D,\ell}\,.
\ee
As indicated, some multiplets can have non-zero spin $\ell$ (which are necessarily odd for the $\BB[0,2]$ multiplets and even for the other multiplets by Bose symmetry). The last type of multiplets are the unprotected or `long' multiplets and can also have arbitrary $\Delta$ (provided it lies above the unitarity bound, $\D > \ell + 6$). The $\BB[0,0]_{\ell}$ multiplets contain higher spin currents and will therefore no longer be considered in this work. The contribution of each multiplet to the four-point function is most easily captured by stating their contribution to $a(z,\zb)$ and $h(z)$; this is given in table \ref{tab:superblocks} with the listed `atomic' contributions given by \cite{Beem:2015aoa}:
\begin{equation}
\label{eq:ahatom}
\begin{split}
a^\text{at}_{\Delta,\ell}(z,\bar z) &= \frac{4}{z^{6} \bar z^{6}(\Delta-\ell-2)(\Delta+\ell+2)} \GG_{\Delta+4}^{(\ell)} (\Delta_{12}=0,\Delta_{34}=-2;z,\bar z)\,, \\
h^{\text{at}}_\beta(z) &=  \frac{z^{\b -1}}{1 - \b} {}_2F_1[\b -1,\b;2\b,z]\,,
\end{split}
\end{equation}
where $\GG_{\Delta+4}^{(\ell)} (\Delta_{12}=0,\Delta_{34}=-2;z,\bar z)$ is an ordinary (non-supersymmetric) six-dimensional conformal block, but for a four-point function of operators with unequal scaling dimension $\Delta_{i=1,\ldots 4}$. We also introduced $\Delta_{ij} \colonequals \Delta_i - \Delta_j$. The explicit form of the block is given in \eqref{eq:6dconfblock}. We also use the notation
\be
\GG_{\D}^{(\ell)}(z,\zb)
\ee
for a block with $\Delta_{12} = \Delta_{34} = 0$. With \eqref{eq:ahatom} in hand one can verify, as was done in \cite{Beem:2015aoa}, that for each line in table \ref{tab:superblocks} the application of the operators in equation \eqref{eq:Ainah} yields a finite sum of ordinary conformal blocks with the expected quantum numbers in each of the six R-symmetry channels --- see the figures in the next section for examples.

\begin{table}[h!t]
\centering
\begin{tabular}{>$l<$ >$c<$ >$l<$ >$l<$ l}
\XX & \D &  a^\XX(z,\bar z) &  h^\XX(z) & \text{comments} \\
\hline \hline
\LL[0,0]_{\D,\ell}  & \D  & a^{\text{at}}_{\D,\ell}(z,\bar z) & 0 &  generic long multiplet, $\D > \ell + 6$\\
\BB[0,2]_{\ell -1} & \ell + 7  & a^{\text{at}}_{\ell + 6, \ell}(z,\bar z) & 0 &  $\ell > 0$\\
\DD[0,4] & 8  & a^{\text{at}}_{6,0}(z,\bar z) & 0 &   \\
\BB[2,0]_{\ell - 2} & \ell + 6 &   a^{\text{at}}_{\ell + 4, \ell}(z,\bar z) & 2^{-\ell} h^{\text{at}}_{\ell + 4}(z) & $\ell > 0$\\
\DD[4,0] & 8 &  a^{\text{at}}_{4,0}(z,\bar z) & h^{\text{at}}_{4}(z) & \\
\BB[0,0]_{\ell} & \ell + 4 & 0 &  h^{\text{at}}_{\ell + 4}(z) & higher spin currents, $\ell \geqslant 0$\\
\DD[2,0] & 4 & 0 &  h^{\text{at}}_2(z) & stress tensor multiplet\\
{\bf 1} & 0 & 0 & h^{\text{at}}_0(z) &identity
\end{tabular}
\caption{\label{tab:superblocks} Superconformal blocks contribution from all superconformal multiplets appearing in the OPE of two stress tensor multiplets. The contributions are determined from the atomic building blocks. Bose symmetry requires that $\ell$ is an even integer. Here $\D$ is the dimension of the superconformal primary.}
\end{table}

\subsection{OPE coefficients from the chiral algebra}
The chiral algebra \cite{Beem:2013sza} underlying the $(2,0)$ theories \cite{Beem:2014kka} completely fixes the function $h'(z)$ in terms of a single parameter which one may take to be the $c$ central charge.\footnote{This central charge is determined by the two-point function of the stress tensor, given, \eg, in \cite{Osborn:1993cr}. We normalize it such that a single free tensor multiplet has $c=1$, and thus it is related to the canonically normalized $C_T$ of \cite{Osborn:1993cr} by $C_T = \frac{84}{\pi^6}c$.} This fixes
\be
\label{eq:h}
h(z) = - \left(\frac{z^3}{3}-\frac{1}{z-1}-\frac{1}{(z-1)^2}-\frac{1}{3 (z-1)^3}-\frac{1}{z}\right)- \frac{8}{c} \left(z-\frac{1}{z-1}+\log (1-z)\right) + \beta\,,
\ee
where  $\beta$  is an unphysical integration constant that does not appear in the correlation function, as is clear from equation \eqref{eq:Ainah}. We fix it as  $\beta = -1/6 + 8/c$ such that the atomic decomposition of $h(z)$ reads
\be
h(z)= h^{at}_{0}(z) + \sum\limits_{\substack{\ell=-2\,,\\ \ell \text{ even}}}^\infty b_\ell \; h^{at}_{\ell+4}(z)\,,
\ee
with\footnote{Here $b_{-2}$ should be thought of as the limit of the given expression as $\ell \to - 2$, which gives $b_{-2}= 8/c$.}
\begin{eqnarray}
b_\ell&=&\frac{(\ell +1) (\ell +3) (\ell +2)^2 \frac{\ell }{2}! \left(\frac{\ell }{2}+2\right)!! \left(\frac{\ell }{2}+3\right)!! (\ell +5)!!}{18 (\ell +2)!! (2 \ell +5)!!} \nn \\
&& + \frac{8}{c}\frac{  \left(2^{-\frac{\ell }{2}-1} (\ell  (\ell +7)+11) (\ell +3)!! \Gamma \left(\frac{\ell }{2}+2\right)\right)}{(2 \ell +5)!!}\,.\nn
\end{eqnarray}
One can now use the block decomposition of $h(z)$ to also completely fix the coefficients of the $\DD[2,0]$, $\DD[4,0]$ and $\BB[2,0]_\ell$ multiplets in terms of $c$. According to table \ref{tab:superblocks} the latter two multiplets also give a contribution to $a(z,\zb)$. It is then useful to split off this `chiral' contribution and write
\be
a(z,\zb) = a^{\chi}(z,\zb) + a^u(z,\zb)\,,
\label{eq:achiplusu}
\ee
with $a^\chi(z,\zb)$ capturing the completely known contribution of the $\DD[4,0]$ and $\BB[2,0]_\ell$ multiplets,
\be
a^\chi(z,\zb) \colonequals \sum\limits_{\substack{\ell = 0\,,\\ \ell \text{ even}}}^\infty 2^\ell b_\ell \, a^{\text{at}}_{\ell + 4, \ell}(z,\zb)\,,
\label{eq:achiinblocks}
\ee
and with an `unknown' part $a^u(z,\zb)$ of the form
\be
a^u(z,\zb) = \sum_{\D \geqslant \ell + 6,\ell} \lambda_{\D,\ell}^2 a^{\text{at}}_{\D,\ell}(z,\zb)\,.
\label{eq:auinblocks}
\ee
The blocks that saturate the inequality correspond to the $\DD[0,4]$ or $\BB[0,2]_\ell$ multiplets, and all other blocks correspond to $\LL[0,0]_{\D,\ell}$ multiplets.

\subsection{Crossing symmetry equations}
The crossing symmetry equations which arise from permuting the external operators lead to a set of linear algebraic relations for the functions $A_R(z,\zb)$. Substituting \eqref{eq:Ainah} into these relations, one finds that all the derivatives can be eliminated and (assuming the above form of $h(z)$) one also finds simple algebraic crossing equations for the function $a(z,\zb)$ \cite{Beem:2015aoa}. These read:
\be
\begin{split}
a(z,\zb) &= \frac{1}{(1-z)^5(1-\zb)^5} a\left(\frac{z}{z-1},\frac{\zb}{\zb - 1}\right)\\
z \zb\, a(z,\zb) &=  (1-z)(1-\zb) a(1-z,1-\zb) + \CC_h(1-z,1-\zb) - \CC_h(z,\zb)
\label{eq:crosssym}
\end{split}
\ee
where
\be
\label{eq:Ch}
\CC_{h}(z,\bar z) = \frac{1}{(z - \bar z)^3}\frac{h\left(z\right)-h(\zb)}{z \bar{z}}\,.
\ee
The first of the crossing equations relates the $t$- and $u$-channel and is solved by demanding that the $s$-channel block decomposition (equations \eqref{eq:achiinblocks} and \eqref{eq:auinblocks}) only contains even spin operators. The second one is less trivial as it relates the block decompositions in different channels. It will be used extensively below.

\section{Regge trajectories and supersymmetry}
\label{sec:reggetrajectoriesAR}
As we mentioned in the introduction, a given superconformal multiplet contains several conformal primary operators whose Regge trajectories are naturally related by supersymmetry. It might then appear natural to focus on the Regge trajectories of the superconformal multiplets as a whole, and to ignore the trajectories of the superconformal descendants. For the $\DD[2,0]$ correlator one can do so by analyzing the Regge trajectories for all spins directly for the function $a(z,\zb)$ instead of those for the six functions $A_R(z,\zb)$. Perhaps surprisingly, doing so would only paint an incomplete picture: for several superconformal Regge trajectories the  arrangement of the conformal descendants, which is dictated by supersymmetry, is \emph{not} automatically correct. Instead, the combined demands of analyticity in spin together with supersymmetry lead to non-trivial constraints for superconformal Regge trajectories.

The aim of this section is to exhibit these constraints and argue for a particular structure of the superconformal trajectories that resolves them. We will first consider the contributions of the short multiplets, which lie on straight trajectories, and then move on to the long multiplets whose twist is not fixed by supersymmetry. The next two subsections detail the issues and are unavoidably a bit technical. The hasty reader may want to skip to subsection \ref{subsec:resolvingissues} for a summary of the issues we uncovered and their potential resolution.

All of the issues we discuss below will happen for low spins where, as mentioned in the introduction, the operators do not always manifestly lie on Regge trajectories. The threshold value $\ell_*$ is claimed to be determined by the Regge behavior of the correlation function. In section~\ref{subsec:Reggegrowth} we will show that the Regge behavior of the six functions $A_R(z,\zb)$ is actually quite a bit softer than in a non-supersymmetric theory, leading to extended  manifest analyticity in spin. Let us here already quote the upshot, which is that:
\be \label{analyticityAR}
\begin{split}
&A_{[4,0]} \text{ has analyticity in spin for }  \ell > -3\,, \\
&A_{[2,2]} \text{ has analyticity in spin for }  \ell > -2\,, \\
&A_{[2,0]} \text{ has analyticity in spin for }  \ell > -1 \,, \\
&A_{[0,4]} \text{ has analyticity in spin for }  \ell > -1\,, \\
&A_{[0,2]} \text{ has analyticity in spin for }  \ell >0\,, \\
&A_{[0,0]} \text{ has analyticity in spin for }  \ell > 1\,.
\end{split}
\ee
If these inequalities are violated then the shape of the Regge trajectories for the six functions under consideration is less tractable, and a priori we cannot even exclude isolated contributions.\footnote{The appearance of isolated contributions (besides the identity operator) would not be generic, but in other inversion formulas they have nevertheless been shown to make an appearance \cite{Lemos:2017vnx,Iliesiu:2018fao}.} However, as long as these inequalities are obeyed we can safely draw conclusions on the shape of the Regge trajectories for the six functions under consideration.

\subsection{Short multiplets and straight trajectories}
\label{subsec:shortstrajectories}

\subsubsection*{Short multiplets with unknown coefficients}
\begin{figure}[t]
\begin{center}
\vskip .1cm
\includegraphics[width=\textwidth]{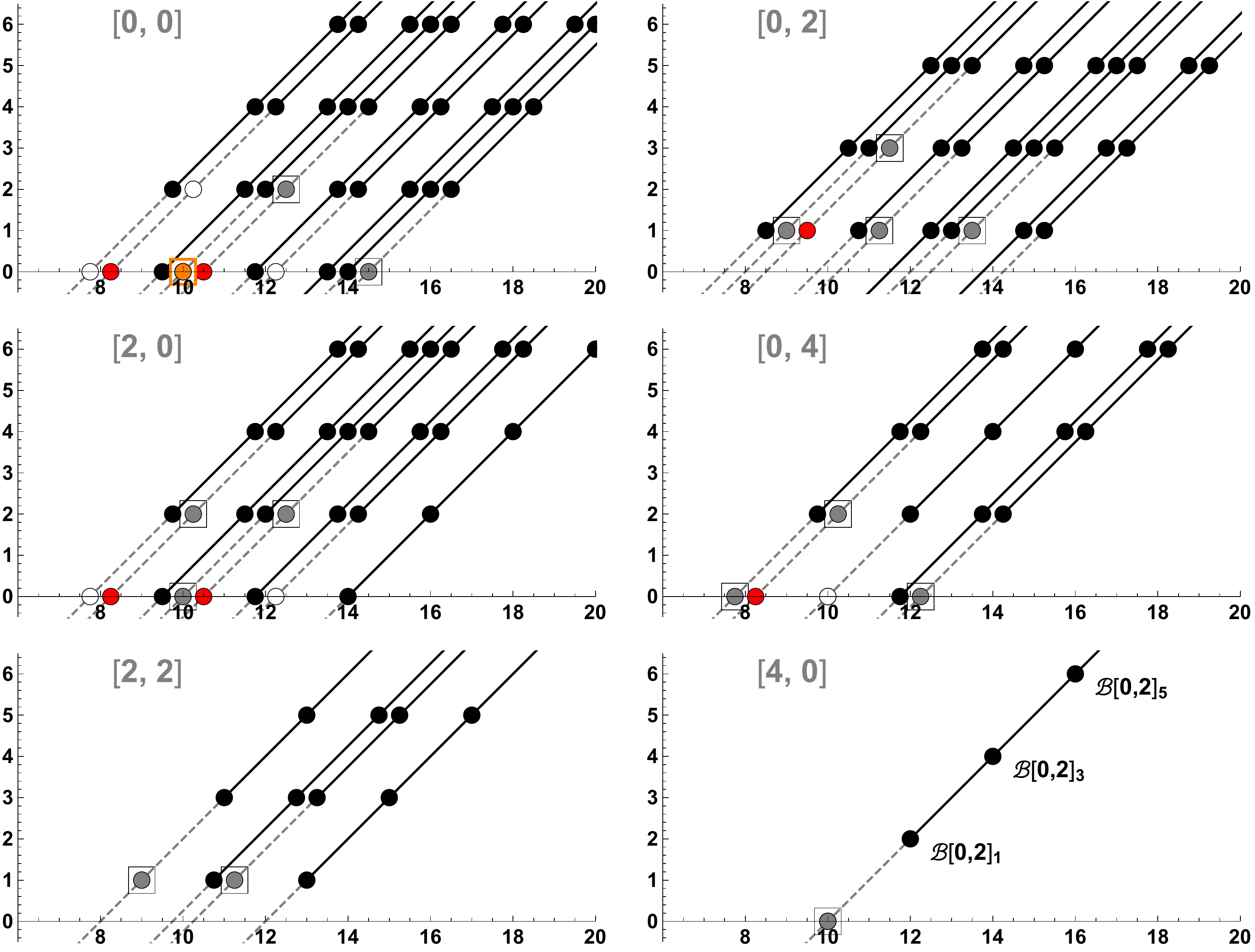}
\caption{$(\D,\ell)$ planes: $\BB[0,2]_{\ell - 1}$ trajectories (lines and dots) and $\DD[0,4]$ operators (squares)}
\label{figure:short at bound trajectories}
\end{center}
\end{figure}

We start our exploration with the $\BB[0,2]_{\ell -1}$ multiplets (for even $\ell > 0$), which are short multiplets whose coefficients are not fixed by the chiral algebra. Since their dimensions are fixed, they lie on straight Regge trajectories. We begin by plotting the conformal primary descendants and their trajectories for a few low-lying spins for each of the R-symmetry channels. This yields the black dots and solid black lines in figure \ref{figure:short at bound trajectories}.

Note that we have added a small horizontal split to overlapping Regge trajectories for presentational purposes: the twist of all trajectories are really even integers. For space reasons we have only added a (partial) legend to the $[4,0]$ channel: the black half-line connects the $\BB[0,2]_{1}$, $\BB[0,2]_{3}$, $\BB[0,2]_{5}$, \ldots multiplets, and the corresponding primary operators are always the first three black dots on this half-line. (In other words, in each $R$-symmetry channel the operator content of the $\BB[0,2]_{1}$ multiplet is given by those black dots that lie on the endpoint of a black half-line, and so on for the remaining multiplets.)

The most interesting feature of these trajectories now follows from the dashed lines, which indicate the continuation of this trajectory to unphysical spins of the superconformal primary. We are forced to draw this continuation because analyticity in spin dictates that Regge trajectories cannot just end in the middle of the $(\D,\ell)$ plane. After going down two units in spin along this dashed line we find the hypothetical ``$\BB[0,2]_{-1}$'' multiplet and going down four units yields the ``$\BB[0,3]_{-3}$'' multiplet. Neither of these multiplets exist, so what are we to make of them?

Let us first consider the ``$\BB[0,2]_{-1}$'' multiplet. In the plots we show gray, orange and white dots at the putative locations of its conformal primary operators. The white dots are easiest to explain: they correspond to the absence of an operator due to a kinematical zero. For example, the twist ten trajectory in the $[0,4]$ channel gives a contribution of the form
\be
\lambda^2_{\BB[0,2]_{\ell - 1}} \frac{\ell (4 + \ell) (5 + \ell) (9 + \ell)}{12 (1 + \ell) (3 + \ell) (6 + \ell) (8 + \ell)} \mathcal{G}_{10 + \ell}^{(\ell)}(z,\zb)\,,
\ee
with $\lambda^2_{\BB[0,2]_{\ell - 1}} $ the contribution of this superconformal multiplet to \eqref{eq:auinblocks}. We see that this contribution vanishes at $\ell = 0$ and this explains the corresponding white dot. The orange and the gray dots are then actual conformal primary blocks that, barring a dynamical zero in the overall OPE coefficient, need to be accommodated by other supermultiplets.

As the reader might have expected from table \ref{tab:superblocks}, the multiplet that comes to the rescue is the $\DD[0,4]$ multiplet. The operator contents of the latter is indicated by the squares in the figure and we see a nice one-to-one match between these and the gray and orange dots of the ``$\BB[0,2]_{-1}$'' multiplet. Although this is not obvious from the figure, \emph{almost} all the coefficients work out as well. For example, in the $[2,0]$ channel there is a twist 8 trajectory of the form:
\be
\lambda^2_{\BB[0,2]_{\ell - 1}} \frac{(\ell+5)^2 (\ell+9)}{3 (\ell+8) (2 \ell+9) (2 \ell+11)} \mathcal{G}_{10 + \ell}^{(2 + \ell)} (z,\zb)\,,
\ee
and in the limit $\ell \to 0$ this precisely matches a contribution to the $\DD[0,4]$ multiplet of the form:
\be
\lambda^2_{\DD[0,4]} \frac{25}{3 \cdot 8 \cdot 11} \mathcal{G}_{10}^{(2)}(z,\zb)\,.
\ee
Similarly we find that the coefficients agree for all the other gray dot/square combinations. It is therefore natural to postulate that
\be \label{B02becomesD04}
\lim_{\ell \to -1} \lambda^2_{\BB[0,2]_{\ell}} = \lambda^2_{\DD[0,4]}\,,
\ee
where taking the limit of course only makes sense because of analyticity in spin.

Surprisingly, assuming equation \eqref{B02becomesD04} does not resolve everything: there is a strange mismatch which occurs for the scalar of dimension 10 in the $[0,0]$ channel. In the $\BB[0,2]_{\ell -1}$ supertrajectory we obtain:
\be
\lim_{\ell \to 0} \lambda^2_{\BB[0,2]_{\ell-1}} \frac{3 (\ell+4) (\ell+5) (\ell (\ell+9) (4 \ell (\ell+9)+59)-360)}{280 (\ell+1) (\ell+3) (\ell+6) (\ell+8) (2 \ell+7) (2 \ell+11)} \mathcal{G}_{10 + \ell}^{(\ell)} (z,\zb) =  -\frac{15}{2156} \lambda^2_{\DD[0,4]} \mathcal{G}_{10}^{(0)} (z,\zb)\,,
\ee
where we used equation \eqref{B02becomesD04}. Although the $\DD[0,4]$ multiplet should indeed have a block with these quantum numbers, its coefficient is different and should actually be:
\be
\frac{1}{308} \lambda^2_{\DD[0,4]} \mathcal{G}_{10}^{(0)}(z,\zb)\,.
\ee
This mismatch is why we colored this combination orange rather than gray in the figure.\footnote{\label{foot:nonanal}One may wonder how the issue arises, given that everything is clearly analytic in spin at the level of the function $a(z,\zb)$. The problem is that the $\BB[0,2]_{\ell -1}$ multiplet induces a block $\mathcal{G}^{(\ell - 4)}_{\ell + 10}(z,\zb)$ in the $[0,0]$ channel which has negative spin for $\ell = 0,2$. As we send $\ell \to 2$ this becomes a spin $-2$ block, which one may check vanishes identically in six dimensions and therefore does not contribute. But for $\ell \to 0$ we find a spin $-4$ block, which happens to be equal to ($16/3$ times) a spin 0 block of the same dimension. This yields an additional contribution to the coefficient of the spin 0 block of dimension 10 in the $\DD[0,4]$ multiplet and causes the non-analyticity in spin.} Let us call this {\bf issue 1}. We note that this issue is only present if we have analyticity down to spin 0 in the $[0,0]$ R-symmetry channel, which does not rigorously follow from the Regge limit analysis summarized in \eqref{analyticityAR}.

Let us now go down two more units in spin and investigate the ``$\BB[0,2]_{-3}$'' multiplet which is indicated by the red dots in the figure. (Perhaps surprisingly, there is no kinematical zero that prevents any of these operators from appearing.) According to \eqref{analyticityAR} we expect to have analyticity of the CFT data at all the red points except the scalars in the $[0,0]$ channel, but these blocks do not correspond to the operator content of any other superconformal multiplet. They are therefore genuinely unwanted contributions, and the fact that they do not vanish automatically is what we call {\bf issue 2}. Of course, this issue would be resolved immediately if the overall coefficient vanishes, so if $\lim_{\ell \to - 2} \lambda^2_{\BB[0,2]_{\ell - 1}} = 0$, but we emphasize that this would be a \emph{dynamical} constraint. We will find another potential dynamical resolution below.

\subsubsection*{Short multiplets with known coefficients}
\begin{figure}[t]
\begin{center}
\vskip .1cm
\includegraphics[width=\textwidth]{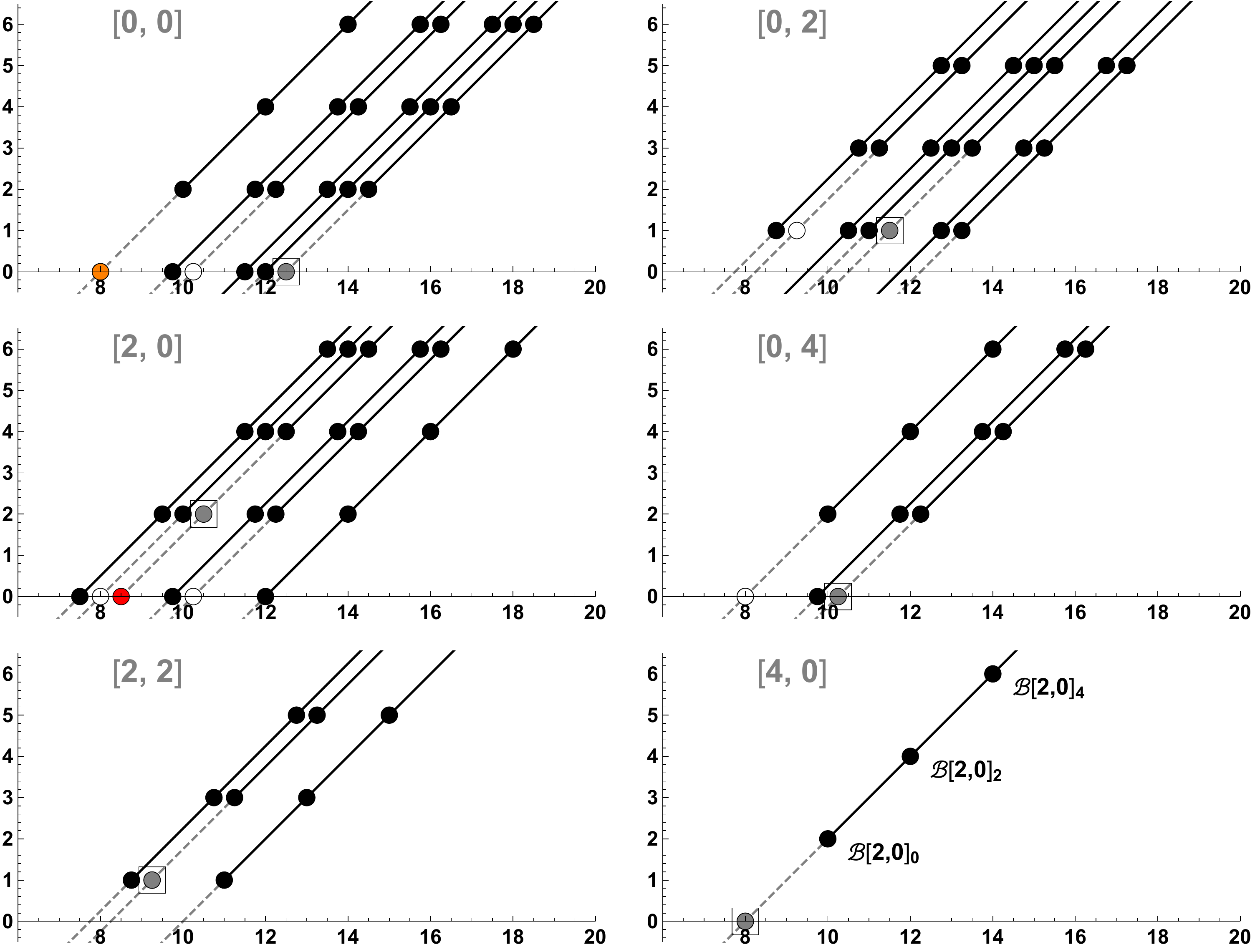}
\caption{$(\D,\ell)$ planes: $\BB[2,0]_{\ell - 2}$ trajectories (lines and dots) and $\DD[4,0]$ operators (squares)}
\label{figure:chrial algebra short trajectories}
\end{center}
\end{figure}

Our next set of multiplets contains short multiplets with a contribution to the chiral algebra: the $\BB[2,0]_{\ell - 2}$ and the $\DD[4,0]$ multiplets. We can repeat the above analysis for these operators, but with the one change that here the OPE coefficients are completely fixed:
\be \label{B20becomesD40}
\lambda^2_{\BB[2,0]_{\ell - 2}} = 2^\ell b_\ell, \qquad \qquad \lambda^2_{\DD[4,0]} = \lim_{\ell \to 0} \lambda^2_{\BB[2,0]_{\ell - 2}} = 2 b_0\,.
\ee
and therefore the analogue of equation \eqref{B02becomesD04} is manifestly true. For the six functions $A_R(z,\zb)$ these multiplets give the picture shown in figure \ref{figure:chrial algebra short trajectories}.

The white and gray dots now indicate the operator content of a ``$\BB[2,0]_{-2}$'' multiplet and, as expected, match almost perfectly with the squares that correspond to a $\DD[4,0]$ multiplet. Yet there is again one exception: the scalar of dimension 8 in the $[0,0]$ channel. For this operator we have
\be
\lim_{\ell \to 0} \lambda^2_{\BB[2,0]_{\ell - 2}}\frac{9 (\ell-1) (\ell+8)}{1400 (\ell+1) (\ell+6)} \mathcal{G}_{8+\ell}^{(\ell)}(z,\zb) = - \frac{3}{350} \lambda^2_{\DD[4,0]} \mathcal{G}_8^{(0)}(z,\zb)\,,
\ee
whereas the $\DD[4,0]$ multiplet does not have such a scalar. We call this {\bf issue 3}, although we must note once again that analyticity is not guaranteed from \eqref{analyticityAR} for spin zero. 

We have two more issues to discuss. First, as for the previous figure, the further continuation down to a ``$\BB[2,0]_{- 4}$'' multiplet induces another unwanted scalar operator of dimension 8 in the $[2,0]$ channel with coefficient
\be
\lim_{\ell \to - 2} \lambda^2_{\BB[2,0]_{\ell - 2}} \frac{(\ell+4)^2 (\ell+5)^2 (\ell+8)}{8 (\ell+6) (2 \ell+7) (2 \ell+9)^2 (2 \ell+11)} \mathcal{G}_{10+\ell}^{(2+ \ell)}(z,\zb) = \lambda^2_{\BB[2,0]_{- 4}} \frac{9}{700} \mathcal{G}_8^{(0)}(z,\zb)\,.
\ee
This isolated operator once more cannot fit in a superconformal multiplet and so the block must somehow cancel: this is {\bf issue 4}. Notice that in this case the OPE coefficient is known,
\be
\lambda^2_{\BB[2,0]_{- 4}}  = \frac{1}{4} b_{-2} = \frac{2}{c}\,.
\ee
The putative resolution of issue 2 therefore cannot work in this case, since $2/c > 0$ in all but the generalized free theory.

The last issue for this set of blocks is the $\DD[2,0]$ multiplet itself. It contributes three conformal blocks:
\be
\begin{split}
[0,0]&: \frac{6}{175} \mathcal{G}^{(2)}_6(z,\zb)\,,\\
[2,0]&: \frac{1}{2} \mathcal{G}^{(0)}_4(z,\zb)\,,\\
[0,2]&: \frac{1}{5} \mathcal{G}^{(1)}_5(z,\zb)\,,
\end{split}
\ee
which we recognize as the contributions from the stress tensor, the superconformal primary, and the R-symmetry current. This block has an OPE coefficient equal to $b_{-2} =8/c$ as determined by the chiral algebra and so is present in any bona fide theory. It is however clear that neither of these three operators fits into any of the Regge trajectories we have drawn so far, and this is {\bf issue 5}.

\subsection{Long multiplets}
\begin{figure}[t]
\begin{center}
\vskip .1cm
\includegraphics[width=\textwidth]{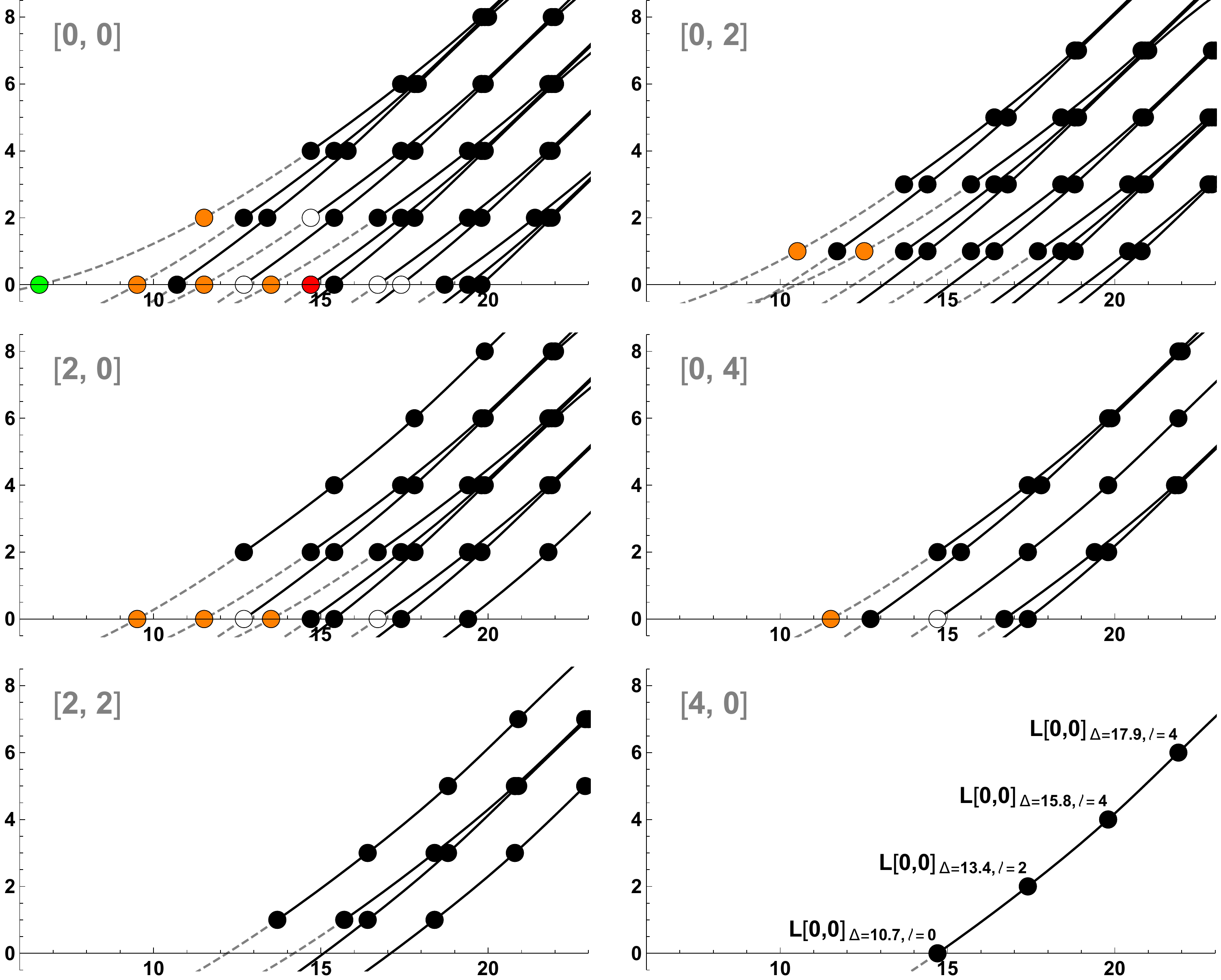}
\caption{$(\D,\ell)$ planes: a randomly chosen $\LL[0,0]_{\Delta,\ell}$ trajectory extending down to spin $-4$.}
\label{figure:generic long trajectories}
\end{center}
\end{figure}

Now let us analyze the Regge trajectories for the long multiplets $\LL[0,0]_{\Delta,\ell}$. For these multiplets the scaling dimensions are unknown, and likewise there is no a priori proof that the Regge trajectories $\ell(\Delta)$ have a particularly simple form. We can nevertheless explore the consequences if we \emph{assume} that there exists a trajectory that extends to low spins. Just as for the protected multiplets, we will find issues corresponding to non-analyticity in the induced trajectories for the $A_R(z,\zb)$ as soon as any Regge trajectory for a supermultiplet crosses the lines with $\ell = 0$, $\ell -2$, or $\ell = -4$. For efficiency of presentation we would like to capture all the issues in one plot, and to do so we simply invented an otherwise random trajectory\footnote{The sketched trajectory is not \emph{entirely} random: we picked a convex shape and an asymptotic twist of 12, meaning we can think of it as the second double-twist trajectory after the leading one (which asymptotes to twist 8).} that crosses all these three lines as shown in figure \ref{figure:generic long trajectories}. We however stress that the issues we are about to list are \emph{local}, in the sense that they hold for any trajectory crossing these low spins, and do not depend on the global shape of the trajectory. For example, in our view it would have been perfectly possible for a trajectory that intersects the $\ell = -4$ line to be disconnected from a trajectory that extends to positive spins.

The legend in the $[4,0]$ channel indicates the scaling dimension that we have chosen for the first four long multiplets on our hypothetical trajectories. (Recall that $\Delta > \ell + 6$ for an $\LL[0,0]_{\Delta,\ell}$ multiplet by unitarity which is easily obeyed here.) Unlike the previous plots there are no exactly overlapping induced trajectories for any of the $A_R(z,\zb)$ and therefore there was no need to add a horizontal split by hand.

At low spins we encounter our first issue at spin $0$. We see some kinematical zeroes, indicated by the white dots, which are analytic in spin and nothing to worry about. However the red dot indicates a problem: for a long supermultiplet with quantum numbers $(\Delta,\ell)$ and $\ell$ generic there exists a conformal primary R-symmetry singlet of the form:
\be
\lambda^2_{\LL[0,0]_{\Delta,\ell}} \xi(\ell,\Delta) \mathcal{G}_{\Delta + 4}^{(\ell)}(z,\zb)\,,
\ee
with $\xi(\ell,\Delta)$ a rational function of $\Delta$ and $\ell$ that is too ugly to include here. In the limit $\ell \to 0$ we find:
\be
\lim_{\ell \to 0} \xi(\ell,\Delta) = -\frac{3 (\Delta -2) (\Delta +4) (\Delta  (\Delta +2)-6)}{448 (\Delta -3) (\Delta +1)^2 (\Delta +5)}\,,
\label{eq:issue6get}
\ee
which is not the correct expression for a spin zero long multiplet, which instead has a contribution of the form:\footnote{The discrepancy is again due to a spin $\ell -4$ block in the supermultiplet --- see the footnote~\ref{foot:nonanal}.}
\be
\lambda^2_{\LL[0,0]_{\Delta,0}} \frac{9 (\Delta -4) (\Delta -2) (\Delta +4) (\Delta +6)}{1792 (\Delta -3) (\Delta -1) (\Delta +3) (\Delta +5)} \mathcal{G}_{\Delta + 4}^{(0)}(z,\zb)\,.
\label{eq:issue6should}
\ee
We can call this {\bf issue 6}. We stress that it exists for \emph{any} long multiplet trajectory that crosses the $\ell = 0$ line, provided analyticity holds down to spin zero in the singlet channel.\footnote{Note that while the difference between \eqref{eq:issue6get} and \eqref{eq:issue6should} vanishes for $\Delta=2$, this value is not allowed for $\ell = 0$ by unitarity.}

Going down in spin, we reach the orange points at spin $-2$, which correspond to {\bf issue 7}. Again, a ``$\LL[0,0]_{\Delta,-2}$'' multiplet does not exist and so generically the combination of orange points should not actually be present in a physical theory. Notice that we have shown the situation for generic $\Delta$; much like the function $\xi(\ell,\Delta)$ vanishes at $\Delta = 2$ 
 there are some zeroes in the coefficients for specific values of $\Delta$ and then some of the orange points may disappear. This will be important below.

Finally if a long supertrajectory hits spin $-4$ then we induce a single scalar of dimension $\Delta + 4$ in the $[0,0]$ channel. This is the green dot and {\bf issue 8}, which once more is only an issue if analyticity holds down to spin zero in this channel.

\subsection{Resolving the issues}
\label{subsec:resolvingissues}
Let us recap. The issues we have collected are:
\begin{enumerate}
	\item The limit $\ell \to 0$ of the $\BB[0,2]_{\ell -1}$ multiplet should be a $\DD[0,4]$ multiplet but gives the wrong coefficient for a dimension $10$ scalar in the $[0,0]$ channel. \label{item:issue1}
	\item The limit $\ell \to - 2$ of the $\BB[0,2]_{\ell -1}$ multiplet gives unwanted operators in several channels.\label{item:issue2}
	\item The limit $\ell \to 0$ of the $\BB[2,0]_{\ell - 2}$ multiplet should be a $\DD[4,0]$ multiplet but gives the wrong coefficient for a dimension $8$ scalar in the $[0,0]$ channel.\label{item:issue3}
	\item The limit $\ell \to -2$ of the $\BB[2,0]_{\ell -2}$ multiplet gives an unwanted dimension $8$ scalar in the $[2,0]$ channel with coefficient $9/(350 c)$.\label{item:issue4}
	\item The operators in the $\DD[2,0]$ multiplet do not fit in a short Regge trajectory.\label{item:issue 5}
	\item The limit $\ell \to 0$ of a generic $\LL[0,0]_{\Delta,\ell}$ multiplet should be an $\LL[0,0]_{\Delta,0}$ multiplet but gives the wrong coefficient for a dimension  $\Delta + 4$ scalar in the $[0,0]$ channel.\label{item:issue6}
	\item The limit $\ell \to -2$ of a generic $\LL[0,0]_{\Delta,\ell}$ multiplet gives unwanted operators in several channels.\label{item:issue7}
	\item The limit $\ell \to -4$ of a generic $\LL[0,0]_{\Delta,\ell}$ multiplet gives an unwanted scalar of dimension $\Delta + 4$ in the $[0,0]$ channel.\label{item:issue8}
\end{enumerate}

We see that issues \ref{item:issue1}, \ref{item:issue3}, \ref{item:issue6} and \ref{item:issue8} all pertain only to scalars in the R-symmetry singlet channel. It is not entirely clear that we need to take them seriously: according to equation \eqref{analyticityAR} analyticity in spin for $A_{[0,0]}(z,\zb)$ is guaranteed only down to spin $\ell>1$. Although this would be one way to resolve the issues (or at least provide us with a license to ignore them), there is another option: issue \ref{item:issue8} has the potential to resolve the other 3 issues. More precisely we simply postulate the existence of otherwise unknown unprotected Regge trajectories that cross the $\ell = -4$ line at $\Delta = 6$ (to resolve issue \ref{item:issue1}), at $\Delta = 4$ (to resolve issue \ref{item:issue3}) and at $\Delta = \hat \Delta$ with $\hat \Delta$ the dimension of any generic long multiplet that crosses the $\ell = 0$ line (to resolve issue \ref{item:issue6}). Whether this is the correct resolution, or whether there is another mechanism at play, is not something we can hope to address with our current knowledge of the $(2,0)$ theories.

More interesting resolutions can be found for the other issues. The first observation is a perfect cancellation between a special case of issue \ref{item:issue7} and issue \ref{item:issue2}: if a long multiplet trajectory crosses $\ell = -2$ exactly at $\Delta = 4$ then it induces precisely the same set of blocks as the ``$\BB[0,2]_{-3}$'' multiplet. We can write that
\be
\LL[0,0]_{\Delta = 4, \ell = -2} = \BB[0,2]_{-3}\,,
\ee
where, as always, the evaluation at negative spin is understood to be defined through analytic continuation. This means that issue \ref{item:issue2} can be resolved not only by demanding that the $\BB[0,2]_{-3}$ multiplet has zero coefficient (as we hypothesized above), but also by a long trajectory hitting spin $-2$ exactly at $\Delta = 4$ with the right coefficient. Just as for the spin $-4$ long multiplets, at present our understanding of the $(2,0)$ theories is insufficient to know which of these two potential resolutions is realized. We do note, however, that hitting spin $-2$ at $\Delta = 4$ would not be entirely unreasonable for the first subleading trajectory. This trajectory asymptotes to $\Delta = \ell + 10$ but according to a large spin analysis is expected to slope towards lower $\Delta$ at lower spins. Again with our current knowledge of $(2,0)$ theories we cannot tell if this or other mechanisms are in place to resolve these issues.

This leaves us with issue \ref{item:issue 5} which is arguably the most interesting. Our suggested resolution comes about by another special case of issue \ref{item:issue7}: an ``$\LL[0,0]_{\Delta = 2, \ell = -2}$'' long supermultiplet, or more precisely the analytic continuation of a regular long multiplet to $\ell = -2$ and $\Delta = 2$, gives the contribution:
\be
\begin{split}
[0,0]&: - \frac{24}{175} \mathcal{G}^{(2)}_6(z,\zb)\,,\\
[2,0]&: - 2 \mathcal{G}^{(0)}_4(z,\zb) + \frac{9}{700} \mathcal{G}^{(0)}_8(z,\zb)\,,\\
[0,2]&: - \frac{4}{5} \mathcal{G}^{(1)}_5(z,\zb)\,.
\end{split}
\ee
Therefore, if we add this multiplet with a (negative) coefficient $- 2 / c$ then we reproduce all the stress tensor multiplet blocks with the right coefficient. The one mismatch is an additional scalar block of dimension 8 in the $[2,0]$ channel, but its coefficient $-9/(350 c)$ is precisely such that it cancels the unwanted conformal block of issue \ref{item:issue4}, which is thereby also resolved! It is therefore entirely natural to conjecture the following observation:

\begin{claim} The leading long $\LL[0,0]_{\Delta, \ell}$ multiplet trajectory extends to $\ell = -2$ where it hits $\Delta = 2$ and has a residue corresponding to an OPE coefficient of $-2/c$. This yields the conformal blocks of the stress tensor multiplet in the different $A_R(z,\zb)$.
\label{claim:ST}
\end{claim}

\begin{figure}[t]
\begin{center}
\vskip .1cm
\includegraphics[width=\textwidth]{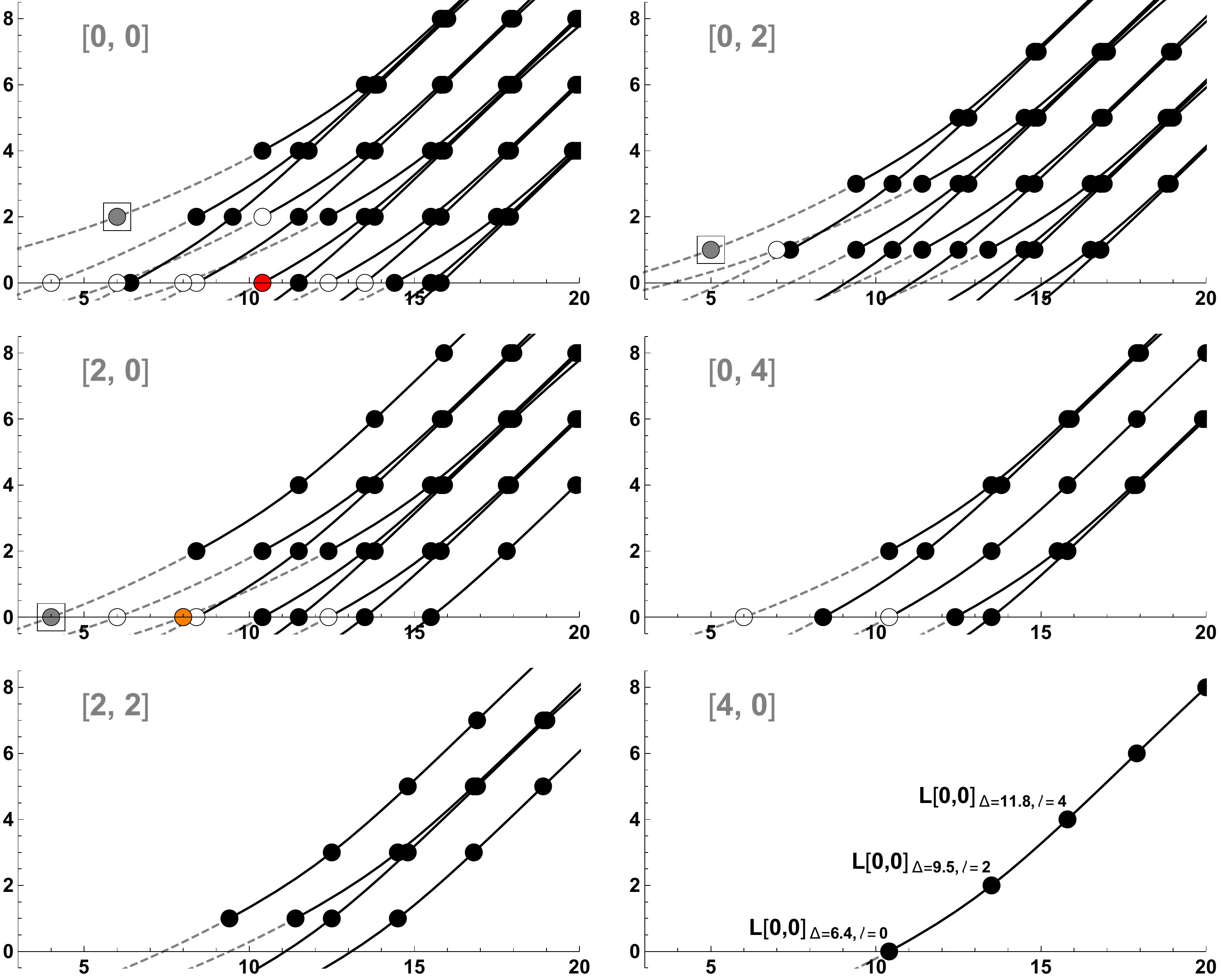}
\caption{$(\D,\ell)$ planes: leading $\LL[0,0]_{\Delta,\ell}$ trajectory (lines and dots) and $\DD[2,0]$ operators (squares)}
\label{figure:leading long resolution}
\end{center}
\end{figure}

\paragraph{}The observation that the stress tensor multiplet lies on an unprotected trajectory implies a remarkable interplay between the long and protected multiplets that we could not have observed without appealing to analyticity in spin of descendants. It leads to the improved picture for the leading long trajectory in the $(2,0)$ theories shown in figure \ref{figure:leading long resolution}.

This picture is again heuristic: to draw the trajectory we took some reasonable guesses for the scaling dimensions of the first few long multiplets based on the numerical bootstrap results of \cite{Beem:2015aoa} at $c = 25$. We then included the point $\Delta = 2$ at spin $-2$ and drew an otherwise arbitrary curve through all these points. As before, the solid part of the curve corresponds to long multiplets with physical spins $\ell \geqslant 0$ and the dotted part is the continuation to negative spin. The gray and orange dots correspond to respectively the resolution of issue \ref{item:issue 5} and \ref{item:issue4}. As for any long multiplet there is still an instance of issue \ref{item:issue8} which is indicated by the red dot.

From the picture we also observe that the very leading Regge trajectory in the $[0,0]$ channel is given by a specific conformal primary descendant of dimension $\Delta + 4$ and spin $\ell + 4$ if the superconformal primary has dimension $\Delta$ and spin $\ell$. (This trajectory asymptotes to the line $\Delta = \ell +8 $ which means that it is indeed more leading than any of the short trajectories.) According to our observation this is the trajectory that the stress tensor lies on. In this sense the $(2,0)$ theories would then be similar to a non-supersymmetric theory like the Ising or $O(N)$ models where there is ample numerical evidence that the stress tensor lies on the leading (non-straight) Regge trajectory.

\subsubsection{Resolution in generalized free field theory}
We can now investigate how the issues are resolved in theories whose spectrum and OPE coefficients we know exactly. There are only two such options: the theory of $N$ free tensor multiplets, and generalized free field theory. However, the former is qualitatively different from the cases considered in this work since it is free, and will thus have an extra family of short multiplets ($\BB[0,0]_{\ell \geqslant 0}$), which contain conserved currents of spin larger than two. These multiplets give rise to various new conformal primary trajectories and change the aforementioned issues. In particular, the stress tensor fits as the continuation to $\ell=-2$ of this trajectory.

We will then consider only the case of generalized free field theory. The corresponding four-point function is obtained just by Wick contractions, and solves the crossing equations with $c = \infty$. The function $a(z,\zb)$ for this theory reads:
\be
a(z,\zb) = \frac{z^3 (\zb-1)^3-3 z^2 (\zb-1)^3+3 z (\zb-1)^3-(\zb-2) ((\zb-1) \zb+1)}{3 (z-1)^3 z^2 (\zb-1)^3 \zb^2}\,.
\label{eq:cinftyazzb}
\ee
An added advantage of this function is its very soft behavior in the Regge limit and in the second-sheet light-cone limit, which per the analysis of section \ref{subsec:Reggegrowth} leads to an improvement over the generic behavior of equation \eqref{analyticityAR}: for this function we have analyticity for all physical spins in all channels. This means all eight issues must be resolved. Issues~\ref{item:issue4} and  \ref{item:issue 5} are automatically resolved since $c=\infty$, and in what follows we will see how the remaining issues are resolved for the leading trajectories. We note that the superconformal block decomposition of this correlator is easily found, and besides the protected multiplets we find towers of unprotected multiplets at the double twist values $\Delta =8+ \ell + 2n$, $n\in \mathbb{N}_{0}$. The corresponding coefficients can be found by applying the supersymmetric inversion formula of section~\ref{sec:inversion}.

Let us start with the leading trajectory of the R-symmetry singlet channel, which has twist $\tau= \Delta - \ell = 8$. Operators on this trajectory get contributions from the two short trajectories as well as the leading long trajectory with $\Delta = 8 + \ell$, $\ell \geqslant 0$.  Taking the OPE coefficients of these operators we can reconstruct the leading trajectory of $A_{[0,0]}(z,\zb)$ using the superconformal blocks. Altogether we find the following conformal primaries
\be 
\begin{split}
A_{[0,0]}(z,\zb)&\Big\vert_{\tau= 8} = \sum\limits_{\substack{\ell \geqslant 0\,,\\ \ell \text{ even}} } 2^{\ell+2} b_{\ell+2} c^{\BB[2,0]_\ell}_{2,2} \mathcal{G}_{\ell +10}^{(\ell +2)}(z,\zb)\\
+&\sum\limits_{\substack{\ell \geqslant 1\,, \\ \ell \text{ odd}}} \lambda^2_{\BB[0,2]_{\ell}}\left( c^{\BB[0,2]_{\ell}}_{1, 1}   \mathcal{G}_{\ell +9}^{(\ell +1)}(z,\zb) + 
c^{\BB[0,2]_\ell}_{3, 3} \mathcal{G}_{\ell +11}^{(\ell +3)}(z,\zb)\right)\\
+&\sum\limits_{\substack{\Delta= 8 + \ell\,,\\ \ell \geqslant 0\,,\\ \ell \text{ even}} } \lambda^2_{\LL[0,0]_{\Delta,\ell}} \left( c^{\LL[0,0]_{\Delta,\ell}}_{4,4}  \mathcal{G}_{\Delta +4}^{(\ell +4)}(z,\zb)    + c^{\LL[0,0]_{\Delta,\ell}}_{ 2,2} \mathcal{G}_{\Delta +2}^{(\ell +2)}(z,\zb) +
 c^{\LL[0,0]_{\Delta,\ell}}_{0,0} \mathcal{G}_\Delta ^ {(\ell)}(z,\zb)\right)\,,
\end{split}
\ee
where $c^{\bullet}_{\bullet,\bullet}$ are the coefficients of the expansion of the superconformal block of the respective supermultiplet in conformal blocks. We chose not to spell them out here, but they are completely known rational functions of $\Delta$ and $\ell$.
Analyticity in spin of the leading Regge trajectory for $\ell \geqslant 0 $ requires the above to be identical to
\be
\begin{split} 
&A_{[0,0]}(z,\zb)\Big\vert_{\tau= 8} = \sum\limits_{\substack{\Delta =8 + \ell,\\ \ell \geqslant0\\ \ell \mathrm{ even}}}   \left(  2^{\ell} b_{\ell} c^{\BB[2,0]_{\ell-2}}_{2,2} +
\lambda^2_{\BB[0,2]_{\ell-1}} c^{\BB[0,2]_{\ell-1}}_{1, 1}  +
\lambda^2_{\BB[0,2]_{\ell-3}} c^{\BB[0,2]_{\ell-3}}_{3, 3}\right.\\
&\left.+ 
\lambda^2_{\LL[0,0]_{\Delta-4,\ell-4}} c^{\LL[0,0]_{\Delta-4,\ell-4}}_{4,4}     + 
\lambda_{\LL[0,0]_{\Delta-2,\ell-2}}^2 
c^{\LL[0,0]_{\Delta-2,\ell-2}}_{2,2}  +\lambda_{\LL[0,0]_{\Delta,\ell}}^2  c^{\LL[0,0]_{\Delta,\ell}}_{0,0} 
\right) \mathcal{G}_\Delta ^ {(\ell)}(z,\zb)\,,
\end{split}
\label{eq:A00leadtwist}
\ee
such that the OPE coefficients are an analytic function of $\ell$.
Indeed these two expressions agree due to the following relations between OPE coefficients and the superconformal block coefficients:
\be 
\lambda_{\LL[0,0]_{6,-2}}=0\,, \qquad
c^{\BB[0,2]_{-1}}_{1, 1}=0\,,\qquad
\lambda_{\BB[0,2]_{-3}}=0\,, \qquad 
c^{\LL[0,0]_{4,-4}}_{4,4} \lambda_{\LL[0,0]_{4, -4}}^2 = -  b_{0} c^{\BB[2,0]_{-2}}_{2,2}  \,.
\ee
The last equation means issues~\ref{item:issue3} and~\ref{item:issue8} (for the leading long trajectory) cancel out, while issue~\ref{item:issue2} and \ref{item:issue7} (for the leading long trajectory) are resolved by the OPE coefficients vanishing. 
The same exercise can be done for the twist 10 conformal primaries in the R-symmetry singlet channel, and this time a relation between (a) the spin $-4$ long multiplet of the sub-leading trajectory ($\Delta = 10 + \ell$), (b) the $\DD[0,4]$ multiplet, and (c) the short trajectory $\BB[2,0]_{\ell}$ at spin $-1$ conspire to solve issues~\ref{item:issue1} and \ref{item:issue8} (for the subleading long trajectory). Carrying out these checks for higher twists, one finds that the OPE coefficients continue to conspire to resolve all issues; for example issue~\ref{item:issue6} for the leading long trajectory  ($\Delta = 8 + \ell$) is canceled by  issue~\ref{item:issue8} for the sub-sub-leading long trajectory  ($\Delta = 12 + \ell$).

\subsection{Shadow symmetry in all channels}
\label{subsec:shadowsy}
We have seen that a single Regge trajectory at the level of supermultiplets induces several Regge trajectories for the $A_R(z,\zb)$ which have integer shifts in $\ell$ and $\Delta$ and OPE coefficients determined by the super-Regge trajectory. What is not at all obvious is then whether \emph{shadow symmetry} is automatically realized. To see that it is, let us focus on the long multiplets. Consider a long trajectory where the spin of the superconformal primary is given by a function $\ell_{s.c.p.}(\Delta)$. To deduce the shadow symmetry of this function consider first the $[4,0]$ channel. There we see that the super-trajectory induces a single regular conformal trajectory that is shifted by four units, so 
\be
\ell_{[4,0]}(\Delta) = \ell_{s.c.p.}(\Delta - 4)\,.
\ee
This is a regular bosonic trajectory and therefore obeys ordinary shadow symmetry: $\ell_{[4,0]} (6 - \Delta) = \ell_{[4,0]}(\Delta)$. The shadow symmetry of the super-trajectory is therefore
\be
\ell_{s.c.p.}(\Delta) = \ell_{s.c.p.}(-2 - \Delta)\,,
\ee
and in particular the shadow-symmetric point at the level of the super-trajectory lies at $\Delta = -1$. Now consider the coefficient of the $[4,0]$ descendant. In our conventions it equals
\be
\lambda^2_{[4,0]}(\Delta + 4,\ell) = \frac{(\Delta -\ell -2) (\Delta +\ell +2)}{(\Delta -\ell -6) (\Delta +\ell -2)} \lambda^2_{s.c.p.}(\Delta,\ell)\,,
\ee
with $\lambda^2_{s.c.p.}(\ell,\Delta)$ denoting the coefficient of the block corresponding to the superconformal primary (in the $[0,0]$ channel). The ordinary shadow symmetry \cite{Caron-Huot:2017vep} reads
\be
\frac{\lambda^2_{[4,0]}(6 - \Delta,\ell)}{K^{0,0}_{6 - \Delta,\ell}} = \frac{\lambda^2_{[4,0]}(\Delta,\ell)}{K^{0,0}_{\Delta,\ell}}\,,
\label{eq:shadow40}
\ee
where $K^{0,0}_{\D,\ell}$ is a kinematical factor defined in equation~\eqref{eq:KDL} below, and implies a slightly modified shadow symmetry for the superconformal primary:
\be \label{scpshadow}
\frac{\lambda^2_{s.c.p.}(\Delta,\ell)}{K^{0,0}_{\D + 4,\ell}} = \frac{(\ell-\Delta ) (-\Delta +\ell+6) (\Delta +\ell-2) (\Delta +\ell+4) }{(-\Delta +\ell-4) (-\Delta +\ell+2) (\Delta +\ell+2) (\Delta +\ell+8)} \frac{\lambda^2_{s.c.p.}(-2 -\Delta,\ell)}{K^{0,0}_{2-\Delta,\ell}}\,.
\ee
Next, let us consider another trajectory, for example the superconformal primary itself in the $[0,0]$ channel. How can the above equations be compatible with ordinary shadow symmetry in this channel, which would send $\D \to 6 - \Delta$ and not $\Delta \to - 2 - \Delta$? The answer is that another superconformal descendant comes to the rescue, as follows. To restore shadow symmetry we need a `shadow superconformal primary' trajectory $\ell_{s.s.c.p.}(\Delta)$ given by the shadow of the shadow:
\be
\ell_{s.s.c.p.}(\Delta) \colonequals \ell_{s.c.p.}(6 - \Delta) = \ell_{s.c.p.}(\Delta - 8)\,.
\ee
This trajectory does indeed exist in the supermultiplet: it is the top component of the long multiplet with dimension $\Delta + 8$. Compared to the superconformal primary, its coefficient is the messy expression:
\be
\begin{split}
\lambda^2_{s.s.c.p.}(\Delta + 8,\ell)& = \frac{\Delta  (\Delta +6) (-\Delta +\ell+2) (\Delta -\ell) (\Delta -\ell+2)^2 (\Delta -\ell+4)}{65536 (\Delta +2) (\Delta +4) (-\Delta +\ell-3) (-\Delta +\ell+1) (-\Delta +\ell+6) (\Delta -\ell+1)^2 }\\ & \times
\frac{(\Delta +\ell+2) (\Delta +\ell+4) (\Delta +\ell+6)^2 (\Delta +\ell+8)}{(\Delta +\ell-2) (\Delta +\ell+3) (\Delta +\ell+5)^2 (\Delta +\ell+7)} \lambda^2_{s.c.p.}(\Delta,\ell)\,,
\end{split}
\ee
but it is precisely the one necessary to recover shadow symmetry between these two multiplets in the $[0,0]$ channel: with a little computation one finds that
\be
\frac{\lambda^2_{s.s.c.p.}(6 - \Delta,\ell)}{K^{0,0}_{6 - \Delta,\ell}} = \frac{\lambda^2_{s.c.p.}(\Delta,\ell)}{K^{0,0}_{\Delta,\ell}}\,.
\ee
We have checked that similar relations exist between other superconformal descendants and that therefore the entire set of Regge trajectories in each of the six R-symmetry channels is shadow symmetric provided the relation \eqref{scpshadow} holds. To the best of our knowledge, the corresponding identities involving the coefficients of the different conformal blocks inside the superconformal multiplet have not been observed before. We expect it to be a very general property, valid for any superconformal algebra, that can for example be used as an additional verification in the computations of superconformal blocks, and perhaps provide extra constraints on the superconformal blocks of non-BPS operators.\footnote{In fact, we naturally would expect that a similar identity holds if we decompose ordinary conformal blocks into lower-dimensional conformal blocks. This would ensure that, if we for example analyze a correlation function of a three-dimensional theory with two-dimensional conformal blocks, the shadow symmetry of each three-dimensional trajectory separately suffices to ensure shadow symmetry of the (generically infinitely many) two-dimensional trajectories it induces. It would be interesting to verify this.} It would be interesting to see if this property can be deduced more directly from properties of the superconformal algebra.

\section{Supersymmetric inversion}
\label{sec:inversion}
The demonstration of analyticity in spin of the OPE data for any CFT proceeds via the so-called Lorentzian inversion formula of \cite{Caron-Huot:2017vep}. For the four-point function under consideration one could in principle apply the formula to the six different functions $A_R(z,\zb)$ and combine the data from these operations in order to get a complete picture for the Regge trajectories of the superconformal multiplets. A more elegant approach is to work directly with the function $a(z,\zb)$. As we will discuss below, this function essentially has all the right properties for the Euclidean and Lorentzian inversion formulas to apply, and its block decomposition will give us direct access to all the OPE coefficients of operators that contribute to it.

\subsection{\texorpdfstring{Inversion formula for $a(z,\zb)$}{Inversion formula for a(z,zb)}}
As can be seen from equations \eqref{eq:ahatom}, \eqref{eq:achiinblocks} and \eqref{eq:auinblocks}, the function $a(z,\zb)$, multiplied by $(z\zb)^6$,  admits a conformal block decomposition as follows:
\be
\begin{split}
(z \zb)^6 a(z,\zb) &= (z \zb)^6 a^\chi(z,\zb) + (z \zb)^6 a^u(z,\zb)  \,,\\
(z \zb)^6 a^\chi(z,\zb)
&= \sum_{\ell=0,\; \ell\, \mathrm{even}}^{\infty} \frac{2^{\ell}b_\ell}{\ell+3} \GG_{\ell+8}^{(\ell)}(\Delta_{12}=0,\Delta_{34}=-2; z,\zb)\,, \\
(z \zb)^6 a^u(z,\zb) 
 &= \sum\limits_{\substack{\Delta \geqslant \ell + 6,\\ \ell \geqslant0,\; \ell\, \mathrm{even}}} 
\frac{4\lambda_{\Delta,\ell}^2}{ (\Delta-\ell-2)(\Delta+\ell+2)} \GG_{\Delta+4}^{(\ell)}(\Delta_{12}=0,\Delta_{34}=-2; z,\zb)\,.
\end{split}
\label{ablockdecomplete}
\ee
It is therefore most natural to apply the inversion formula of \cite{Caron-Huot:2017vep} to $(z \zb)^6 a(z,\zb)$. The inversion formula results in a function $c(\Delta,\ell)$ whose meromorphic structure captures the OPE data. Normally one evaluates $c(\Delta,\ell)$ at integer $\ell$ and, for generic values of $\Delta$, a pole in $c(\Delta,\ell)$ at some value $\Delta^*$ signifies the presence of a conformal block in the OPE decomposition, whose coefficient is simply given by (minus) the residue of the pole. In our case there is a small offset to take into account: a pole at generic $\Delta^*$ signifies the presence of a supermultiplet contributing as $a^\text{at}_{\Delta^*-4,\ell}$  (for a long multiplet this means the primary has dimension $\Delta_{\scp} = \Delta^* - 4$ according to table~\ref{tab:superblocks}), and whose OPE coefficient is related to the residue by the factor $4/(\Delta^* - \ell -6)/(\Delta^* + \ell - 2)$.

The decomposition \eqref{ablockdecomplete} shows that we should treat $(z\zb)^6 a(z,\zb)$ as a four-point function of non-identical scalar operators with $\Delta_{12} = 0$ and $\Delta_{34} = -2$.
Furthermore, all the spins in its block decomposition are even integers and therefore there is no distinction between the $t$- and the $u$-channel contributions.
The singularities of $(z \zb)^6 a(z,\zb)$ are also those of a four-point function  of non-identical scalars, as can be seen from its conformal block decomposition \eqref{ablockdecomplete} and its crossing equation \eqref{eq:crosssym}.
All in all, the associated spectral density should then be computable through the Lorentzian inversion formula \cite{Caron-Huot:2017vep} as\footnote{The factor $2^{\D-5+\ell}$ follows from our normalization of conformal blocks. It is spelled out in appendix \ref{app:confblock} and differs from that of \cite{Caron-Huot:2017vep} by a factor $2^{\ell}$.}
\begin{align}
c(\Delta,\ell) 
= 
2^{\D-5+\ell}
\frac{(1 + (-1)^\ell)}{4}   \kappa^{0,-2}_{\Delta+\ell} \int\limits_0^1 dz d\zb \, \mu(z,\zb)\,  \GG_{\ell+5}^{(\Delta -5)}(0,-2;z,\zb) \, \mathrm{dDisc}_t\left[ (z \zb)^6  a(z,\zb)\right]\,,
\label{eq:inversionformula}
\end{align}
where the most important factor involves the \emph{double discontinuity} operation, which for a generic four-point function $g(z,\zb)$ reads:
\begin{align}
\dDisc_t\left[g(z,\zb) \right] 
&= \cos\left( \pi \frac{\Delta_{34}-\Delta_{12} }{2}\right)  g(z,\zb)
-\frac{1}{2} e^{\ii  \pi \frac{\Delta_{34}-\Delta_{12} }{2}}  g(z,1-(1-\zb)e^{2 \pi \ii})\nonumber\\ 
&\qquad-\frac{1}{2} e^{-\ii  \pi \frac{\Delta_{34}-\Delta_{12} }{2}}  g(z,1-(1-\zb)e^{-2 \pi \ii})\,.\label{ddiscdef}
\end{align}
where the $e^{2 \pi i}$ factors indicate an analytic continuation onto a secondary sheet, in this case around $\bar z = 1$. The other factors are given by:
\begin{align}
\kappa_{\D+\ell}^{\D_{12}, \D_{34}} &=
\frac{\Gamma(\frac{\D+\ell+\D_{12}}{2}) \Gamma(\frac{\D+\ell-\D_{12}}{2}) \Gamma(\frac{\D+\ell+{\D_{34}}}{2}) \Gamma(\frac{\D+\ell-\D_{34}}{2})}{2 \pi^{2} \Gamma(\D+\ell-1) \Gamma(\D+\ell)}\,,\\
\mu(z, \bar{z})&=
\left|\frac{z-\bar{z}}{z \bar{z}}\right|^{d-2} \frac{((1-z)(1-\bar{z}))^{\frac{\D_{34}-\D_{12}}{2}}}{(z \bar{z})^{2}}\,,
\end{align}
and we recall that $\GG_\Delta^{(\ell)}(\Delta_{12},\Delta_{34};z,\zb)$ is an ordinary bosonic conformal block. For our case one should set $\Delta_{12} = 0$ and $\Delta_{34} = -2$ in all of the above formulas.

\subsection{Single-valuedness}
In the derivations of the Lorentzian inversion formula \cite{Caron-Huot:2017vep,Simmons-Duffin:2017nub} it is usually assumed that the function to be `inverted' is a proper CFT four-point function. Here this is not exactly the case: although $a(z,\zb)$ has a nice decomposition into $s$-channel conformal blocks, it has slightly awkward $t$-channel behavior. To see this, recall the crossing symmetry equation \eqref{eq:crosssym} and take $z, \zb \to 1$. The functions $a(1-z,1-\zb)$ and $\CC_h(1-z,1-\zb)$ are nicely behaved in that limit, but $\CC_h(z,\zb)$ has a logarithmic term not seen in an ordinary four-point function. Loosely speaking we can write that:\footnote{This equation attempts to highlight the logarithmic singularity but in doing so is a bit misleading, because it also shows an apparent singularity at $z = \zb$. The latter cancels in the full $a(z,\zb)$.}
\be
(z \zb)^6 a(z,\zb) \supset \frac{8 (z \zb)^4}{c (z - \zb)^3}  \left(\log(1-z) - \log(1-\zb)\right)\,.
\ee
The problem with this term is that it spoils Euclidean single-valuedness of $a(z,\zb)$: when setting $\zb = z^*$ and taking $z$ around $1$ in the complex plane the function does not return to itself.

It would be nice to investigate the true importance of Euclidean single-valuedness around $1$ (and around $\infty$) for the validity of the Lorentzian inversion formula more generally. We can however show that it is unimportant in our case in a simpler way. We consider the function 
\be
\begin{split}
a^\text{free}(z,\zb) & \colonequals  \frac{ \left(z ^3 (\bar z-1)^3-3 z ^2 (\bar z-1)^3+3 z  (\bar z-1)^3-(\bar z-2) ((\bar z-1) \bar z+1)\right)}{3 (z -1)^3 z ^2 (\bar z-1)^3 \bar z^2}\\ & \,\, +\frac{4(-(z -\bar z) (z +\bar z-2)+2 (z -1) (\bar z-1) \log (1-z )-2 (z -1) (\bar z-1) \log (1-\bar z))}{c (z -1) z ^2 (\bar z-1) \bar z^2 (z -\bar z)^3}\,.
\end{split}
\ee
This function was introduced in \cite{Heslop:2004du} as the four-point function obtained by simple Wick contractions in the theory of $N$ free tensor multiplets, with $c=N$ (and it is also a fairly natural object from the holographic perspective). The conformal block decomposition of this function is rather easy to find. Furthermore, if we ignore its non-single-valuedness and blindly substitute it into the Lorentzian inversion formula \eqref{eq:inversionformula} we get the right answer.\footnote{We have explicitly checked this for the operators of twist 4, 6, 8 and 10 but strongly suspect this to be true for all other operators as well.} So at least for $a^\text{free}(z,\zb)$ the non-singlevaluedness is not an issue.

The argument for the validity of the Lorentzian inversion formula for generic $a(z,\zb)$ is then as follows. Let us define:
\be
\tilde a(z,\zb) = a(z,\zb) - a^\text{free}(z,\zb)\,.
\ee
Then $(z\zb)^6 \tilde a(z,\zb)$ is single-valued and has a good conformal block decomposition (albeit without positive coefficients), so the Lorentzian inversion formula should be applicable. Of course we get a different density $\tilde c(\Delta,\ell)$ which includes the spectrum of $a^\text{free}(z,\zb)$ in addition to the physical spectrum of $a(z,\zb)$. But the Lorentzian inversion formula is fundamentally a linear operation that in itself does not require single-valuedness. So we can split $\tilde a(z,\zb)$ again and apply the Lorentzian inversion formula to each term separately. Since we verified that the inversion of $a^\text{free}(z,\zb)$ works and the corresponding $c^\text{free}(\D,\ell)$ yields the correct conformal block decomposition, it must be that the inversion of $a(z,\zb)$ gives
\be
\tilde c(\D,\ell) + c^\text{free}(\D,\ell)\,,
\ee
and this function has the right analyticity properties to recover the conformal block decomposition of $a(z,\zb)$ and nothing more.

\subsection{Behavior on the second sheet}
\label{subsec:Reggegrowth}
As we stated above, the results of \cite{Caron-Huot:2017vep} tie the Regge behavior of correlation functions to the minimal spin for which the OPE data is guaranteed to be analytic in spin. Non-analyticity happens when the contributions to the integrals from `arcs at infinity' are non-zero \cite{Caron-Huot:2017vep,Simmons-Duffin:2017nub,Rutter:2020vpw}. These arcs correspond to the case where $\zb \to 0$ on the secondary sheets reached by going around $\zb = 1$. If one sends $z\to 0$ at the same rate as $\zb$ then this is the Regge limit. It is customary to parameterize this limit as:
\be 
z = w\, \sigma\,, \quad  \zb =\frac{w}{\sigma }\,, \quad w \to 0\,, \quad \sigma \text{ fixed}\,.
\label{eq:Reggelimit}
\ee
For a general function $g(z,\zb)$ the arguments of \cite{Caron-Huot:2017vep} imply that with a Regge behavior of the form
\be \label{Reggespinconnection}
g \sim w^{1 - \ell_*}\,,
\ee
the `arcs at infinity' contributions from the Regge limit vanish if the integral $\oint_0 dw\, w^{\ell - \ell_* - 1}$ converges, so if $\ell > \ell_*$. In this way softer Regge behavior implies a larger domain of analyticity in spin.

Let us investigate the behavior of our correlation function in this limit. We begin by noting that the functions $A_R(z,\zb)$ have a decomposition into ordinary bosonic blocks, including an identity operator for the singlet channel. The non-supersymmetric arguments of \cite{Caron-Huot:2017vep} then go through: the $A_R(z,\zb)$ must be bounded by a constant in the Regge limit and analyticity in spin holds at least for any spin $\ell > 1$. Supersymmetry allows us to do better. We can see this by eliminating the derivatives in \eqref{eq:Ainah} to find an expression for $a(z,\zb)$ in terms of ($z$ and $\zb$ dependent) linear combinations of the $A_R(z,\zb)$ and $h(z)$. 
Since we know $h(z)$ exactly and can also bound the $A_R(z,\zb)$, it is not hard to deduce that
\be 
(z \zb)^6 a(z,\zb) \sim \frac{3 A_{[2,0]}+A_{[0,4]}}{3} w^2 \,, \qquad \text{as } w \to 0\,.
\label{eq:ainReggenaive}
\ee
in the Regge limit. This is a softer behavior in the Regge limit with respect to a bosonic correlator that grows as $w^0$, ensuring analyticity in spin down to spin $\ell >-1$. However, we expect analyticity to hold all the way down to spin $\ell >-3$. One partial argument for this are the conformal primary trajectories in the $R$-symmetry singlet channel: as shown in section \ref{sec:reggetrajectoriesAR} a superconformal primary of spin $\ell$ has a descendant of spin $\ell + 4$ in this channel. The $\ell > -3$ result for the superconformal primaries then immediately follows from the non-supersymmetric result that analyticity in spin holds down to spin $1$ for any conformal family.

This expectation can be proven by bounding the growth of $a(z,\zb)$ in the Regge limit directly, as done in \cite{Caron-Huot:2017vep}, using the fact that the $t$- and $u$-channel OPEs converge in that limit. In appendix~\ref{app:reggebound} we use the $t$- and $u$-channel OPEs in an expansion in the $z$ and $\bar{\zb}$ variables to bound the growth of $a(z,\zb)$ as
\begin{equation}
(z \zb)^6 a(z,\zb) \sim  w^4 \,; \qquad \text{as } w \to 0\,,
\label{eq:ainRegge}
\end{equation}
along almost any direction in the complex $w$-plane, confirming analyticity in spin  of $c(\Delta,\ell)$ all the way for all spins greater than $-3$!\footnote{To be precise, the $z$ and $\zb$ expansion can be used to show the given behavior along any direction with $\arg(w)\neq \pi/2$. Using the $\rho$ variables it might be possible to show the bound is valid also exactly along the imaginary axis. See appendix~\ref{app:reggebound} for details.}

We note that the non-single-valuedness of $a(z,\zb)$ is not important: since $a^\text{free}(z,\zb)$ behaves like $w^5$ in this limit, equation \eqref{eq:ainRegge} also gives the leading behavior for the single-valued correlator $\tilde a(z,\zb)$.

Let us finally discuss the Regge behavior of the six functions $A_R(z,\zb)$. A naive estimate arises by plugging the behavior \eqref{eq:ainRegge} into equation \eqref{eq:Ainah}, but the latter contains derivatives and if a function vanishes at least as fast $w^4$ then it is not mathematically guaranteed that its derivative vanishes at least as fast as $w^3$. In appendix \ref{app:reggebound} we therefore explain how to bound the derivatives of $a(z,\zb)$ directly, leading to the Regge behavior:
\be
\begin{split}
 A_{[4,0]} \sim w^{4}\,, \qquad
A_{[2,2]} \sim w^{3}\,, \qquad
A_{[2,0]} \sim w^{2}\,, \\
A_{[0,4]} \sim w^{2}\,, \qquad
A_{[0,2]} \sim w^{1}\,, \qquad
A_{[0,0]} \sim w^{0} \,.
\end{split}
\label{eq:AinRegge}
\ee
This directly leads to the results quoted in equation \eqref{analyticityAR} at the beginning of section \ref{sec:reggetrajectoriesAR}.

\subsubsection*{The lightcone limit}
Another potential contribution from the `arcs at infinity' comes from a lightcone limit on the second sheet that corresponds to $\zb \to 0$ holding $z$ fixed. In this limit the convergence of the Lorentzian inversion formula depends on $\Delta$ and $\ell$; more precisely, if $g(z,\zb) \sim \zb^{\tau^*/2}$ in that limit, then we need
\be
\Delta - \ell < \tau^*\,, \qquad\text{and} \qquad d - \Delta - \ell < \tau^*\,.
\ee
It is not immediately clear what the correct value of $\tau^*$ could be. On the first sheet $\tau^*$ is equal to the lowest twist in the $s$-channel spectrum. In that case $\tau^* \geq d/2 - 1$ would be appropriate for a generic CFT, because the $s$-channel identity operator needs to be subtracted, while in our case $\tau^* = 8$. As discussed in \cite{Simmons-Duffin:2017nub}, this value again leads to analyticity in spin for $\ell > 1$ because we can set $\Delta = d/2$, and in our case it would not spoil $\ell>-3$. On the secondary sheets this is not necessarily true because the $s$-channel block decomposition no longer converges -- instead one uses the $t$-channel conformal block decomposition (and unitarity) to bound the full correlator by its behavior on the first sheet. The best possible bound that can be rigorously proven in this way corresponds to $\tau^* \geq 0$. This value is however unlikely: it would imply non-convergence of the Lorentzian inversion formula for $\Delta < \ell$ and invalidate the analyses of the first Regge trajectory in the literature. (In our case we can show that $\tau^* \geq 4$ instead of zero from \eqref{eq:asecondt} and \eqref{eq:asecondu}, but the effect is similar.) It would be interesting to have a better handle on the lightcone behavior on the secondary sheets, but we will not consider possible lightcone contributions further in the remainder of this paper. See \cite{Hartman:2015lfa,Simmons-Duffin:2017nub,Kologlu:2019bco,Kravchuk:2021kwe} for previous discussions on the second sheet lightcone limit.

\subsection{\texorpdfstring{Analyticity properties of $c(\Delta,\ell)$}{Analyticity properties of c(D,l)}}
\label{subsec: Analyticity properties of c_Delta_ell}
As we explained in the previous subsection, conformal blocks in the decomposition of $a(z,\zb)$ correspond to poles in the function $c(\Delta,\ell)$. These are however not the only poles: there are also kinematical singularities. It is therefore worthwhile to take a look at the analytic structure of $c(\Delta,\ell)$, which we will do in this subsection. In contrast with most other studies we will also be interested in what happens at \emph{negative} spins $\ell$. The result of our analysis is summarized in figure \ref{fig:analyticitycdeltaj}.

\subsubsection{Kinematical singularities}
Recall that the shadow symmetry of the OPE density is embodied in the following equation
\begin{align}
\frac{c(\D,\ell)}{K^{\D_{12},\D_{34}}_{\D,\ell}} = \frac{c(d-\D,\ell)}{K^{\D_{12},\D_{34}}_{d-\D,\ell}}\,,
\end{align}
where
\begin{align}
K^{\D_{12},\D_{34}}_{\D,\ell}=\frac{\G(\D-1)}{\G(\D-d/2)}\k^{\D_{12},\D_{34}}_{\D+\ell}\,.
\label{eq:KDL}
\end{align}
The shadow symmetry for $c(\D,\ell)$ follows automatically from $c(\Delta,\ell)$ as defined through  eqs.~\eqref{eq:inversionformula} and \eqref{ablockdecomplete}, and from shadow symmetry in the $[4,0]$ channel given in \eqref{eq:shadow40}.
For this reason we choose to analyze the kinematical singularities of $c(\D,\ell)/K_{\D,\ell}^{0,-2}$ and one can recover the kinematical singularities of $c(\D,\ell)$ by removing the $K$ factor. From \eqref{eq:inversionformula} we see that these singularities are encoded in the kernel block times a prefactor, more precisely $2^{\D-5}\frac{\G(\D-3)}{\G(\D-1)}\GG_{\ell+5}^{(\D-5)}(0,-2;z,\zb)$. Using \eqref{eq:pure blocks} it can be easily checked that this combination is shadow symmetric.

The generic pole structure of conformal blocks was discussed in \cite{Kos:2013tga,Kos:2014bka}, but it does not apply to even spacetime dimensions where we can also encounter double poles. In our case $d=6$ and thus we will analyze the pole structure using the closed form of the six-dimensional conformal blocks directly.

We are only interested in the pole locations and below in table~\ref{tab:poles and residues of the kernel block} we tabulate both the simple and double poles in $\frac{\G(\D-3)}{\G(\D-1)}\GG_{\ell+5}^{(\D-5)}(0,-2; z,\zb)$.
{\renewcommand{\arraystretch}{1.5}
\begin{table}[h!]
	\begin{center}
	\begin{tabular}{c | c| c}
		pole locations in ${\ell}$& simple poles & double poles \\\hline\hline
		$(\D-3)-4-2n$ & $n\in\mathbb{Z}_{\geq0}$ & 
		$\D\in\mathbb{Z}_{\leq3},n\in\mathbb{Z}_{\geq0}$\\ 
		${-3}$ &  & $\D=2,4$ \\
		${-(\D-3)-4-2n}$ & $n\in\mathbb{Z}_{\geq0}$ &  $\D\in\mathbb{Z}_{\geq4},n\in\mathbb{Z}_{\geq0}$
	\end{tabular}
	\end{center}
	\begin{center}
	\caption{\label{tab:poles and residues of the kernel block}Simple and double poles of $\frac{\G(\D-3)}{\G(\D-1)}\GG_{\ell+5}^{(\D-5)}(0,-2; z,\zb)$}
	\end{center}
\end{table}
}
We have written these as poles in spin because we will soon be interested in the setup with fixed $\D$ and negative $\ell$.

\subsubsection{Dynamical poles and analyticity in spin}
Finally we have to take into account the dynamical poles in $c(\Delta,\ell)$. The conformal block decomposition of $(z\zb)^6 a(z,\zb)$ was given in equation \eqref{ablockdecomplete}. The short multiplets induce two straight Regge trajectories: one with known OPE coefficients at $\Delta = \ell + 8$ and one with unknown OPE coefficients at $\Delta = \ell + 10$. The long multiplets have unknown scaling dimensions, but as we discussed in the previous subsection we expect the leading long trajectory to smoothly continue to the stress tensor. If we take the various offsets into account this means that it must cross through the point $\Delta = 6$ and $\ell = -2$ in $c(\Delta,\ell)$. The numerical results of \cite{Beem:2015aoa} further suggest that this trajectory asymptotes to the leading double twist trajectory, which here implies that its asymptotic twist equals $12$.

In section \ref{subsec:Reggegrowth} we analyzed the Regge behavior of the function $a(z,\zb)$ and found that it behaves rather smoothly; if $z = \sigma w$ and $\zb = w / \sigma$ then $(z \zb)^6  a(z,\zb) \sim w^4$ as $w \to 0$ on the secondary sheets. The integral in \eqref{eq:inversionformula} therefore converges in that region as long as $\ell > -3$, and the contributions from the `arcs at infinity' in the Regge limit must also vanish for this range of spins.

\begin{figure}[!t]
\centering{}
\includegraphics[width=15cm]{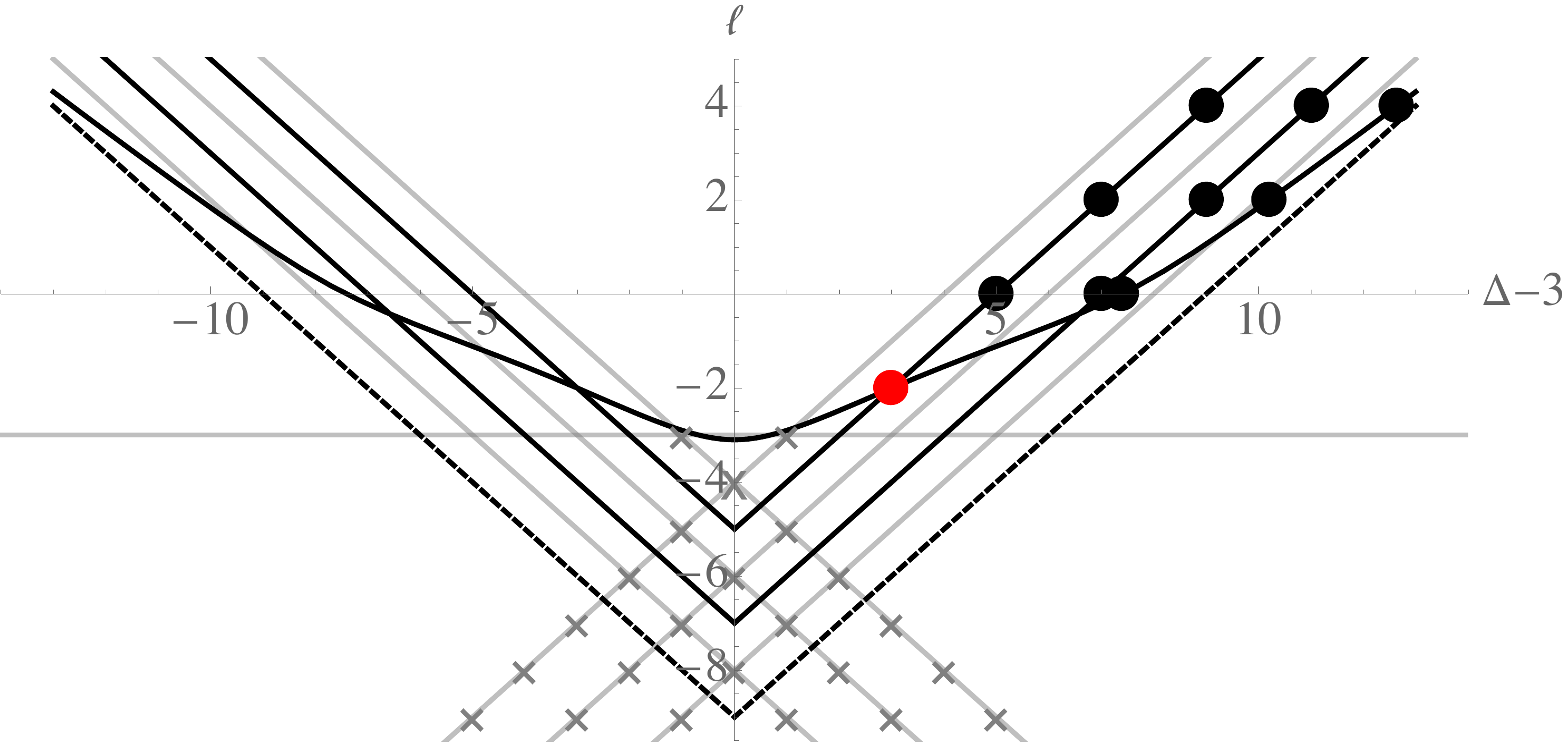}
\caption{\label{fig:analyticitycdeltaj}Regge trajectories (black) and kinematical poles (gray) of $c(\Delta,\ell)/K_{\D,\ell}^{0,-2}$ for $a(z,\zb)$. The dots indicate physical operators with the red one corresponding to the the stress tensor multiplet. The dashed line indicates the asymptotic double twist behavior of the leading long multiplets. The gray crosses indicate double poles.}
\end{figure}

An impression of the interplay of kinematic and dynamical poles is shown in figure~\ref{fig:analyticitycdeltaj}. We omitted the subleading unprotected Regge trajectories in the figure. Note that the claim~\ref{claim:ST} implies the short $\BB[2,0]$ trajectory and the leading long trajectory intersect at spin $-2$ and $\Delta=6$ in figure~\ref{fig:analyticitycdeltaj}. The OPE coefficients of the two trajectories are such that the residue of $c(\Delta,-2)$ at $\Delta =4$ vanishes.

\section{Practical supersymmetric inversion}
\label{sec:practicalities}
In this section we will discuss some practical aspects of working with the inversion formula \eqref{eq:inversionformula} for $a(z,\zb)$. We will first discuss the use of the crossing equation and the simplifications that occur after substitution of the $t$-channel conformal block decomposition. This provides sufficient background for a small (and crude) numerical test of convergence for negative spins which we perform in subsection \ref{shadowconvergencecheck}. It will be convenient to consider the small $z$ expansion of the integrand of equation \eqref{eq:inversionformula}, and in subsection \ref{subsec:inversion setup} we explain how this works and use it to recover the OPE coefficients of the short multiplets that contribute to the chiral algebra. This all provides sufficient background for more serious numerical experiments in section \ref{sec:numerics}.

One subtle but important point of the inversion formula \eqref{eq:inversionformula} is relegated to appendix \ref{app:regulation}. In that appendix we discuss how to regulate divergences that arise in the $z \to 1$ limit of the integral. Roughly speaking these divergences arise because the scaling dimensions are integers, and in practice we find badly divergent integrals of the type $\int^1 dz \, (1-z)^{-n}$. Although such divergences are invisible in the small $z$ expansion discussed in subsection \ref{subsec:inversion setup}, it is of course of fundamental importance that we are able to tame them since otherwise the entire inversion formula would stop making sense.

\subsection{\texorpdfstring{The $t$-channel decomposition}{The t-channel decomposition}}
\label{sec:practicaltchanel}
The double discontinuity \eqref{ddiscdef} vanishes for each $s$-channel block separately and the integral in \eqref{eq:inversionformula} does not commute with the decomposition into these blocks. Instead one can employ crossing symmetry and consider the $t$-channel block decomposition. In typical `experiments' one approximates the four-point function on the right-hand side of \eqref{eq:inversionformula} with a finite sum of $t$-channel blocks and investigates the resulting approximation to $c(\Delta,\ell)$. 

Let us recall that the crossing equation \eqref{eq:crosssym} for $a(z,\zb)$ reads:
\be
\begin{split}
z \zb\, a(z,\zb) &=  (1-z)(1-\zb) \left(a^u(1-z,1-\zb) + a^\chi(1-z,1-\zb)\right) \\
& \qquad+ \CC_h(1-z,1-\zb) - \CC_h(z,\zb)\,,
\label{eq:crosssym2}
\end{split}
\ee
where
\be
\label{eq:Ch2}
\CC_{h}(z,\bar z) = \frac{1}{(z - \bar z)^3}\frac{h\left(z\right)-h(\zb)}{z \bar{z}}\,.
\ee
Notice that we now split $a(1-z,1-\zb) = a^\chi(1-z,1-\zb) + a^u(1-z,1-\zb)$ on the right-hand side to highlight that one part of this function is known and fixed by the chiral algebra as discussed in section \ref{sec:4ptfn}.

The Lorentzian inversion formula instructs us to take the double discontinuity of the right-hand side of equation \eqref{eq:crosssym2}. It is well-known that, in general conformal field theories, the contribution to the double discontinuity vanishes for the $t$-channel operators with double-twist quantum numbers. In our case even more cancellations occur than the ``double-twists'' of the inverted correlator $(z \zb)^6 a(z,\zb)$, although they do correspond to double-twists in the different R-symmetry channels.

First consider $a^u(1-z,1-\zb)$. Its block decomposition was given in equation \eqref{eq:auinblocks} which we repeat here:
\be
a^u(1-z,1-\zb) = \sum_{\D \geqslant \ell + 6,\ell} \lambda_{\D,\ell}^2 a^{\text{at}}_{\D,\ell}(1-z,1-\zb)\,.
\label{eq:auinblocks2}
\ee
It is not hard to verify that the blocks with $\D = \ell + 6$ that saturate the inequality, which we recall correspond to the $\DD[0,4]$ multiplet (for $\ell = 0$) and the $\BB[0,2]_\ell$ multiplets (for $\ell > 0$), have vanishing integrated double discontinuity. Therefore only the long multiplets can contribute, and:
\be
\dDisc_t\left[(1-z)(1-\zb) a^u(1-z,1-\zb)\right] = \sum_{\D > \ell + 6,\ell} \lambda_{\D,\ell}^2 \, \dDisc_t\left[ (1-z)(1-\zb) a^{\text{at}}_{\D,\ell}(1-z,1-\zb)\right]\,.
\label{eq:dDisc auinblocks2}
\ee
so the inequality constraint for $\Delta$ has become strict. The double discontinuity of a single $t$-channel conformal block is
\begin{align}
\begin{aligned}
\operatorname{dDisc}_t\left[\GG_{\D}^{(\ell)}(0,-2;1-z,1-\zb)\right]
=&-2\sin^2\left(\frac{\D+\ell}{2}\pi\right)\GG_{\D}^{(\ell)}(0,-2;1-z,1-\zb),
\end{aligned}
\end{align}
where we have used the fact that $\ell$ is an even integer.

Next we consider $a^\chi(1-z,1-\zb)$. This is a known function whose block decomposition was given in equation \eqref{eq:achiinblocks}. For these blocks the double discontinuity is not vanishing, but instead it cancels almost entirely against the contribution from $\CC_h(1-z,1-\zb)$. In fact we may write:
\begin{multline}
\dDisc_t \left[(1-z)(1-\zb) a^\chi(1-z,1-\zb) + \CC_h(1-z,1-\zb) \right] = \\
\dDisc_t \left[ \frac{\tilde h(1-z) - \tilde h(1-\zb)}{(\zb-z)^3(1-z)(1-\zb)}\right]\,,
\end{multline}
with $\tilde h(z)$ defined as the `truncation' of $h(z)$ to just the identity and the stress tensor block:
\be
\tilde h(z) = h^{\text{at}}_0(z) + b_{-2}h^{\text{at}}_{2}(z) 
= 
-\frac{1}{2}+\frac{1}{z}+\frac{24}{c} \left(1-\frac{2}{z}+\frac{2 \log (1-z)}{z}-\frac{2 \log (1-z)}{z^2}\right).
\ee
Notice that the identity and the stress tensor are the only multiplets that do not contribute to $a^\chi(1-z,1-\zb)$.

The previous two paragraphs imply that, out of the all $t$-channel data, \emph{only the identity, the stress tensor multiplet, and the long multiplets contribute to the double discontinuity}. The contribution of all the other short multiplets simply drops out. Notice that there is furthermore a contribution to the double discontinuity from $\CC_h(z,\zb)$ which is non-vanishing and not subject to the above cancellations. In equations:
\be
\begin{split}
&\dDisc_t\left[(z \zb)^6 a(z,\zb) \right] \\
&= \sum_{\D > \ell + 6,\ell} \lambda_{\D,\ell}^2 \dDisc_t\left[(z\zb)^5 (1-z)(1-\zb) a^{\text{at}}_{\D,\ell}(1-z,1-\zb)\right] \\
&\qquad \qquad + \dDisc_t \left[ \frac{(z\zb)^5\left( \tilde h(1-z) - \tilde h(1-\zb)\right)}{(\zb-z)^3(1-z)(1-\zb)}  - (z\zb)^5\CC_h(z,\zb)\right]\\
&=\sum_{\D > \ell + 6,\ell} 
\frac{-8\lambda_{\Delta,\ell}^2}{ (\Delta-\ell-2)(\Delta+\ell+2)}
\left(\frac{z\zb}{(1-z)(1-\zb)}\right)^5
\sin^2\left(\frac{\D+\ell}{2}\pi\right)\GG_{\D+4}^{(\ell)}(0,-2;1-z,1-\zb)
\\
&\qquad\qquad + \operatorname{dDisc}_{t}\Bigg[
\left(\frac{2 z^4}{(1-z)^3}+\frac{8 z^4 \left(1-6z+3 z^2+2 z^3-6 z^2 \log (z)\right)}{c (1-z)^6}\right) \frac{1}{1-\zb}\\
&\qquad\qquad\qquad\qquad
-\frac{4 z^4}{3 (1-z)^3} \frac{1}{(1-\zb)^2}
+\frac{z^4}{3 (1-z)^3} \frac{1}{(1-\zb)^3}
+\frac{8 z^4}{c (1-z)^3} \log(1-\zb)
\Bigg],
\label{ddiscfromtchannelstuff}
\end{split}
\ee
where in passing to the final expression we have only kept the terms with non-vanishing double discontinuity.

\subsection{Convergence along shadow-symmetric line}
\label{shadowconvergencecheck}
As a first experiment we can try to see if we can numerically observe convergence for negative spins as predicted by the Regge behavior discussed in section \ref{subsec:Reggegrowth}. To do so we can start with \eqref{ddiscfromtchannelstuff} and plug in increasingly many $t$-channel blocks. Since we do not know the exact spectrum (and are only interested in a quick numerical check of the convergence properties anyway) we took a maximally crude approximation: apart from the leading Regge trajectory, for which we used some input from the numerical bootstrap results of \cite{Beem:2015aoa}, we took a `shifted' mean-field-like spectrum of scalar operators where all the dimensions were shifted from their mean field values by 1. This was done to maximize their contribution to the dDisc, because with the shift the prefactor $\sin^2(\pi(\D+\ell)/2)=1$. For the OPE coefficients we took the values obtained from the supergravity computation of \cite{Heslop:2004du} extrapolated to finite $c$.  We did not add spinning operators since these numerically would have contributed much less to the double discontinuity anyway.  We then inverted this spectrum using \eqref{eq:inversionformula} and plot the resulting function $c(\Delta,J)$ at the shadow-symmetric point where $\Delta = d/2+10^{-5}$, taking care to regulate the $z\to 1$ divergences as discussed in appendix \ref{app:regulation}. The results are shown in figure \ref{fig: c(Delta,J) on shadow symmetric line}.

\begin{figure}[!t]
\centering
\begin{subfigure}{.45\textwidth}
\centering
\includegraphics[height=5cm]{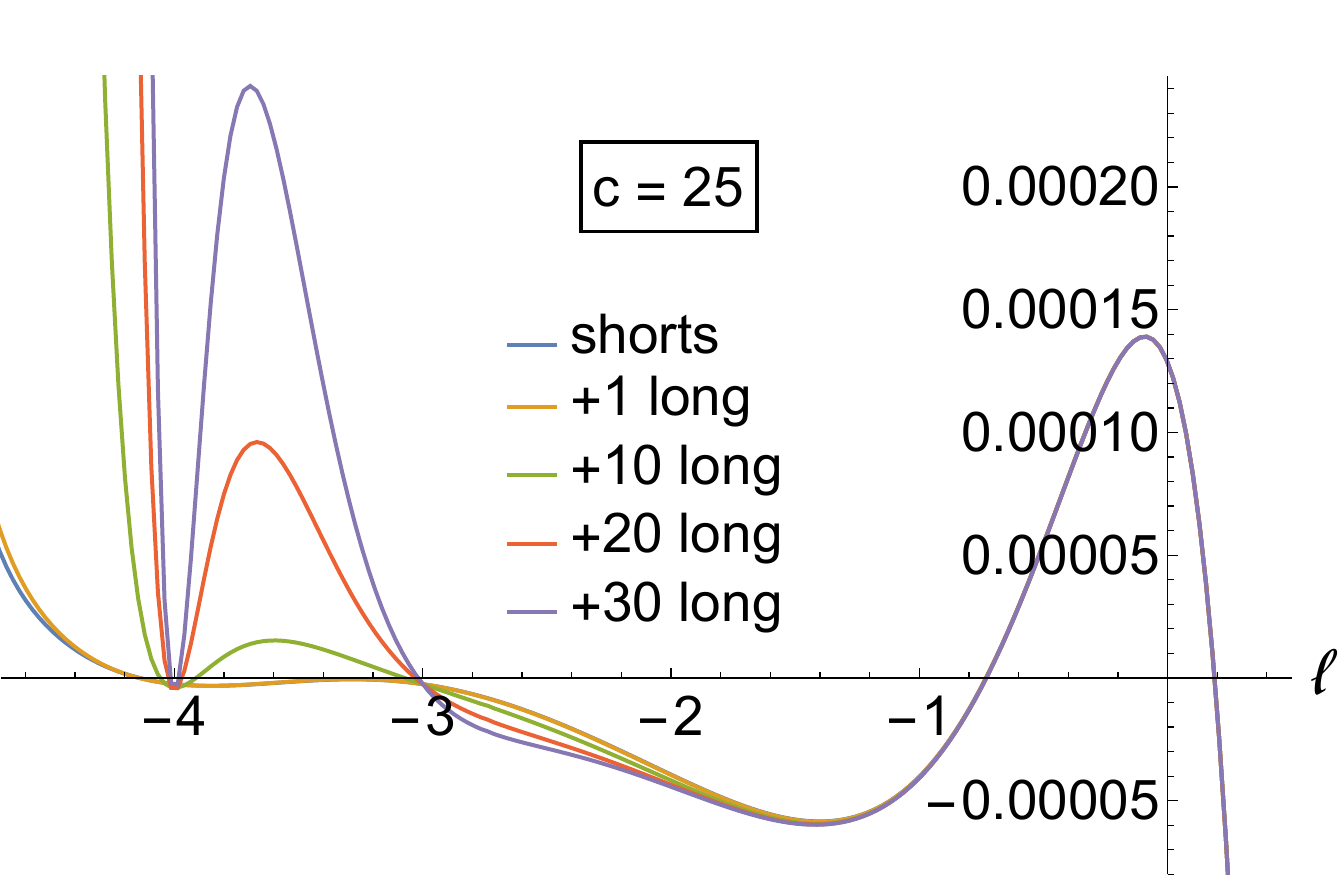}\\
\end{subfigure}\hfill%
\begin{subfigure}{.45\textwidth}
\centering
\includegraphics[height=5cm]{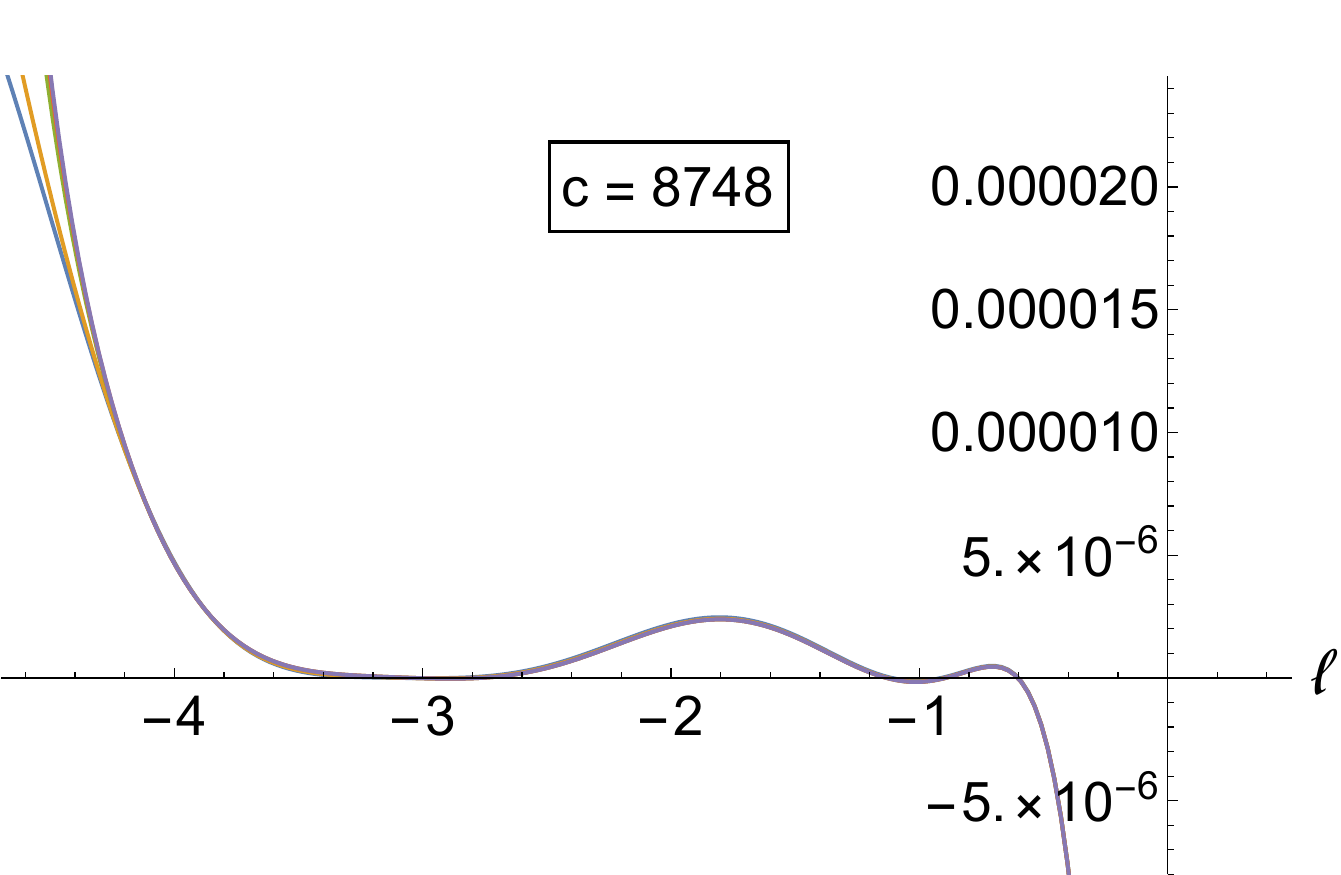}
\end{subfigure}\hfill%
\caption{We plot a crude estimate of $c(\D,\ell)\vert_{\text{phys}}$ as a function of $\ell$ with $\D=3+10^{-5}$ and with $c=25$ (left) and $c= 8748$ (right). The slight displacement from the shadow symmetric line at $\D = 3$ is simply for faster numerics. Here $c(\D,\ell)\vert_{\text{phys}}=(\ell+3)\,c(\D,\ell) / \left[ \k_{\D+\ell}\G((\ell+\D+1)/2)\G((\ell+6-\D+1)/2)\right]$ is just the $c(\D,\ell)$ with all spurious poles divided out.  The long blocks are scalars from increasingly large twist trajectories ($8, 10, 12,...$). Their OPE coefficients are taken from supergravity results for $c=25$ and mean field theory results for $c=8748$. The anomalous dimensions are all set to $-1$.}
\label{fig: c(Delta,J) on shadow symmetric line}
\end{figure}

Note that we have divided out unphysical poles analyzed in subsection \ref{subsec: Analyticity properties of c_Delta_ell}. Although our estimate is extremely crude, we do see convergence for negative spins and a breakdown not until spin $-3$ for $c=25$ and around $-4$ for $c=8748$. Note that the pole at $\ell=-5$ is expected: it is the intercept of the leading short trajectory which we also show in figure \ref{fig:analyticitycdeltaj}.

\subsection{\texorpdfstring{Small $z$ expansion}{Small z expansion}}
\label{subsec:inversion setup}
We would like to use the inversion formula \eqref{ddiscfromtchannelstuff} to extract anomalous dimensions and OPE coefficients. A practical way to do so is to use the small $z$ expansion \cite{Caron-Huot:2017vep}, which works as follows. We first define
\be
h=\frac{\D-\ell}{2},\qquad
\hb=\frac{\D+\ell}{2},
\ee
and rewrite \eqref{eq:inversionformula} as
\be
c(h,\hb) 
= 
\int\limits_0^1 \frac{dz}{2z}z^{-h}
\left(2^{2\hb-5}\,
2\kappa^{0,-2}_{2\hb} z^{h+1}
\int_z^1 d\zb \, 
\mu(z,\zb)\,  \GG_{\hb-h+5}^{(\hb+h-5)}(0,-2;z,\zb) \, \mathrm{dDisc}_t\left[ (z \zb)^6  a(z,\zb)\right]
\right).
\label{eq:c(h,hb)}
\ee
Here we have restricted the range of $\zb$ integration to $z<\bar z <1$ which brings in a factor of $2$. We are also considering only the even spin $c(h,\hb)$ and thus replaced $1+(-1)^\ell=2$. In the limit $z\to 0$ the bracketed expression above can be expanded in powers of $z$ and then the outer integral simply converts these powers into poles of $h$. The OPE coefficients are then given by the coefficients of this power series expansion --- up to a Jacobian to transform from $(h,\hb)$ back to $(\D,\ell)$.

When away from shadow symmetric line we can simplify \eqref{eq:c(h,hb)} by discarding ``half'' of the kernel block. This follows from the fact that one can split a conformal block into two parts:
\begin{align}
\GG_{\ell+d-1}^{(\Delta+1-d)}(\D_{12},\D_{34};z, \bar{z})
&=(\GG^{\rm{pure}})_{\ell+d-1}^{(\Delta+1-d)}(\D_{12},\D_{34};z, \bar{z})\nonumber\\
&\quad+2^{d-2\Delta} \frac{\Gamma(\Delta-1) \Gamma\left(-\Delta+\frac{d}{2}\right)}{\Gamma\left(\Delta-\frac{d}{2}\right) \Gamma(-(\Delta+1-d))} 
(\GG^{\rm{pure}})_{\ell+d-1}^{(-\Delta+1)}(\D_{12},\D_{34};z, \bar{z}),
\label{eq:pure blocks}
\end{align} 
where each of the $\GG^{\rm{pure}}$ can be expanded into pure power terms in the limit $z\ll\zb\ll1$ \cite{Caron-Huot:2017vep}. For example, the contribution from the first $\GG^{\text{pure}}$ to the $\zb$ integral kernel in \eqref{eq:c(h,hb)} has the following expansion
\begin{align}
\begin{split}
2^{\hb+h+1-d}z^{h+1}\mu(z,\zb)(\GG^{\rm{pure}}&)_{\hb-h+d-1}^{(\hb+h+1-d) }(\D_{12},\D_{34};z, \bar{z})
\\
&=
\frac{(1-\zb)^{\frac{\D_{34}-\D_{12}}{2}}}{\zb^2}
\sum_{n=0}^{\infty}\, z^n
\sum_{j=-n}^{n}
B^{\D_{12},\D_{34}}_{n,j}(h,\hb) k^{\D_{12},\D_{34}}_{2(\hb+j)}(\zb)\,,
\end{split}
\label{eq:expansion of zb integral kernel}
\end{align}
where the constant prefactor is inserted to simplify notation, and
\be
k_{2\hb}^{\D_{12},\D_{34}}(z)\colonequals z^{\hb}\, _2F_1\left(\hb-\frac{\D_{12}}{2},\hb+\frac{\D_{34}}{2},2\hb; z\right)\,.
\ee
Similar manipulations can be done for the second pure power term, with one notable difference that the second expansion scales as $z^{\D-d/2}\left(1+O(z)\right)$ which we can ignore when we are to the right of the shadow symmetric line.

The expansion coefficients in \eqref{eq:expansion of zb integral kernel} are easy to fix\footnote{For this it is useful to use the identity \cite{Caron-Huot:2017vep}
\begin{align}
\frac{1}{\bar{z}} k_{2\hb}(\bar{z})
=k_{2\hb-2}(\bar{z})
+\left(\frac{1}{2}+\frac{ \D_{12} \D_{34}}{8\hb(\hb-1)}\right) k_{2\hb}(\bar{z})
+\frac{\left(\D_{12}^{2}-4\hb^{2}\right)\left(\D_{34}^{2}-4\hb^{2}\right)}{64\hb^{2}\left(4\hb^{2}-1\right)} k_{2\hb+2}(\bar{z})\,.
\end{align}}
and the first few coefficients with $d=6,\,\D_{12}=0,\,\D_{34}=-2$ are
\begin{align}
\begin{split}
B_{0,0}(h,\hb)&=1\,,\\
B_{1,1}(h,\hb)&=-\frac{\left(\hb^2-1\right) (h-\hb-2)}{2 (2 \hb-1) (2 \hb+1) (h-\hb-3)}\,,\\
B_{1,0}(h,\hb)&=1-\frac h2 \,,\\
B_{1,-1}(h,\hb)&=-\frac{2 (h+\hb-3)}{h+\hb-4}\,.
\end{split}
\end{align}

After substituting the inverted block with $\GG^{\text{pure}}$ and using the expansion \eqref{eq:expansion of zb integral kernel}, we can define a generating function as
\begin{align}
C(z,\hb+j)=
2^{\hb-h+1}\,
\kappa^{0,-2}_{2\bar h} 
\int_{z}^{1} \frac{d\bar{z}}{\bar z^2 (1-\zb)}
k^{0,-2}_{2(\hb+j)}(\bar{z}) \operatorname{dDisc}_{t}\left[(z \zb)^6  a(z,\zb)\right]\,,
\label{eq:generatingfunction}
\end{align}
such that \eqref{eq:c(h,hb)} can be rewritten as
\begin{align}
\begin{split}
c(h,\bar h) 
&=\int_{0}^{1} \frac{dz}{2z} z^{-h} 
\left(\sum_{n,j} z^n B_{n,j}(h,\hb) C(z,\bar h+j)\right)\\
&\equiv\int_{0}^{1} \frac{dz}{2z} z^{-h}\, \widetilde C(z,h,\hb)\,.
\end{split}
\label{eq:definition of C tilde}
\end{align}

In the existing literature the small $z$ expansion of the inversion formula is generally used to extract CFT data of the leading trajectory from the generating function. For such a case we would only need the leading ($n=j=0$) term. We may write
\begin{align}
C(z,\hb) \approx P(\hb)z^{h(\hb)}\,,
\end{align}
and the OPE data for the leading trajectory can then be extracted as
\begin{align}
h(\hb)=\lim_{z\to0}\frac{z \partial_z C(z,\hb)}{C(z,\hb)}\,,
\qquad
P(\hb)=\lim_{z\to0} \frac{C(z,\hb)}{z^{h(\hb)}}\,,
\label{eq: leading twist data extraction}
\end{align}
where $h(\hb)=h_{\infty}+\delta h(\hb)$ with $h_{\infty}$ the double-trace value and $\delta h(\hb) \to 0$ as $\hb \to \infty$. The anomalous dimension can be determined by solving the equation
\begin{align}
\d h(\hb)=\hb-\ell-h_{\infty}\,.
\label{eq:equation for delta-h}
\end{align}
and the squared OPE coefficient can be calculated through
\begin{align}
\l^2_{\D,\ell} = \left(1-\frac{\partial h(\hb)}{\partial \hb}\right)^{-1} \cdot P(\hb)\,,
\label{eq:Jacobian for OPE coefficient}
\end{align}
where the Jacobian factor is needed because $\l^2_{\D,\ell}$ is the residue of $c(\D,\ell)$ with respect to $\D$ at fixed $\ell$.

In our case the setup is slightly different. A quick look at figure \ref{fig:analyticitycdeltaj} shows that for all non-negative spins the leading two trajectories are expected to be straight: they correspond respectively to the short multiplets belonging to the chiral algebra (with known coefficients) and the short multiplets at the unitarity bound of the long multiplet (with unknown coefficients). After that we find the leading unprotected trajectory. In equations this means that, if we write the generic power expansion ansatz,
\be
\widetilde{C}(z,h,\hb)\approx\sum_{k} P_k(h,\hb)z^{h_k(\hb)}\,,
\label{eq:power expansion ansatz for higher twist}
\ee
then in our case
\be
\label{power expansion ansatz with protected terms}
\widetilde{C}(z,h,\hb) \approx P_4(h,\hb) z^4 + P_5(h,\hb) z^5 + \sum_{k} P_k(h,\hb)z^{h_k(\hb)}\,,
\ee
with the sum running over all the unprotected Regge trajectories.

\subsubsection*{Consistency with the conformal block decompositions}

Let us now offer some comments on the consistency of \eqref{power expansion ansatz with protected terms} with the conformal block decompositions. Firstly, it agrees (as it should) with the $s$-channel expansion of $a(z,\zb)$: taking all the prefactors into account it is easy to see that the contributions from $a^{\text{at}}_{\ell + 4, \ell}$ begin at $z^4$, those from $a^{\text{at}}_{\ell + 6, \ell}$ at $z^5$, and those from the long multiplets at even higher powers; for example, in mean field theory the leading unprotected trajectory begins to contribute at the $z^6$ term.

Next, let us try to discern the origin of the various powers in \eqref{power expansion ansatz with protected terms} from the $t$-channel perspective. (In the footnote below we discuss the obvious issue of convergence.) To do so we have to remember the inhomogeneous crossing symmetry equation for $\dDisc_t\left[ (z\zb)^6 a(z,\zb) \right]$, which concretely speaking leads to the extra terms displayed in equation \eqref{ddiscfromtchannelstuff}. These (known) terms already result in $z^4$ and $z^5$ terms (without logarithms) at small $z$, and in the next subsection we will show that the $z^4$ term precisely reproduces the OPE coefficients of the short multiplets that contribute to the chiral algebra. This leaves us with the sum over $t$-channel blocks corresponding to the unknown operators. At small $z$, a single $t$-channel block looks like:\begin{multline} \label{tchannelblocksmallz}
\left(\frac{z \zb}{(1-z)(1-\zb)}\right)^5 \GG_{\D}^{(\ell)}(0,-2;1-z,1-\zb)
=\\
\sum_{n=0}^{\infty} z^{n + 5} H_{\D,\ell}^{(1), n}(1-\bar{z})
+
\sum_{n=0}^{\infty} z^{n + 6} \log(z) H_{\D,\ell}^{(2), n}(1-\bar{z})\,,
\end{multline}
where the prefactor on the left-hand side matches that of the crossing equation \eqref{ddiscfromtchannelstuff} and the explicit form of the $H$ functions will not be important for us. Remarkably, this is precisely the structure that we would expect from equation \eqref{power expansion ansatz with protected terms}: no contribution to the $z^4$ term, which is therefore determined entirely by the inhomogeneous terms in the crossing symmetry equation, a non-zero $z^5$ term, which means a non-zero contribution to the short multiplets with undetermined coefficients, and a logarithm signifying an anomalous dimension starting only at order $z^6$.\footnote{An important caveat to the above analysis is that an infinite sum over $t$-channel blocks need not have the same behavior as a single block, and so using equation \eqref{tchannelblocksmallz} in the crossing symmetry equation does not constitute a completely reliable estimate for the small $z$ limit of $a(z,\zb)$. In fact, a comparison of the (known) $z^4$ term on both sides of the crossing equations does show that the $t$-channel blocks must sum up to give non-zero $z^4$ and $z^5$ terms in $(z\zb)^6 a(z,\zb)$. However, it is also easy to see that these terms have vanishing double discontinuity and are therefore unimportant in the Lorentzian inversion formula. Also, each $t$-channel block only contributes with $\log(z)$ and integer $z$ powers, while finite anomalous dimensions (with respect to exact double-twist dimensions) require higher powers of the $\log(z)$.}

In the next subsection we will first invert the  $z^4$ term in order to recover the known OPE coefficients of the chiral algebra. In section \ref{sec:numerics} we will numerically estimate some of the unknown OPE data corresponding to the $z^5$ term and beyond. 

\subsubsection{Recovering the chiral algebra shorts}
\label{subsec: recovering shorts}
In agreement with the previous discussion we will take the coefficient of $z^4$ from the inhomogeneous terms in the crossing equation \eqref{ddiscfromtchannelstuff} and ignore any contribution from the unprotected multiplets. Substituting this into equation \eqref{eq:generatingfunction} for the generating function, the integral to do is then:
\begin{align}
\begin{aligned}
C(z,\bar{h})\Big|_{z^4}=&
2^{\hb-h+1}\,
\kappa^{0,-2}_{2\bar h} \int_{0}^{1} \frac{d\bar{z}}{\bar z^2 (1-\zb)}
k^{0,-2}_{2\bar{h}}(\bar{z}) \\
&\times
\operatorname{dDisc}_{t}\left[
\frac{1}{3} \frac{1}{(1-\zb)^3}-\frac{4}{3}\frac{1}{(1-\zb)^2}+\left(2+\frac{8}{c}\right)\frac{1}{1-\zb}+\frac{8}{c} \log (1-\zb)
\right]\,,
\end{aligned}
\end{align}
Using the factor $4/(\D_{\text{s.c.p.}}-\ell-2)(\D_{\text{s.c.p.}}+\ell+2)$ to convert to physical OPE coefficient, we have
\begin{align}
C(z,\bar{h}=\ell+4)\Big|_{z^4}=
\frac{\lambda_{\BB[2,0]_{\ell-2} }^2}{\ell+3}\,.
\end{align}
and with a bit of algebra one then finds
\be 
\lambda_{\BB[2,0]_{\ell-2} }^2= 
\frac{\sqrt{\pi }  2^{-\ell -7} (\ell +1) (\ell +2) (\ell +3) \Gamma (\ell +7)}{9 \Gamma \left(\ell +\frac{7}{2}\right)}
+
\frac{\sqrt{\pi }  2^{-\ell -2} \left(\ell ^2+7 \ell +11\right) \Gamma (\ell +4)}{c \; \Gamma \left(\ell +\frac{7}{2}\right)}\,.
\label{eq:chiral algebra OPE coef}
\ee
It is easy to check that the result matches $2^\ell b_\ell$ for all even spins, but is a nicer function of $\ell$.

\section{Numerical approximations}
\label{sec:numerics}

In this section we will use the small $z$ expansion of subsection \ref{subsec:inversion setup} for some numerical experiments. Our main question is whether the input of some limited $t$-channel OPE data can produce a reliable approximation of the $s$-channel OPE data. We will mostly focus on the low-spin operators in the leading Regge trajectory which will allow us to make comparisons to the numerical bootstrap results of \cite{Beem:2015aoa}.

\subsection{Inversion for higher-twist trajectories}
\label{subsec: iterative inversion procedure}
In this section we generalize \eqref{eq: leading twist data extraction}, which states how to extract the leading-twist data from the inversion formula, to higher-twist trajectories. This amounts to extracting the coefficient and powers of the $z^5$ and higher terms in $\widetilde C(z,h,\hb)$ shown in equation \eqref{power expansion ansatz with protected terms}. These terms get contributions from the long multiplets, which we do not know exactly, and so we use an approximation scheme where we input only finitely many long $t$-channel blocks. For the $z^5$ term the scheme is obvious: we just sum the coefficients of all the $z^5$ terms of whatever $t$-channel blocks we put into the inversion formula. The resulting coefficient in the generating function directly provides an estimate for the OPE coefficients of the $\DD[0,4]$ and $\BB[0,2]_\ell$ short multiplets. The integrals to be done are detailed in subsection \ref{subsubsec:shortsatbound} below.

For the unprotected multiplets the ansatz \eqref{power expansion ansatz with protected terms} in principle dictates that we should compute:
\begin{align}
h_n(\hb)=\lim_{z\to0}\frac{z \partial_z \widetilde{C}^{\text{sub.}}_n(z,h,\hb)}{\widetilde{C}^{\text{sub.}}_n(z,h,\hb)}\,,
\qquad
P_n(h,\hb)=\lim_{z\to0} \frac{\widetilde{C}^{\text{sub.}}_n(z,h,\hb)}{z^{h_n(\hb)}}\,.
\label{eq: higher twist data extraction}
\end{align}
where the `sub.' superscript indicates that we should subtract the contribution of the more leading terms in the small $z$ expansion before taking the limit. This prescription however runs into the familiar problem that each $t$-channel block contributes only a $\log(z)z^6$ term as $z \to 0$, and such terms need to exponentiate to recover the expected $z^{\hb(h)}$ behavior, signaling that the $t$-channel sum does not commute with taking $z \to 0$ term by term. In order to remedy this we will evaluate the right-hand sides of \eqref{eq: higher twist data extraction} at finite but small $z$.

\subsubsection{Finite $z$ inversion}
The essence of the ``finite-$z$ inversion'' method \cite{Simmons-Duffin:2016wlq,Albayrak:2019gnz,Caron-Huot:2020ouj} is to replace equation \eqref{eq: higher twist data extraction} with an estimate at small but finite $z_0$
\begin{align}
\begin{split}
h_n(\hb)&\simeq h_{z_0,n}(h,\hb)=\frac{z \partial_z \widetilde{C}^{\text{sub.}}_n(z,h,\hb)}{\widetilde{C}^{\text{sub.}}_n(z,h,\hb)}\Bigg|_{z=z_0}\,,
\\
P_n(h,\hb)&\simeq P_{z_0,n}(h,\hb)= \frac{\widetilde{C}^{\text{sub.}}_n(z,h,\hb)}{z^{h_{z_0,n}(h,\hb)}}\Bigg|_{z=z_0}\,.
\end{split}
\label{eq: higher twist data extraction finite z}
\end{align}
Notice that at finite $z_0$ higher order terms in the expansion \eqref{power expansion ansatz with protected terms} also contribute and this introduces $h$-dependence into the approximation of $h_n(\hb)$ which is therefore now denoted as $h_{z_0,n}(h,\hb)$. Equation \eqref{eq:equation for delta-h} is modified into
\begin{align}
h_{z_0,n}(h(\hb),\hb)=h_{z_0,n}(\hb-\ell,\hb)=\hb-\ell\,.
\label{eq:h approximation finite z}
\end{align}
Since the exact answer is independent of $h$, a weak $h$-dependence of $h_{z_0,n}$ should be a sign of its good approximation to $h(\hb)$. Note also that now we have distinct equations to solve for each spin $\ell$ -- a significant difference compared to the more common (practical) analyses of the leading Regge trajectory.

The value of $z_0$ in \eqref{eq: higher twist data extraction finite z} is crucial to the approximation results we get, but before discussing the determination of $z_0$, let us first introduce another variable $y$ which turns out to be useful for the inversion calculation. The $y$ variable is defined such that the inversion of the ``generalized free'' part in \eqref{ddiscfromtchannelstuff}, i.e., the protected part of $a(z,\zb)$ in the limit $c\to\infty$, gives exactly zero anomalous dimensions \emph{even when} using the finite-$z$ inversion \eqref{eq: higher twist data extraction finite z}. Denoting the inversion result as $\widetilde{C}_n^{\text{sub.}}(z,h,\hb)\vert_{\text{gf}}$, we have
\begin{align}
y^n(z,h,\hb)\colonequals \widetilde{C}_n^{\text{sub.}}(z,h,\hb)\vert_{\text{gf}}\,,
\label{eq: definition of y variable}
\end{align}
and by construction this gives
\begin{align}
\begin{split}
\frac{y \partial_y \widetilde{C}_n^{\text{sub.}}(z,h,\hb)\vert_{\text{gf}}}
{\widetilde{C}_n^{\text{sub.}}(z,h,\hb)\vert_{\text{gf}}}
=n\,,
\end{split}
\end{align}
for any $z_0$. This also gives reliable results for anomalous dimensions of operators with large spin which asymptote to generalized free operators. The switch from $z$ to $y$ introduces a Jacobian factor and \eqref{eq: higher twist data extraction finite z} becomes
\begin{align}
\begin{split}
h_{y_0,n}(h,\hb)
&=\frac{y \partial_y \widetilde{C}^{\text{sub.}}_n(z(y),h,\hb)}{\widetilde{C}^{\text{sub.}}_n(z(y),h,\hb)}\Bigg|_{y=y_0}
=\left(\frac{y \partial_y z} {z} \bigg|_{z=z_0}\right)
\cdot h_{z_0,n}(h,\hb)\\
&=\frac{n y^n(z,h,\hb)}{\partial_z y^n(z,h,\hb)}
\frac{\partial_z \widetilde{C}^{\text{sub.}}_n(z,h,\hb)}{\widetilde{C}^{\text{sub.}}_n(z,h,\hb)}\Bigg|_{z=z_0}
\,,
\end{split}
\label{eq: higher twist anomalous dimension extraction finite y}
\end{align}
where the Jacobian is calculated by taking a derivative on both sides of \eqref{eq: definition of y variable}, and\footnote{Notice that we extract the coefficient of $z^{h_{y_0,n}}$ instead of $y^{h_{y_0,n}}$, because, as can be seen from \eqref{eq: definition of y variable}, $y$ is defined to include the ``generalized free'' OPE coefficient. One could also have defined $y$ dividing the right-hand side of \eqref{eq: definition of y variable} by the OPE coefficient, and then we would extract the coefficient of $y^{h_{y_0,n}}$.}
\begin{align}
P_{y_0,n}(h,\hb)&= \frac{\widetilde{C}^{\text{sub.}}_n(z,h,\hb)}{z^{h_{y_0,n}(h,\hb)}}\Bigg|_{z=z_0}\,.
\label{eq: higher twist OPE extraction finite y}
\end{align}
Although $y$ is conceptually a nicer variable, from \eqref{eq: definition of y variable} we see that it depends on $z,h,\hb$, which makes it practically more complicated. Therefore, we will write $y_0=y(z_0,h,\hb)$ and all the finite-$y$ inversions in this paper will be performed at fixed $z_0$ rather than $y_0$.

\begin{figure}[!t]
\centering
\includegraphics[width=\textwidth]{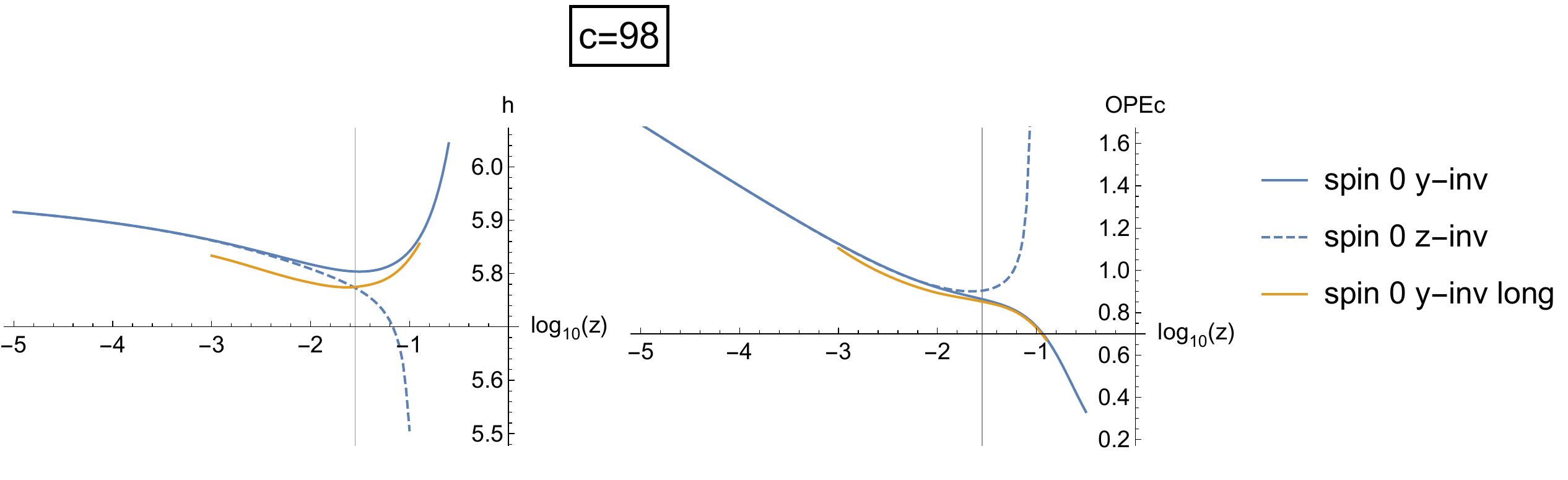}
\caption{Short and long (up to spin 16) inversion results for spin 0 long multiplet with $c=98$. The vertical lines on both plots pass through $\log_{10} z=-1.55$, and this is our choice for the finite $y$ inversion. The orange curves indicate that the plateau regions remain largely unchanged after adding the long multiplet contribution. We also included plots using the $z$ variable for comparison.}
\label{fig: determine the finite z}
\end{figure}

To find the suitable value of $z_0$ (or $y_0$), we will plot both $h_{z_0,n}(h,\hb)$ (or $h_{y_0,n}(h,\hb)$) and $P_{z_0,n}(h,\hb)$ (or $P_{y_0,n}(h,\hb)$) as functions of $z$ and look for a plateau, similar to the approach in \cite{Caron-Huot:2020ouj}. Because of the relatively small amount of unprotected OPE data available to us, we will first pick a value for $z_0$ based on inversion of only the protected part of $a(z,\zb)$ (called \emph{short inversion}). After extracting the unprotected CFT data (see next paragraph for details) we will check if the selected $z_0$ still sits on a plateau when inverting both protected and unprotected part of $a(z,\zb)$ (called \emph{long inversion}). This is illustrated in figure \ref{fig: determine the finite z} where we consider the leading long scalar multiplet with $c=98$. By repeating this procedure for other values of central charge and spin, one will find different suitable values for $z_0$.

\subsubsection{Iterative inversion procedure}
To test whether we can bootstrap the CFT data out of the protected sector we will adopt an iterative procedure. First we invert only the short multiplets to obtain a set of estimated data, denoted as $iter_0$, for the multiplets on the leading long trajectory up to spin $\ell_{\text{max}}=16$.\footnote{It is straightforward to extend to higher $\ell_{\text{max}}$, but we find that this does not noticeably change the results. We do expect, however, that including higher-twist long trajectories can improve the results significantly (see subsection \ref{subsec:numerical results}).} Then we invert these long multiplets from $iter_0$ to obtain a new set of CFT data, $iter_1$, again for the leading long trajectory again up to $\ell_{\text{max}}$. Then we iterate for several times until the data converges  and $iter_n$ will be our final numerical estimate. In practice we found that $2\leqslant n\leqslant 5$ is sufficient, depending on the central charge. As mentioned before the value of $z_0$ for finite-$y(z)$ inversion should be determined separately for each spin. However, for large spins the anomalous dimensions are highly suppressed and vary little as we change $z_0$, thus in practice we can use the same $z_0$ for spins $\ell\geqslant 2$ or 4 depending on the central charge.

\subsubsection{Extract the OPE coefficients of the shorts at bound}
\label{subsubsec:shortsatbound}
With an estimate for the long multiplets in hand we can invert for the OPE coefficients of the non-chiral algebra short multiplets.\footnote{Recall that the $t$-channel non-chiral algebra short multiplets have vanishing double discontinuity and therefore they do not participate in the iteration procedure, as discussed in section~\ref{sec:practicaltchanel}.} This can be done similarly as in section \ref{subsec: recovering shorts}. The main differences are that we need to extract the coefficient of $z^5$ term rather than the $z^4$ term, and invert both protected and unprotected operators.

In equations we have
\begin{align}
\widetilde{C}(z,h=5,\bar{h}=\ell+5)\Big|_{z^5}=
\frac{\l^2_{\BB[0,2]_{\ell-1}}}{2\ell+8}
=
\frac{\big(\l^{\text{short}}\big)^2_{\BB[0,2]_{\ell-1}}+\big(\l^{\text{long}}\big)^2_{\BB[0,2]_{\ell-1}}}{2\ell+8}\,,
\end{align}
and $\l^2_{\DD[0,4]}=\lim_{\ell\to -1}\l^2_{\BB[0,2]_{\ell}}$.
Inverting the protected part in a similar manner as before yields the contribution from the short multiplets in the $t$-channel:
\begin{align}
\begin{split}
\big(\l^{\text{short}}\big)^2_{\BB[0,2]_{\ell-1}}=
&\frac{\sqrt{\pi } 2^{-\ell-9} (\ell+1) (\ell+2) (\ell+4) (\ell+7) (\ell+8) \Gamma (\ell+6)}{9 \Gamma \left(\ell+\frac{9}{2}\right)}\\
&
-\frac{5 \sqrt{\pi } 2^{-\ell-4} (\ell+4) (\ell (\ell+9)+17) \Gamma (\ell+6)}{c (\ell+3) (\ell+6) \Gamma \left(\ell+\frac{9}{2}\right)}\,.
\end{split}
\label{eq:OPE short and bound from shorts}
\end{align}
As for the long multiplets, using the double discontinuity expression \eqref{ddiscfromtchannelstuff} and the expression generating function \eqref{eq:generatingfunction}, we find that they contribute:
\begin{align}
\begin{split}
\big(\l^{\text{long}}\big)^2_{\BB[0,2]_{\ell-1}}=
&2^{\ell}(2\ell+8)
\sum_{\D'_{\text{s.c.p.}}>\ell'+6,\ell'}
\frac{-8\l^2_{\D'_{\text{s.c.p.}},\ell'}\sin^2\left((\D'_{\text{s.c.p.}}+\ell')\pi/2\right)}{(\D'_{\text{s.c.p.}}-\ell'-2)(\D'_{\text{s.c.p.}}+\ell'+2)}
\\
&\times
2\k^{0,-2}_{2(\ell+5)}\int_0^1 d\zb \frac{1}{\zb^2(1-\zb)} k_{2(\ell+5)}^{0,-2}(\zb)\left(\frac{\zb}{1-\zb}\right)^5 \GG_{\D'_{\text{s.c.p.}}+4}^{(\ell')}(0,-2;1,1-\zb)\,,
\end{split}
\label{eq:OPE short and bound from longs}
\end{align}
where we have used primed notation for $t$-channel quantum numbers. Note that the $t$-channel block is evaluated at $z=0$, which gives a finite linear combination of eight hypergeometric functions $_2F_1(\ldots,1-\zb)$. Therefore, for each 6d conformal block the $\zb$ integral breaks down into eight atomic integrals and they can be performed analytically using the method in \cite{Albayrak:2019gnz}. 

Equation \eqref{eq:OPE short and bound from longs} presents a kind of `supersymmetric sum rule' for the OPE data. We see no reason why it could not be absolutely convergent for all spins. One interesting application could then be the vanishing OPE coefficient of the $\DD[0,4]$ multiplet for the $A_1$ theory with $c = 25$ which was discussed already in \cite{Beem:2015aoa}. In that case the positive contribution from equation \eqref{eq:OPE short and bound from shorts} needs to be offset by the uniformly negative contributions from the long multiplets in equation \eqref{eq:OPE short and bound from longs}. More generally, taking only finitely many $t$-channel blocks into account would provide an upper bound on the OPE coefficients.

\subsection{Numerical results}
\label{subsec:numerical results}
In this subsection we present the numerical estimates for the following unknown CFT data:\footnote{Estimates for higher spins (up to $\ell=16$) are available from the authors on request.}
\begin{itemize}
	\item conformal dimensions of the leading long multiplets $\D_{\ell},\ \ell=0,2,4,6$,

	\item OPE coefficients of the leading long multiplets $\lambda^2_{\LL[0,0]_{\D_{\ell},\ell}},\ \ell=0,2,4,6$,

	\item OPE coefficients of non-chiral algebra short multiplets $\lambda^2_{\DD[0,4]}$ and $\lambda^2_{\BB[0,2]_{\ell-1}}$ with $\ell=2,4,6$.
\end{itemize}
We will plot all of this data as a function of the central charge $c$. As a reminder, we note that for the $A_{N-1}$ and $D_N$ theories
\begin{align}
c(A_{N-1})=4N^3-3N-1\,,
\qquad
c(D_{N})=16N^3-24N^2+9N\,.
\end{align}

In each of the three cases we will take as our initial input the inhomogeneous contribution to the crossing equation, \emph{i.e.}, the last two lines on the right-hand side of \eqref{ddiscfromtchannelstuff}. At first order in the iteration this yields something that bears qualitative similarities to the supergravity answer, but with the important difference that we work at finite $c$ and finite $z$ -- so the familiar derivatives of conformal blocks now get replaced with an approximation consisting of regular conformal blocks. We then feed the answers for the leading Regge trajectory into the right-hand side of \eqref{ddiscfromtchannelstuff} and iterate a few times as described in the previous subsection.

Besides this straightforward iteration scheme, we have also attempted to improve our results for the $A_1$ theory with $c = 25$ with a small variation where we input the following numerical bootstrap data \cite{Beem:2015aoa}:
\begin{align}
\D_0\lesssim 6.4\,,\quad 
\lambda^2_{\LL[0,0]_{\ell=0}}\lesssim 1.3\,,\quad
\D_2\lesssim9.4\,.
\label{eq:numerical upper bounds}
\end{align}
These are respectively the dimension and coefficient of the first scalar and the dimension of the first spin 2 operator. These values are rough extrapolations of the bootstrap bounds of \cite{Beem:2015aoa} which are believed to be saturated for the $(2,0)$ theory with $c = 25$. To incorporate these values we have simply replaced the corresponding OPE data in the output of an iteration with \eqref{eq:numerical upper bounds} before feeding it back into the next iteration.
 
\subsubsection{Dimensions of leading long multiplets}
\begin{figure}[!th]
\centering
\includegraphics[width=1\textwidth]{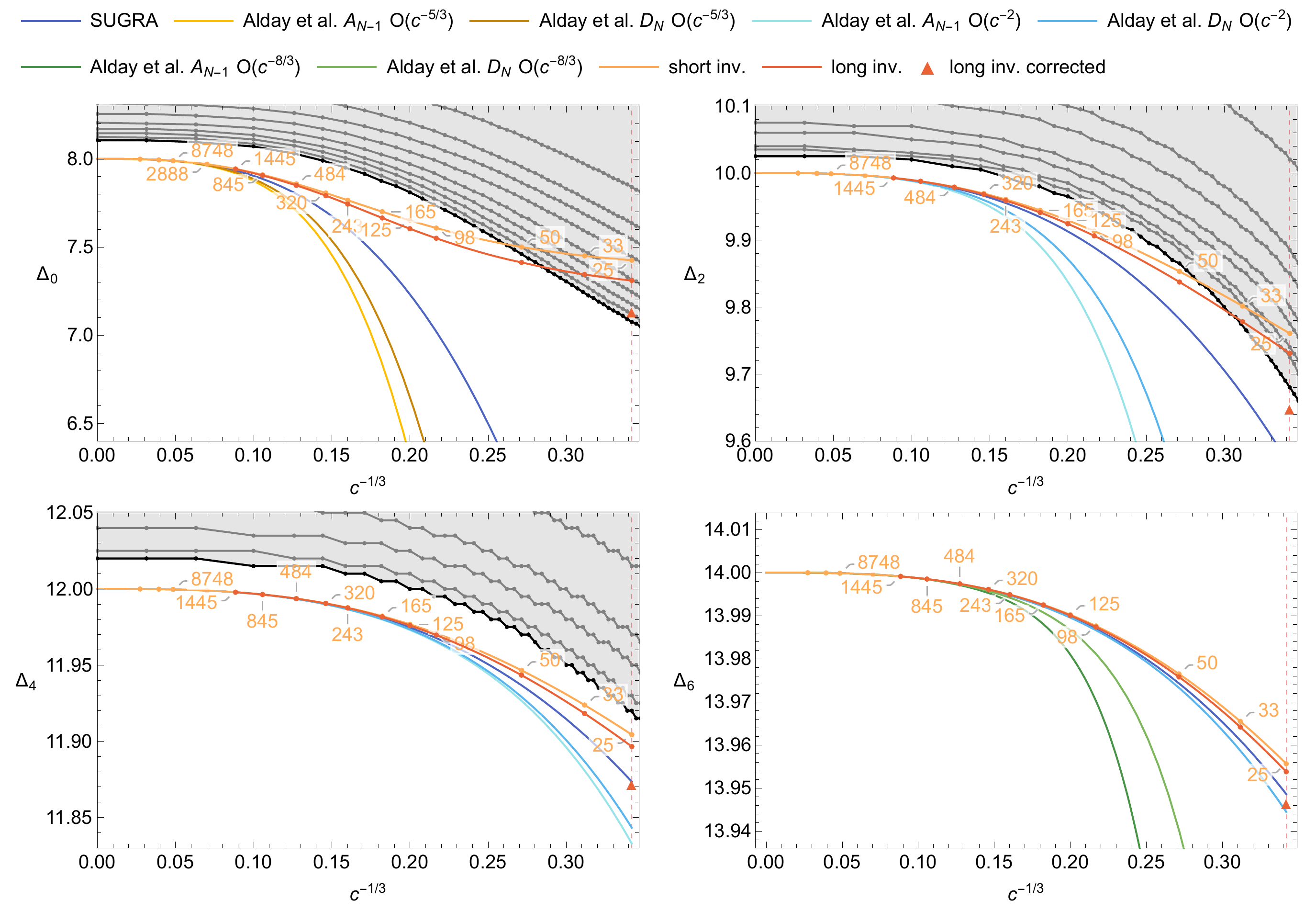}
\caption{Dimensions of the leading long multiplets with spin 0, 2, 4, 6, against $c^{-1/3}$. The gray shaded regions are numerical upper bounds (available for spin 0, 2, 4) from \cite{Beem:2015aoa}. The orange numbers are central charges (a strict subset of the chosen values correspond to physical theories). The orange curves correspond to short inversions and the red curves correspond to stabilized iterative long inversions up to spin 16 (see text for a detailed explanation). For $c=25$, we also perform a corrected iterative long inversion, taking into account the numerical results of \cite{Beem:2015aoa}, and the result is indicated by a triangular point. We also show $1/c$ results from two-derivative supergravity and higher order $c^\a$ corrections (denoted as $O(c^\a)$ in the plot legend) from \cite{Chester:2018dga,Alday:2020tgi} whenever available.}
\label{fig: long dimensions}
\end{figure}

In figure \ref{fig: long dimensions} we present the conformal dimensions of the multiplets in the leading long Regge trajectory for the first few spins as functions of $c^{-1/3}$.\footnote{The reason of using $c^{-1/3}$ is that this helps separate the large $c$ results further than using $c^{-1}$, and $c^{-1/3}$ is proportional to $N^{-1}$ in the large $N$ limit.} We include both the results from \emph{short inversion} (the starting point of the iteration) and \emph{long inversion} (the fixed point of the iterated inversion for the leading long trajectory).

The shaded regions in the plots correspond to the numerical upper bounds from \cite{Beem:2015aoa} which are available for $\ell=0,2,4.$\footnote{Note that the upper bound for leading long scalar at $c=25$ in figure \ref{fig: long dimensions} is around $7.1$ instead of $6.4$ as shown in \eqref{eq:numerical upper bounds}, because these gray shaded bounds are obtained with $\DD[0,4]$ multiplet present while the upper bound $\D_{0}\lesssim6.4$ is obtained by removing $\DD[0,4]$ by hand. See \cite{Beem:2015aoa} for more details. Also, the steps occurring in the spin 4 bound are simply a numerical artifact due to an early termination of the binary search.\label{foot:binarysearch}} We also plot some holographic results: firstly the two-derivative supergravity solution \cite{Arutyunov:2002ff,Heslop:2004du}, which agrees with large-spin perturbation theory, secondly the $c^{-5/3}$ correction for the $A_n$ and $D_n$ theories obtained in \cite{Alday:2020tgi} which arises from higher-derivative terms in the bulk, and thirdly the $c^{-2}$ and $c^{-8/3}$ corrections which arise from bulk loops and which were determined from the tree-level data also in \cite{Alday:2020tgi}.\footnote{The computation in \cite{Alday:2020tgi} left some coefficients undetermined. We only show a comparison with this data for higher spins where these coefficients have no effect.}  Although these curves only correspond to the first few terms in the large $c$ expansion, we have nevertheless plotted them down to finite values of $c$.\footnote{We do not expect a large $c$ expansion to converge, and the regime where finite $c$ answers qualitatively agree with (non-resummed) holographic computations might end up being very small.} 

Our best results are those for intermediate values of $c$ where we differ significantly from the supergravity results and more closely trace the numerical bounds. For example, for $10^2 \lesssim c \lesssim 10^3$ the numerics indicate the existence of an extremal solution to the crossing equations with a fairly large $\Delta_0$. The holographic results significantly underestimate this gap, but our repeated inversion formula tracks it much more reliably. Our iteration scheme is less successful for the only non-trivial theory with $c < 98$, which is the $A_1$ theory with $c = 25$: at this point our estimates sometimes even exceed the numerical bounds. This likely happens because the anomalous dimensions are too large and recovering them correctly in the $s$-channel requires the input of (many) more conformal blocks in the $t$-channel.\footnote{We have also attempted to include one more Regge trajectory in the $t$-channel but this did not significantly change the results. At a superficial level this is because anomalous dimensions and OPE coefficients go to zero quickly for higher trajectories, yielding suppressions in the Lorentzian inversion formula. This leads us to suspect that one may need to include more than just the double-twist Regge trajectories (of the external operators) in the $t$-channel, but we have not investigated this further.} For comparison: in the three-dimensional Ising model the largest anomalous dimension for a physical operator on the leading Regge trajectory is approximately 0.036 (the stress tensor), whereas for $c = 25$ the the scalar (with dimension $\Delta_0$) has an anomalous dimension of about 1.6.

As shown in the plot, the $c^{-5/3}$ corrections are different for the $A_n$ and $D_n$ series of theories. This raises the interesting question whether our iterative inversion scheme could potentially also recover this difference. There might, for example, be different fixed points of our procedure which one could try to find by starting with different initial conditions, for example by an additional input of the $c^{-5/3}$ corrections. In this way we could hope to iterate towards solutions that the numerical bootstrap cannot find. It would be very interesting to try this in future work.

\begin{figure}[!t]
\centering
\includegraphics[width=\textwidth]{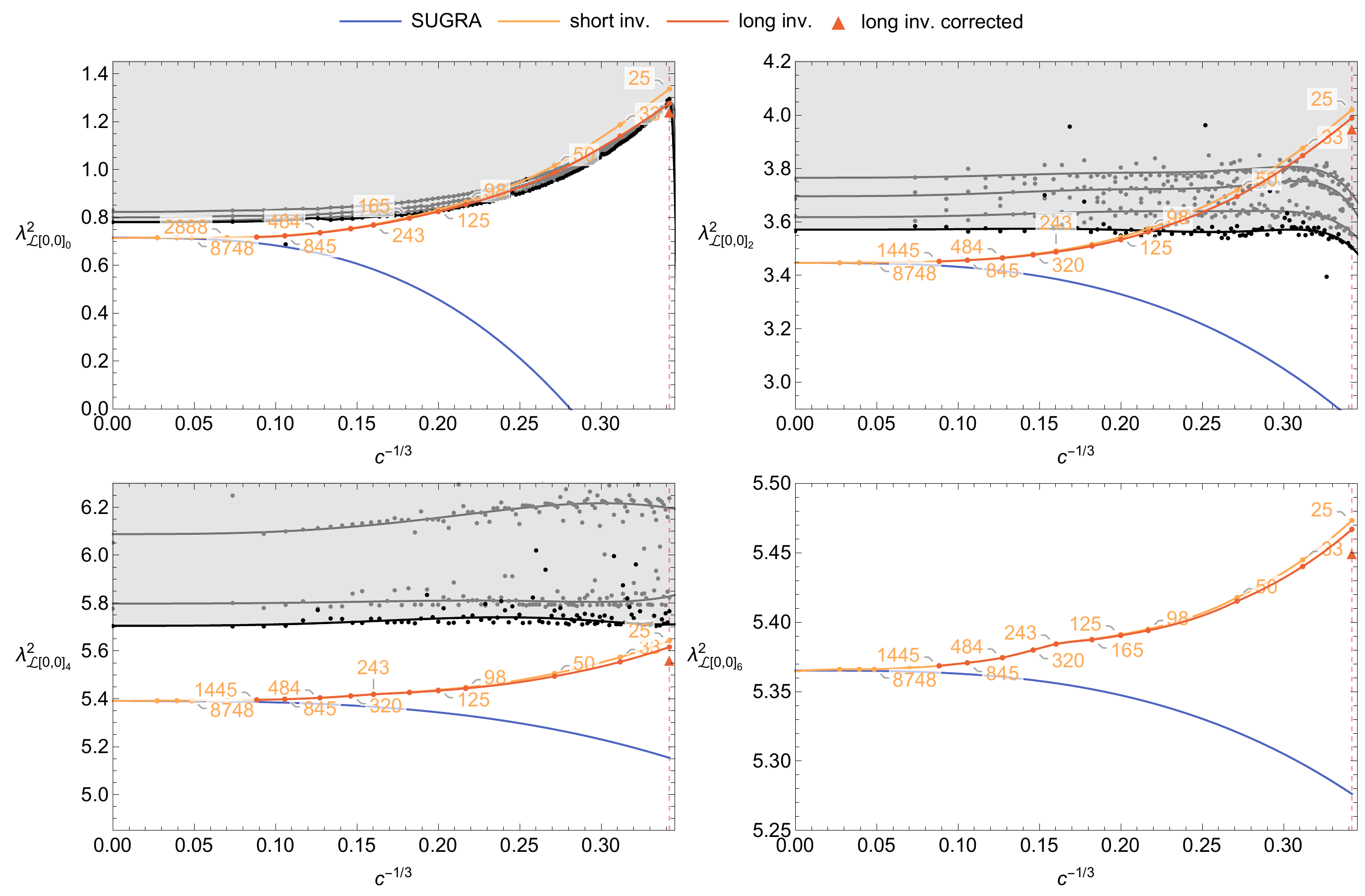}
\caption{OPE coefficients of leading long multiplets with spin 0, 2, 4, 6, against $c^{-1/3}$. The gray shaded regions are unpublished numerical upper bounds (available for spin 0, 2, 4) from \cite{Beem:2015aoa}. The orange numbers are central charges (a strict subset of the chosen values correspond to physical theories). The orange curves correspond to short inversions and the red curves correspond to stabilized iterative long inversions up to spin 16 (see text for detailed explanation). For $c=25$, we also perform a corrected (triangular point) iterative long inversion using the numerical results of  \cite{Beem:2015aoa}.}
\label{fig: OPE long}
\end{figure}

\subsubsection{OPE coefficients of leading long multiplets}

In figure \ref{fig: OPE long} we present the OPE coefficients of the leading long multiplets for the first few lowest spins as functions of $c^{-1/3}$. Similar to before, we show results from short inversion, the corrected and uncorrected long inversions, together with supergravity results and numerical estimates.\footnote{The numerical results are unpublished results from \cite{Beem:2015aoa}. They correspond to upper bounds for the squared OPE coefficients \emph{under the assumption that the corresponding conformal dimensions saturate their own bounds}. They are therefore similar to those shown in figure 12 of \cite{Beem:2015aoa}. Since the inversion procedure does not produce exactly the same conformal dimensions, these upper bounds on OPE coefficients are not entirely applicable, but we do expect them to provide decent estimates. The bounds appear jittery due to an imprecise determination of the long dimensions following from an early termination of the binary search -- see footnote~\ref{foot:binarysearch}, and also due to insufficient numerical precision as noted in \cite{Beem:2015aoa}.} The picture is rather similar: for intermediate central charges we much more accurately trace the numerical bounds (especially in the scalar sector) than the supergravity result. This corroborates our viewpoint that our iterative procedure converges towards the extremal solution.

\begin{figure}[!t]
\centering
\includegraphics[width=\textwidth]{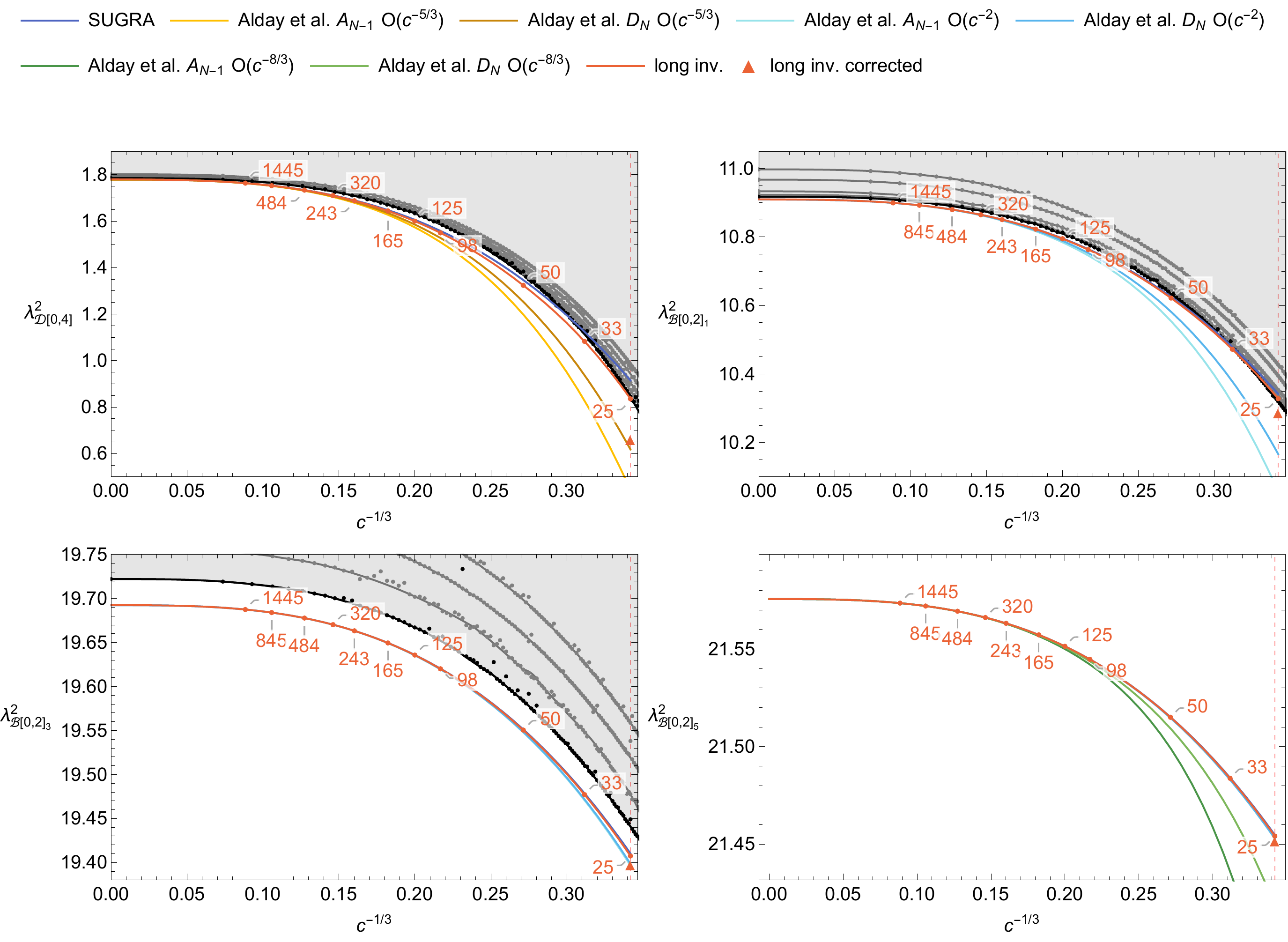}
\caption{OPE coefficients of non-chiral algebra short multiplets with spin 0, 2, 4, 6, against $c^{-1/3}$. The gray shaded regions are numerical upper bounds (available for spin 0, 2, 4) from \cite{Beem:2015aoa}. The red numbers are central charges (a strict subset of the chosen values correspond to physical theories). Short inversions produce the same results as the supergravity ones. The red curves correspond to stabilized iterative long inversions up to spin 16 (see text for detailed explanation). For $c=25$, we also perform a corrected (triangular point) iterative long inversion using numerical data from \cite{Beem:2015aoa}. We also show higher order $c^\a$ corrections (denoted as $O(c^\a)$ in the plot legend) from \cite{Chester:2018dga,Alday:2020tgi} whenever available.}
\label{fig: OPE short at bound}
\end{figure}

\subsection{OPE coefficients of non-chiral algebra short multiplets}
In figure \ref{fig: OPE short at bound} we present estimates for the OPE coefficients of the short multiplets not fixed by the chiral algebra,  for the first few lowest spins and as functions of $c^{-1/3}$.\footnote{The numerical bounds displayed appear jittery due to ``failed searches'' which occur because the numerical precision used is barely sufficient to obtain these bounds as noted in \cite{Beem:2015aoa}.} Here the short inversion produces the same results as the one from supergravity and therefore we only need to show the long inversion results. We also show the results from \cite{Alday:2020tgi} whenever applicable. The relative normalizations between that work and ours are as follows:
\begin{align}
\frac{\left(\lambda^{\text{there}}\right)_{\DD[0,4]}^2}{\left(\lambda^{\text{here}}\right)_{\DD[0,4]}^2}=\frac{3}{8},\qquad
\frac{\left(\lambda^{\text{there}}\right)_{\BB[0,2]_{\ell-1} }^2}{\left(\lambda^{\text{here}}\right)_{\BB[0,2]_{\ell-1} }^2}=\frac{2^{-\ell-1} \ell (\ell+3)}{(\ell+1)(\ell+4)}\,.
\end{align}
Here we find only marginal improvement over the supergravity answers.

Note that for $A_1$ theory ($c=25$), $\lambda^2_{\DD[0,4]}$ should vanish \cite{Beem:2015aoa}, but we were not able to recover this result from any estimate that involves only a single trajectory. Indeed, if we use (i) the numerical upper bounds on $\D_{0,2,4}$, $\lambda^2_{\LL[0,0]_{\ell=0,2,4}}$ from \cite{Beem:2015aoa}, (ii) the assumption that large spin perturbation theory is reliable for $\ell \geqslant 6$, and (iii) convexity of the leading long trajectory then one cannot recover $\lambda^2_{\DD[0,4]} = 0$ at $c = 25$ by just inverting the data on the leading Regge trajectory. The difficulty of recovering this vanishing OPE coefficient is also confirmed by an analysis we present in appendix \ref{app:crudeestimates}, which shows that we need unrealistically large OPE coefficients if we only use a few $t$-channel blocks.\footnote{On the other hand, as explained in subsection \ref{subsubsec:shortsatbound}, the fact that $\lambda^2_{\DD[0,4]} = 0$ translates into a specific sum rule for the other OPE data and one can ask whether this datum can be used as an input in order to improve our other estimates for this theory. We leave this as an interesting open question for future work.} 

Finally, let us mention that a similar negative result holds for the recovery of the stress tensor multiplet. Recall that we argued in section \ref{sec:reggetrajectoriesAR} that the leading unprotected Regge trajectory should cross the stress tensor point at $(\Delta,\ell) = (2,-2)$, which would correspond to an anomalous dimension equal to 4. In practice we observed that the finite $y$ estimates rapidly stopped being sensible already at smaller negative spins: for example, we find a zero in the function $y(z,h,\hb)$ which leads to a singularity in $h_{y_0}(h,\hb)$. It is then not sensible to try even lower spins without adding many more $t$-channel blocks.
\section{Outlook}
\label{sec:conc}
We have explored the consequences of analyticity in spin for the six-dimensional $(2,0)$ theories and found an interesting interplay between supersymmetry and Regge trajectories. Some numerical experiments allowed us to approximately bootstrap the four-point function of the stress tensor multiplet. Let us mention a few possibilities for further explorations.

First, our numerical experiments can be extended. We can certainly consider other correlators or other theories, in different dimensions and with varying amounts of supersymmetry. Other possibilities include the incorporation of subleading Regge trajectories for the correlator at hand, or perhaps a multi-correlator study to improve the estimates. A related direction is the incorporation of known results into the numerics. For example, for the leading trajectory we claim to know both its location and coefficient at spin $-2$ and it would be nice to somehow use this information.

It would be especially nice if we could use the $c^{-5/3}$ results because it would allow us to distinguish between the $A$- and $D$-type theories at large $c$. Perhaps this can be done by finding different fixed points for very large $c$ and then tracing them as $c$ gradually decreased.

Second, it would be interesting to have a better idea of the Regge trajectories in supersymmetric theories for negative or non-integer spins. There should also be supersymmetric versions of the light-ray operators of \cite{Kravchuk:2018htv} and it would be interesting to study their properties. Also, is there a more direct argument for the improved Regge behavior (and larger analyticity in spin) in supersymmetric theories or do we have to analyze individual multiplets for each correlator separately? Furthermore, the softer Regge behavior of supersymmetric correlators means the dispersion relations of \cite{Carmi:2019cub,Caron-Huot:2020adz}, which reconstruct a correlator from its double-discontinuity should apply directly without any subtractions. It would be interesting to consider the convergent sum-rules of \cite{Caron-Huot:2020adz} in the case of the $\NN=(2,0)$ theories, and to solidify our understanding of the Regge intercept by resolving the questions we discussed in section \ref{sec:inversion}.

\subsubsection*{Different amounts of supersymmetry and different dimensions}

Much of what was discussed here in the context of six-dimensional $\NN=(2,0)$ SCFTs is not unique to these theories. Four-point functions of BPS operators often enjoy an extended analyticity in spin due to  a softer Regge behavior of the correlator that is inverted. Roughly, in dimensions $d>2$ one expects analyticity in spin for $\ell > \ell^\star$, with $\ell^\star = 1- \frac{d-2}{2}\mathcal{N}$ for $d \neq 5$, and $\ell^\star=-1$ for $d=5$, from the fact that supersymmetry relates conformal primaries of spin one to conformal primaries with spin $\ell^\star$.\footnote{This can be seen from the structure of the superconformal multiplets in \cite{Cordova:2016emh}. However, a precise statement requires an analysis of the Regge behavior of the correlator whose conformal block decomposition is being inverted. As far as we know this has been only done in four dimensions for chiral operators in the non-chiral channel \cite{Cornagliotto:2017snu}, and $\NN=4$ half-BPS operators \cite{Caron-Huot:2018kta}.}
As a result, Regge trajectories of the superprimaries exchanged in such OPEs are more constrained than their bosonic counterparts, with analyticity in spin imposing constraints on short and long trajectories. We leave a detailed study of the interplay between analyticity in spin and supersymmetry for future work, and simply conclude with a few general comments on how the stress tensor fits into superconformal primary Regge trajectories in interacting theories.

In the case of SCFTs with more than eight supercharges we expect the stress tensor to fit into the leading long trajectory at a \emph{negative spin}, similarly to what was discussed here. For maximally supersymmetric theories it corresponds to the spin $-2$ continuation of the leading long superprimary trajectory, which should have dimension $0$ ($-1$) in four (three) dimensions. This follows from examining the half-BPS superblocks for the stress tensor supermultiplet correlator given in, \eg, \cite{Dolan:2003hv,Nirschl:2004pa,Dolan:2004mu,Beem:2016wfs,Chester:2014fya}, and we expect this structure to hold for other half-BPS correlators.
This fits well with the results for planar $\NN=4$ Super-Yang-Mills, where complete Regge trajectories can be obtained numerically, for any value of the 't Hooft coupling, from the quantum spectral curve approach \cite{Gromov:2015wca}. The Regge trajectory obtained in \cite{Gromov:2015wca} is that of the leading unprotected single-trace conformal primary operators, in the $\mathbf{20}'$ R-symmetry representation, that appear in the stress tensor superprimary's self-OPE.\footnote{Note, however, that the leading singe-trace trajectory is not necessarily the leading Regge trajectory of the full non-perturbative CFT. This trajectory is also analyzed in \cite{Costa:2012cb}.} For spin greater than two, the operators in this trajectory are  genuinely unprotected long operators, but at spin zero one finds the superprimary of the stress tensor multiplet itself. As pointed out in \cite{Gromov:2015wca}, the fact that the spin is an even function of the dimension, for this Regge trajectory, follows from shadow symmetry of the Regge trajectories in the $\mathbf{105}$ R-symmetry channel. We expect shadow symmetry in this channel (and all others) will follow from the structure of the superconformal blocks, as we observed in section \ref{subsec:shadowsy} for the $(2,0)$ theories.

Just like in the $\NN=(2,0)$ case, the stress tensor supermultiplet OPE in $\NN=4$ features various short multiplets whose OPE coefficients are completely fixed from the chiral algebra of \cite{Beem:2013sza}, see \cite{Beem:2016wfs} for details. As argued in \cite{Caron-Huot:2018kta}, the superprimary Regge trajectories are analytic for all spins greater than $-3$, and we expect other connections between long and short trajectories to follow from imposing simultaneously analyticity in spin and supersymmetry.

Analyticity in spin demands SCFTs with eight supercharges, or less, have a structure of superconformal Regge trajectories that looks closer to the non-supersymmetric case -- the stress tensor supermultiplet should fit at \emph{non-negative} spin in the leading long trajectory. For SCFTs with eight supercharges, we expect that the stress tensor superprimary (a scalar and singlet \cite{Cordova:2016emh}) should fit into the leading super-Regge trajectory, which will have the first \emph{unprotected long} operator at spin two. The stress tensor itself would fall in the leading Regge conformal primary trajectory, which has the first unprotected operator at spin four. At the level of kinematics this expectation can be checked from the known superblocks of half-BPS flavor current multiplets which were written down explicitly, for arbitrary dimensions, in~\cite{Chang:2017xmr,Bobev:2017jhk}.
The chiral algebra of \cite{Beem:2013sza} provides a check at the level of \emph{dynamics} for the four-point function of flavor current multiplets in four-dimensional $\NN=2$ SCFTs. Requiring the stress tensor to fit into the leading unprotected Regge trajectory fixes the OPE coefficient and scaling dimension of this trajectory at zero spin -- see the blocks given in \cite{Dolan:2003hv,Beem:2014zpa}. These values turn out to be precisely what is needed to ensure analyticity down to spin zero of the function $(z \zb)^2 \GG(z,\zb)$ defined in \cite{Beem:2014zpa}, which admits a block decomposition and has a softer Regge growth (ensuring analyticity for $\ell>-1$) similarly to $(z \zb)^6 a(z,\zb)$ here. Once again, more interconnections between short and long trajectories will likely follow from analyticity in spin.


\acknowledgments

We are grateful for discussions with Chris Beem, Eric Perlmutter and Leonardo Rastelli which initiated this work, and for their contributions in the early stages of this project. We would also like to thank Simon Caron-Huot and Marco Meineri for discussions.

BvR and XZ are supported by the Simons Foundation grant \#488659 (Simons Collaboration on the non-perturbative bootstrap). ML is supported in part by STFC grant ST/T000708/1. ML also acknowledges FCT-Portugal under grant PTDC/MAT-OUT/28784/2017. Part of this work was carried out at  Perimeter Institute for Theoretical Physics during the ``Bootstrap 2019'' conference. Research at Perimeter Institute is supported by the Government of Canada through Industry Canada and by the Province of Ontario through the Ministry of Economic Development \& Innovation. Part of this work was also performed at the Aspen Center for Physics, which is supported by National Science Foundation grant PHY-1607611.

\appendix
\section{(Super)conformal blocks and projectors}
\label{app:superblock}
\label{app:Rsym}

We consider the four-point function of the superprimary of the stress tensor multiplet $\Phi^{\{IJ\}}(x)$, where $I\,,J$ are fundamental $so(5)$ indices. Contracting the $\so(5)$ indices with null vectors $Y^I$ as $\Phi(x,Y)= Y_I Y_J \Phi^{\{IJ\}}(x)$, we can write the four-point function as
\begin{equation}
\langle \Phi(x_1,Y_1) \Phi(x_2,Y_2) \Phi(x_3,Y_3) \Phi(x_4,Y_4) \rangle = 16 \frac{\left(Y_1 \cdot Y_2\right)^2 \left(Y_3 \cdot Y_4\right)^2}{x_{12}^8 x_{34}^8} \sum_R A_{R}(z,\bar z) P^R(\a,\bar{\a})\,,
\end{equation}
with
\be
\begin{aligned}
\frac{1}{\a \bar \a} &:= \frac{\left(Y_1 \cdot Y_2\right) \left(Y_3 \cdot Y_4\right)}{\left(Y_1 \cdot Y_3\right)\left(Y_2 \cdot Y_4\right)}~, \qquad  &\frac{(\a -1)(\bar \a - 1)}{\a \bar \a} &:= \frac{\left(Y_1 \cdot Y_4\right)\left(Y_2 \cdot Y_3\right)}{\left(Y_1 \cdot Y_3\right)\left(Y_2 \cdot Y_4\right)}\,,\\
 \quad z \zb &:= \frac{x_{12}^2 x_{34}^2}{x_{13}^2 x_{24}^2} \,, \qquad &(1-z)(1-\zb)&:=\frac{x_{14}^2 x_{23}^2}{x_{13}^2 x_{24}^2}\,. 
\end{aligned}
\ee
Here $P^R$ are the projectors onto each of the representations $R$ appearing in the tensor product of two $[2,0]$ representations in \eqref{eq:tensorprod} and are given by \cite{Dolan:2004mu}
\be 
\begin{split}
Y^{[4,0]}(\alpha , \bar \alpha) &=(\alpha \bar \alpha)^2 +(\alpha -1 )^2(\bar \alpha - 1)^2 + 4 \alpha \bar \alpha  (\alpha -1 )(\bar \alpha - 1) -\frac{8 (\alpha \bar \alpha +(\alpha -1 )(\bar \alpha - 1) )}{9}+\frac{8}{63}\,,\\
Y^{[2,2]}(\alpha , \bar \alpha) &= (\alpha \bar \alpha)^2 -(\alpha -1 )^2(\bar \alpha - 1)^2 -\frac{4 (\alpha \bar \alpha -(\alpha -1 )(\bar \alpha - 1) )}{7}\,,\\
Y^{[0,4]}(\alpha , \bar \alpha) &= (\alpha \bar \alpha)^2 +(\alpha -1 )^2(\bar \alpha - 1)^2 - 2 \alpha \bar \alpha  (\alpha -1 )(\bar \alpha - 1) -\frac{2 (\alpha \bar \alpha +(\alpha -1 )(\bar \alpha - 1) )}{3}+\frac{1}{6}\,,  \\
Y^{[0,2]}(\alpha , \bar \alpha) &= \alpha +\bar \alpha -1\,,\\
Y^{[2,0]}(\alpha , \bar \alpha) &= \alpha \bar \alpha+(\alpha -1 )(\bar \alpha - 1) -\frac{2}{5}\,, \\
Y^{[0,0]}(\alpha , \bar \alpha) &= 1 \,.
\end{split}
\ee

\label{app:confblock}
We denote the six-dimensional (non-supersymmetric) conformal blocks appearing the de decomposition of a four-point function of operators with dimensions $\Delta_{i=1,\ldots4}$ by $\GG_{\D}^{(\ell)}(\D_{12},\D_{34};z,\bar z)$, where $\Delta_{ij}= \D_i-\D_j$. Here and throughout the paper we omit the first two arguments ($\Delta_{12}\,,$ $\Delta_{34}$) whenever they are vanishing. Their explicit form reads \cite{Dolan:2003hv,Dolan:2011dv}	\footnote{With respect to \cite{Dolan:2003hv,Dolan:2011dv,Beem:2015aoa} we removed an overall factor of $(-1)^\ell$.}
\be 
\begin{split}
\GG_{\D}^{(\ell)}(\D_{12},\D_{34};z,\bar z) &= \FF_{00} - \frac{\ell+3}{\ell+1} \FF_{-1 1} + \frac{(\Delta-4)(\ell+3)}{16(\Delta-2)(\ell+1)} \\
& \frac{(\Delta-\ell-\Delta_{12}-4)(\Delta-\ell+\Delta_{12}-4)(\Delta-\ell+\Delta_{34}-4)(\Delta-\ell-\Delta_{34}-4)}{(\Delta-\ell-5)(\Delta-\ell-4)^2(\Delta-\ell-3)}\FF_{02} \\
&- \frac{\Delta -4 }{\Delta -2}\frac{(\Delta+ \ell-\Delta_{12})(\Delta+\ell+\Delta_{12})(\Delta+\ell+\Delta_{34})(\Delta+\ell-\Delta_{34})}{16 ( \Delta+ \ell-1)(\Delta+\ell)^2(\Delta+\ell+1)} \FF_{11} \\
&+\frac{2( \Delta-4)(\ell+3)\Delta_{12} \Delta_{34}}{(\Delta+\ell)(\Delta+\ell-2)(\Delta-\ell-4)(\Delta-\ell-6)} \FF_{01}\,,
\end{split}
\label{eq:6dconfblock}
\ee
where
\be
\begin{split}
\FF_{nm}(z, \bar{z}) &= \frac{(z \bar z)^\frac{\Delta-\ell}{2}}{(z-\bar{z})^3}\left(\left(\frac{z}{2}\right)^\ell z^{n+3} \bar{z}^{m} {}_2F_1\left(\frac{\Delta +\ell-\Delta_{12}}{2}+n,\frac{\Delta +\ell+\Delta_{34}}{2}+n,\Delta+\ell+2n,z \right) \right. \\
&\left. {}_2F_1\left(\frac{\Delta -\ell-\Delta_{12}}{2}-3+m,\frac{\Delta -\ell+\Delta_{34}}{2}-3+m,\Delta-\ell-6+2m,\bar z \right) - \left( z \longleftrightarrow \bar{z} \right)\right)\,.
\end{split}
\label{eq:Ffor6dconfblock}
\ee
\section{\texorpdfstring{Regge bound of $a(z,\zb)$}{Regge bound of a(z,zb)}}
\label{app:reggebound}

In this appendix we show that the function $a(z,\zb)$ is bounded in the Regge limit as \eqref{eq:ainRegge}. This limit corresponds to going around the branch cut starting at $\zb=1$ and sending $w$, defined in \eqref{eq:Reggelimit}, to zero along any direction in the complex-$w$ plane.\footnote{Compared to \cite{Caron-Huot:2017vep} we have re-scaled $w$ such that $4 w_{\text{there}} = w_{\text{here}}$.} Following \cite{Caron-Huot:2017vep}, we do so by bounding the function on the secondary sheets (obtained by going around the branch cut in either direction according to the phase of $w$) by its (positive) value on the first sheet with real values of the cross-ratios.

The function $a(z,\zb)$ admits a decomposition in powers of $z$ and $\zb$ following from its $s$-channel OPE, given in equations \eqref{eq:achiplusu}, \eqref{eq:achiinblocks} and \eqref{eq:auinblocks}, as
\begin{equation}
(z \zb)^6 a(z,\zb) = \sum\limits_{\Delta'\geqslant 8 + |\ell|,\ell} b_{\Delta', \ell}	\, z^{\frac{\Delta'- \ell}{2}} \zb^{\frac{\Delta'+ \ell}{2}}\,,
\label{eq:s-chan}
\end{equation}
where $\Delta'\geqslant 8 + |\ell|$ follows from the decomposition of $a(z,\zb)$ in the blocks given in eq.~\eqref{eq:ahatom}, with the prime in $\Delta$ to remind us the sum runs over both conformal primaries and descendants, and where $\ell$ runs over positive and negative spins of any parity.
To bound $a(z,\zb)$ on the second sheet we now need to show the coefficients $b_{\Delta',\ell}$ are non-negative. The function $a(z,\zb)$ is only related to a physical correlator through the inverse of the differential operator $\Delta_2$ as given in equation \eqref{eq:diffops}, so we do not know of a direct way to use reflection positivity to show positivity of the coefficients $b_{\Delta',\ell}$. However, reflection positivity would also follow from positivity of the coefficients of the decomposition of $a(z,\zb)$ in blocks as given in \eqref{eq:achiinblocks} and \eqref{eq:auinblocks}, so all that remains is to show the blocks $a^\text{at}_{\Delta,\ell}(z,\bar z)$ in \eqref{eq:ahatom} have a positive decomposition in powers of $z$ and $\zb$. We have checked this to be the case for all relevant scaling dimensions and for various spins in a series expansion in small $z$ and $\zb$, but do not have a proof, due to the convoluted form of six-dimensional conformal blocks \eqref{eq:6dconfblock}. We will proceed using the $z$ and $\zb$ coordinates and not the $\rho$ variable of \cite{Hogervorst:2013sma} since the expansion of $z$ and $\zb$ in powers of $\rho$ and $\bar{\rho}$ 
is not positive.\footnote{While $(z \zb)^6 a(z,\zb)$ does not have an expansion in positive powers of $\rho$ and $\bar{\rho}$ it is still possible that $(z \zb)^6 a(z,\zb)/(1-z)/(1-\zb)$ does, which given the form of the crossing equations would be a sufficient condition. However, due to the convoluted form of the blocks, we could only check this to a rather low order in the $\rho$ and $\bar{\rho}$ expansion. If the expansion is indeed positive, then following the same reasoning as below would yield the bound \eqref{eq:ainRegge} for all values of $\arg(w)$.}

\subsubsection*{Bounding $a(z,\zb)$}

The expansion \eqref{eq:s-chan} converges for $|z|,\,|\zb|<1$.  Using the second crossing symmetry equation \eqref{eq:crosssym} we can use instead the $t$-channel decomposition to compute
\begin{equation}
a(z,\zb) = \frac{1}{((1-z)(1- \zb))^5 z \zb}\sum\limits_{\Delta'\geqslant 8 + |\ell|,\ell} b_{\Delta', \ell} (1-z)^{\frac{\Delta'-\ell}{2}} (1-\zb)^{\frac{\Delta'+ \ell}{2}} + \frac{\CC_h(1-z,1-\zb) - \CC_h(z,\zb)}{z \zb}\,,
\end{equation}
where the sum now converges for $|1-z|,\, |1-\zb|<1$. This representation of $a(z,\zb)$ allows us to continue to the secondary sheets by going around the cut starting at $\zb=1$, provided $|1-\zb|<1$. For $|1-\zb|>1$ we will use the $u$-channel below.
Going around the branch cut amounts to introducing phases in the scaling-block expansion above, and we get
\be 
\begin{split}
|\asecond(z,\zb)| &\leqslant \frac{|1-z||1-\zb|}{|z \zb|} \frac{z' \zb'}{(1-z')(1-\zb')} a(z', \zb') + \left|\frac{\Csecond_h(1-z,1-\zb) - \Csecond_h(z,\zb)}{z \zb}\right| \\
&- \frac{\CC_h(1-z',1-\zb') - \CC_h(z',\zb')}{(1-z') (1-\zb')} \frac{|1-z||1-\zb|}{|z \zb|}\,,
\label{eq:asecondt}
\end{split}
\ee
where we used the second crossing equation again, and defined
\begin{equation}
1-z' \equiv |1-z| <1\,, \qquad 1-\zb' \equiv |1-\zb| <1
\end{equation}
We can now take the limit of $w \to 0 $, using
\begin{equation}
z \sim \sigma w\,,\qquad
\zb \sim \frac{1}{\sigma} w\,, \qquad
z' \sim \sigma |w| \cos(\arg(w))\,,\qquad
\zb' \sim \frac{1}{\sigma} |w| \cos(\arg(w))\,.
\end{equation}
The behavior of $a(z', \zb')$ as $w \to 0 $ is controlled by the lowest dimensional operator appearing in its decomposition. which is the superprimary of the $\DD[4,0]$ supermultiplet with $\Delta'=8$, while the behavior of the remaining terms can be found from the explicit form of  $\CC_h(z,\zb)$ given in eq.~\eqref{eq:Ch}. All in all we find
\be 
|\asecond(z,\zb)| \leqslant  \frac{\sec ^6(\arg(w) )+1}{3 |w|^8}\,, \quad \text{as } w \to 0\,,
\ee
where the leading contribution comes from the terms involving $\CC_h$ only. This bound is valid for $|\arg(w)| <\pi/2$ which corresponds to the region of convergence of the $t$-channel decomposition. To obtain a bound for $|\arg(w)| > \pi/2$ we now turn to a $u$-channel decomposition.

Combining both equations in \eqref{eq:crosssym} we write now an expansion valid for $|1-z|,\, |1-\zb|>1$
\be 
\begin{split}
a(z,\zb) &= \frac{(1-z)(1-\zb)}{z \zb} \sum\limits_{\Delta'\geqslant 8 + |\ell|,\ell} b_{\Delta', \ell} \left(\frac{1}{1-z}\right)^{\frac{\Delta'-\ell}{2}} \left(\frac{1}{1-\zb}\right)^{\frac{\Delta'+ \ell}{2}} \\
&+ \frac{1}{z(1-z)^4}\frac{1}{\zb(1-\zb)^4} \left(\CC_h\left(\frac{1}{1-z},\frac{1}{1-\zb}\right) - \CC_h\left(\frac{z}{z-1},\frac{\zb}{\zb-1}\right)\right)\,.
\end{split}
\ee
Going around the branch-cut starting at $\zb=1$ in either direction once again introduces just phases in the expansion, while $\CC_h$ is a known function, and so we write
\be 
\begin{split}
|\asecond(z,\zb)| &\leqslant \frac{(1-z')^5 (1-\zb')^5 (z' \zb')}{|1-z|^5 |1-\zb|^5 |z \zb|} a(z', \zb') \\
&+ \left|\frac{1}{z(1-z)^4}\frac{1}{\zb(1-\zb)^4} \left(\Csecond_h\left(\frac{1}{1-z},\frac{1}{1-\zb}\right) - \Csecond_h\left(\frac{z}{z-1},\frac{\zb}{\zb-1}\right)\right)\right| \\
&- \left(\CC_h\left(\frac{1}{1-z'},\frac{1}{1-\zb'}\right) - \CC_h\left(\frac{z'}{z'-1},\frac{\zb'}{\zb'-1}\right)\right) \frac{(1-z')^6(1-\zb')^6}{|1-z|^5 |1-\zb|^5 |z \zb|}\,.
\label{eq:asecondu}
\end{split}
\ee
Just as before we find that the Regge behavior of $a(z,\zb)$ is bounded by 
\be 
|\asecond(z,\zb)| \leqslant  \frac{\sec ^6(\arg(w) )+1}{3 |w|^8}\,, \quad \text{as } w \to 0\,,
\ee
thus showing eq.~\eqref{eq:ainRegge} for $|\arg(w)|\neq \pi/2$.
 
\subsubsection*{Bounding $A_R(z,\zb)$}

Finally, we need to bound the behavior of the $A_R(z,\zb)$ in the Regge limit. These functions are obtained from  $a(z,\zb)$ and its first $z$ and $\zb$ derivatives through equation \eqref{eq:Ainah}. Taking a derivative with respect to $z$ and/or $\zb$ in equation \eqref{eq:s-chan} does not spoil positivity of the expansion coefficients. We can then can repeat the above computation bound directly $\partial_z a(z,\zb)$, $\partial_\zb a(z,\zb)$ and $\partial_\zb \partial_z a(z,\zb)$. The bounds obtained then imply the Regge behavior of $A_R(z,\zb)$ quoted in \eqref{eq:AinRegge}, for $|\arg(w)|\neq \pi/2$.

\section{\texorpdfstring{Regulating the divergence of $c(\D,\ell)$ near $z\to1$}{Regulating the divergence of c(D,l) near z -> 1}}
\label{app:regulation}
In this appendix we discuss the regularization of integrals in the Lorentzian inversion formula \eqref{eq:inversionformula} and explicitly work out the case of four-dimensional mean field theory as an example.

The integrals over $z$ (and $\zb$) in $c(\D,\ell)$ diverge when the integrand scales in the limit $z\to1$ as $O\left((1-z)^{-p}\right)$ with $p\geqslant 1$. In general this divergence can be resolved by analytic continuation of $p$ from the convergent region $p<1$ to its actual value, but when $p$ is an integer such analytic continuation fails because of factors like $\G(-p+n)$ where $n$ is a positive integer. In the $\zb$ variable there are additional sine functions from the double discontinuity operation and they produce compensating zeros to yield finite answers; this is why we encountered no divergences in the inversion of the $z^4$ term that we performed in section \ref{subsec: recovering shorts}. This leaves us with the divergence from the $z$-integral when $p$ is a large enough integer. As was already discussed in \cite{Caron-Huot:2017vep,Simmons-Duffin:2017nub}, one can either regulate the $z$-integral by setting a cutoff $1-\epsilon$ and then drop the divergent terms in the limit $\epsilon\to0$; or one can keep $p$ generic for integration, then series expand $p$ around the actual integral value and discard divergent terms. Here we apply the second approach.

Let us consider the split of the double discontinuity as in equation \eqref{ddiscfromtchannelstuff}. It is easy to verify that the contribution from the long multiplets always converges, but the contribution from the short multiplets behaves as $(1-z)^{-4}$ and needs to be regulated. The way to do so is to write $c(\D,\ell)$ as 
\begin{align}
c(\D,\ell)=\frac{c^{(-2)}(\D,\ell)}{(p-4)^2}+\frac{c^{(-1)}(\D,\ell)}{p-4}+c^{(0)}(\D,\ell)+\mathcal{O}(p-4)\,,
\end{align}
and then simply drop $c^{(-2)}(\D,\ell)$ and $c^{(-1)}(\D,\ell)$.

Two follow-up checks are needed: first, $c^{(0)}(\D,\ell)$ should give the correct residues at physical poles; second, the subtracted parts which are (the principal series integral of) $c^{(-2)}(\D,\ell)$ and $c^{(-1)}(\D,\ell)$ should not have a conformal block decomposition to make sure that nothing physical is subtracted.

The first check is straightforward. We simply calculate $c^{(0)}(\D,\ell)$ for the short multiplets and find that the residues at twist 8 and 10 in the sense of $(z \zb)^6 a(z,\zb)$ match \eqref{eq:chiral algebra OPE coef} and \eqref{eq:OPE short and bound from shorts} respectively. The second check is however technically more involved because conformal blocks in 6d are complicated functions. Therefore we will instead do a simpler check to illustrate the essential idea. 

\subsection*{Example: mean field theory in 4d}
Let us now consider the four-point function $\langle\phi\phi\phi\phi\rangle$ of mean field theory in 4d. We first use the inversion formula to reproduce the CFT data, and then focus on the cases with integral external dimensions. In this section the normalization of the conformal blocks is the same as that in \cite{Caron-Huot:2017vep}.

For MFT we only need to invert the $t$-channel identity operator, and we get
\begin{align}
\begin{split}
c(\D,\ell)&=
\frac{1+(-1)^\ell}{4} \kappa_{\Delta+\ell} \int\limits_0^1 dz d\zb \mu(z,\zb) \GG_{\ell+3}^{(\Delta -3)}(z,\zb) 
\mathrm{dDisc}\left[ \left(\frac{z \zb}{(1-z)(1-\zb)}\right)^{\D_\phi} \right]\\
&=
\frac{(1+(-1)^\ell)(\Delta -2) (\ell+1)}{4 \pi ^2 }
\sin ^2(\pi  \D_\phi)\Gamma (1-\D_\phi)^2 \Gamma (2-\D_\phi)^2\\
&\quad\times
\frac{
\Gamma \left(\frac{1}{2}(\D+\ell)\right)^2 
\Gamma (-\Delta+\ell+4) 
\Gamma \left(\frac{1}{2}(-\Delta+\ell+2\D_{\phi})\right) 
\Gamma \left(\frac{1}{2}(-4+\Delta+\ell+2\Delta_{\phi})\right)}
{
\Gamma\left(\frac{1}{2} (-\Delta+\ell+4)\right)^2 
\Gamma (\D+\ell-1) 
\Gamma \left(\frac{1}{2} (-\Delta+\ell-2 \Delta_{\phi}+8)\right) 
\Gamma \left(\frac{1}{2} (\Delta+\ell-2 \Delta_{\phi}+4)\right)
}\,.
\end{split}
\label{eq: MFT c Delta ell}
\end{align}
It it straightforward to check that when $\ell$ is an even integer, the residue of $-c(\D,\ell)$ at $\D=2\D_\phi+\ell+2n$ indeed gives the OPE coefficient of MFT given in \cite{Fitzpatrick:2011dm}. 

When $\D_\phi\in\mathbb{Z}_{\geqslant2}$ it can be seen from the second line of \eqref{eq: MFT c Delta ell} that $c(\D,\ell)$ is divergent, since there is a fourth-order pole and only a second-order zero. To regulate $c(\D,\ell)$ for any integer $\D_\phi$ we expand
\begin{align}
c(\D,\ell;\D_{\text{ex}})=
\frac{c^{(-2)}(\D,\ell;\D_\phi)}{(\D_{\text{ex}}-\D_{\phi})^2}
+\frac{c^{(-1)}(\D,\ell;\D_\phi)}{\D_{\text{ex}}-\D_{\phi}}
+c^{(0)}(\D,\ell;\D_\phi)
+O(\D_{\text{ex}}-\D_\phi)\,,
\end{align}
where we have kept the dependence on external dimension explicit. The residues are
\begin{align}
\begin{split}
&c^{(-2)}(\D,\ell;\D_\phi)
=
\frac{(1+(-1)^\ell)(\Delta-2)(\ell+1)}{2\Gamma(\Delta_{\phi}-1)^2\Gamma(\Delta_{\phi})^2 }
\\
&\qquad\times
\frac{  
\Gamma \left(\frac{1}{2}(\D+\ell)\right)^2 
\Gamma (-\Delta+\ell+4) 
\Gamma \left(\frac{1}{2}(-\Delta+\ell+2\D_{\phi})\right) 
\Gamma \left(\frac{1}{2}(-4+\Delta+\ell+2\Delta_{\phi})\right)}
{\Gamma\left(\frac{1}{2} (-\Delta+\ell+4)\right)^2 
\Gamma (\D+\ell-1) 
\Gamma \left(\frac{1}{2} (-\Delta+\ell-2 \Delta_{\phi}+8)\right) 
\Gamma \left(\frac{1}{2} (\Delta+\ell-2 \Delta_{\phi}+4)\right)}\,,
\end{split}
\label{eq:c_minus2}
\end{align}
and
\begin{align}
\begin{split}
c^{(-1)}(\D,\ell;\D_\phi)=
&\bigg[-2 \psi(\D_{\phi}-1)-2 \psi(\D_{\phi})
+\psi\left(\frac{-\Delta+\ell+2\D_{\phi}}{2}\right)
+\psi\left(\frac{-4+\Delta+\ell+2\D_{\phi}}{2}\right)\\
&\quad+\psi\left(\frac{-\Delta+\ell-2 \D_{\phi}+8}{2}\right)
+\psi\left(\frac{\Delta+\ell-2 \D_{\phi}+4}{2}\right)\bigg]
\times c^{(-2)}(\D,\ell;\D_\phi)\,.
\end{split}
\end{align}

{\renewcommand{\arraystretch}{2.2}%
\begin{table}[t!]
	\begin{center}{}
	\begin{tabular}{c |c| c}
		{\renewcommand{\arraystretch}{1}%
		\begin{tabular}{c}
		poles in ${\D}$\\
		$(n=0,1,2,\ldots)$
		\end{tabular}
		}& origin & comment   \\
		\hline
		\hline
		$\D=-\ell-2n$
		& ${\Gamma \left(\frac{1}{2}(\D+\ell)\right)^2 }$ & 
		{\renewcommand{\arraystretch}{1}%
		\begin{tabular}{c}
		$\k$-factor poles,\\
		not picked up
		\end{tabular}
		}
		\\
		$\D=\ell+5+2n$ & ${\frac{\G(-\D+\ell+4)}{\G\left(\frac{1}{2} (-\Delta+\ell+4)\right)^2 }}$ & 
		{\renewcommand{\arraystretch}{1}%
		\begin{tabular}{c}
		kernel block poles,\\
		canceled by block poles
		\end{tabular}
		}
		\\
		${\D=2\D_\phi+\ell+2n}$ & ${\Gamma \left(\frac{1}{2}(-\Delta+\ell+2\D_{\phi})\right) }$ & 
		{\renewcommand{\arraystretch}{1}%
		\begin{tabular}{c}
		double-twist poles,\\
		zero residue
		\end{tabular}
		}
		\\
		${4-\D=2\D_\phi+\ell+2n}$ & ${\Gamma \left(\frac{1}{2}(-4+\Delta+\ell+2\Delta_{\phi})\right)}$ & 
		{\renewcommand{\arraystretch}{1}%
		\begin{tabular}{c}
		shadow double-twist poles,\\
		not picked up
		\end{tabular}
		}
	\end{tabular}
	\end{center}
	\begin{center}
	\caption{\label{tab:poles in cminus2}Poles in $\D$ of $c^{(-2)}(\D,\ell;\D_\phi)$}
	\end{center}

\end{table}
}

Now we check the poles of $\D$ in $c^{(-2)}(\D,\ell;\D_\phi)$. There are four sets of poles and they all come from the $\G$-functions in the numerator of \eqref{eq:c_minus2}. From left to right, the first set of poles originate from the $\k$-factor. In general this factor reads
\begin{align}
\k_{\D+\ell}^{\D_{12},\D_{34}}=\frac{\G(\frac12(\D+\ell+\D_{12}) \G(\frac12(\D+\ell-\D_{12}) \G(\frac12(\D+\ell+\D_{34}) \G(\frac12(\D+\ell-\D_{34})}{2\pi^2\G(\D+\ell)\G(\D+\ell-1)}\,,
\end{align}
and has poles at
\begin{align}
\D=-\ell\pm\D_{12}-2n,\quad \D=-\ell\pm\D_{34}-2n, \quad
n=0,1,2,\ldots\,.
\end{align}
As discussed in \cite{Simmons-Duffin:2017nub} these poles are on the left hand side of the $\D$-integration contour when the \emph{external} dimensions are in the principal series and therefore not picked up. When we analytically continue external dimensions to physical (and real) values, although some of the poles may move to the right of contour integration, the contour should be deformed such that the poles remain to the left and not picked up. The second set of poles are the spurious poles similar to the ones listed in the last row of table \ref{tab:poles and residues of the kernel block}. These poles will be picked up in the principal-series integration
\begin{align}
g(z,\zb) = \int_{d/2-i\infty}^{d/2+i\infty} \frac{d\D}{2\pi i}\, c(\D,\ell) \GG_{\D}^{(\ell)}(z,\zb)+(\text{non-norm.})\,,
\end{align}
but as shown in \cite{Simmons-Duffin:2017nub} they will be canceled exactly by the poles of the block $\GG_{\D}^{(\ell)}(z,\zb)$, thus nothing physical is subtracted. The third and fourth set of poles are physical double-twist poles and their shadows. One can check that the residues of the third set of poles vanish when $\D_\phi\in\mathbb{Z}_{\geq2}$ and the fourth set of poles are never picked up. We summarize these results in table \ref{tab:poles in cminus2} and note in passing that the check related to physical poles is the most important one.

Turning to $c^{(-1)}(\D,\ell)$, it is straightforward to check that it does not contain any poles other than those in $c^{(-2)}(\D,\ell)$. Therefore, we conclude that $c^{(-2)}(\D,\ell)$ and $c^{(-1)}(\D,\ell)$ do not have a block decomposition and can be safely subtracted.

\section{\texorpdfstring{Exploring the required $t$-channel contributions}{Exploring the required t-channel contributions}}
\label{app:crudeestimates}

In this appendix we explore the issues with the $A_1$ theory a bit further. In particular, we would like to know whether the vanishing of the $\DD[0,4]$ OPE coefficient at $c = 25$ can be recovered at all from a sum over $t$-channel blocks, and if so what properties such a block decomposition has. Therefore we take the rather crude ansatz where we pack the contribution of the entire unprotected part of $a(z,\zb)$ into the first few $t$-channel long blocks with some (unrealistically) large anomalous dimensions and OPE coefficients. Demanding then that $\l^2_{\DD[0,4]}=0$ might give us an idea of what this implies for the unprotected data.

Concretely we experimented with approximating the entire unprotected contribution in the $t$-channel by the following three groups of data:
\begin{align}
&\text{1 block:}  & \D_0&=6.4, & \l^2_{\LL[0,0]_0}&=11.4071, & & & & \nonumber \\
\label{eq:crude estimation data}
&\text{2 blocks:} & \D_0&=6.4, & \l^2_{\LL[0,0]_0}&=1.59017, & \D_2&=8.4, & \l^2_{\LL[0,0]_2}&=4.21943  \,,\\
&\text{2+2 blocks:} & \D_0&=6.4, & \l^2_{\LL[0,0]_0}&=1.44136, & \D_2&=8.4, & \l^2_{\LL[0,0]_2}&=3.84518, \nonumber \\
& & \D_{0'}&=8.86, & \l^2_{\LL[0,0]_{0'}}&=6.08945, & \D_{2'}&=10.86, & \l^2_{\LL[0,0]_{0'}}&=8.74061, \nonumber
\end{align}
where the first two groups consider only multiplets on the leading long trajectory while the last one also takes into account of the subleading long trajectory, which we distinguish by adding a prime in the spin. The inversion results are shown in table \ref{tab:crude estimation}. For all of these sets we have imposed $\l^2_{\DD[0,4]}=0$. From the table we can draw some qualitative conclusions: 
\begin{itemize}
	\item The OPE coefficient of the non-chiral algebra short multiplets are relatively stable across all three sets of input data.
	\item The dimensions of the long multiplets in all columns are lower than the inversion results in figure \ref{fig: long dimensions}.
	\item Distributing the contributions to double discontinuity into more blocks on the leading long trajectory lowers the dimensions and OPE coefficients of the leading long multiplets. However, notice that in the ``2 blocks'' column $\D_0$ is still higher while $\D_2$ is already lower than the numerical bootstrap's prediction $\D_0\simeq6.4,\D_2\simeq9.4$. All the OPE coefficients decrease after the redistribution.
	\item Distributing the contribution to double discontinuity into both leading and subleading long trajectories further slightly lowers $\D_0$ and increases $\D_2$, which moves the results closer to numerical predictions. From the OPE coefficients, $\l^2_{\LL[0,0]_0}$ receives the most significant change, a decrease that is also consistent with our expectation. Therefore, we again see evidence that we should include contributions from the subleading long trajectory to fully recover the unprotected CFT data.
\end{itemize}

{\renewcommand{\arraystretch}{1.5}%
\begin{table}[!ht]
	\begin{center}
	\begin{tabular}{c | c| c | c}
		  & 1 block  & 2 blocks  & 2+2 blocks \\\hline\hline
		\multirow{3}{*}{
		{\renewcommand{\arraystretch}{1}%
		\begin{tabular}{c}
		$\l^2_{\BB[0,2]_{\ell-1}}$\\
		$\ell=2,4,6$
		\end{tabular}
		 }
		}
		 & 10.1005 & 10.1009 & 10.1127 \\
		 & 19.3475 & 19.3476 & 19.3518 \\
		 & 21.4387 & 21.4387 & 21.4399 \\\hline
		\multirow{4}{*}{
		{\renewcommand{\arraystretch}{1}%
		\begin{tabular}{c}
		$\D_{\ell}$\\
		$\ell=0,2,4,6$
		\end{tabular}
		 }
		} 
		 & 7.0681 & 6.9434  & 6.9405  \\
		 & 9.4951 & 9.3667  & 9.3783 \\
		 & 11.8086 & 11.7476 & 11.7559 \\
		 & 13.9220 & 13.8990 & 13.9029 \\ \hline
		\multirow{4}{*}{
		{\renewcommand{\arraystretch}{1}%
		\begin{tabular}{c}
		$\l^2_{\LL[0,0]_\ell}$\\
		$\ell=0,2,4,6$
		\end{tabular}
		 }
		} 
		 & 1.9259 & 1.5510 & 1.5282 \\
		 & 4.1565 & 3.8376 & 3.8343 \\
		 & 5.5590 & 5.2950 & 5.3085 \\
		 & 5.4229 & 5.3212 & 5.3312 \\
	\end{tabular}
	\end{center}
	\begin{center}
	\caption{\label{tab:crude estimation}Crude estimates of unprotected CFT data of $A_1$ theory for the first few lowest spins. The results are obtained by imposing $\l^2_{\DD[0,4]}=0$ and approximating the entire unprotected part of $a(z,\zb)$ by: a single scalar block on the leading long trajectory (``1''), one block of spin $0$ and one block of spin $2$ on the leading long trajectory (``2''), and blocks of spin $0$ and spin $2$ both on the leading and subleading long trajectories (``2+2''). The input CFT data are listed in \eqref{eq:crude estimation data}.}
	\end{center}
\end{table}
}

\bibliography{./aux/biblio}
\bibliographystyle{./aux/JHEP}

\end{document}